\documentclass[pra,onecolumn,floatfix,a4paper,superscriptaddress]{revtex4}
\usepackage{bm,color,graphicx,amsmath,txfonts}
\usepackage{here}
\usepackage{array}
\usepackage{graphicx}
\usepackage[colorlinks, citecolor=blue,linkcolor=blue]{hyperref}
\usepackage{braket}
\usepackage{dsfont}
\newcommand{\e}{{e}}
\begin{document}

\title{Quantum teleportation, entanglement, LQU and LQFI in $e^{+}e^{-} \to \text{Y}\bar{\text{Y}}$ processes at BESIII through noisy channels}

%$e^{+}e^{-} \to \text{Y}\bar{\text{Y}}$ processes at BESIII
\author{Elhabib Jaloum}
\affiliation{LPTHE-Department of Physics, Faculty of Sciences, Ibnou Zohr University, Agadir, Morocco}

\author{Mohamed Amazioug}
\email{m.amazioug@uiz.ac.ma}
\affiliation{LPTHE-Department of Physics, Faculty of Sciences, Ibnou Zohr University, Agadir, Morocco}

\date{\today}

\begin{abstract}

Quantum teleportation, a protocol that has received extensive and intensive attention in quantum information processing, allows a quantum state to be transferred from one particle to another. In this study, we analytically investigate fidelity ($F$), logarithmic negativity (LN), local quantum uncertainty (LQU) and local quantum Fisher information (LQFI) as a discord-like measure of quantum correlations in $e^{+}e^{-} \to \text{Y}\bar{\text{Y}}$ processes at BESIII through noisy channels, using experimental feasible parameters, where $\text{Y}$ and $\bar{\text{Y}}$ refer to the spin-$1/2$ hyperon and its antihyperon, respectively. We discuss the dependence of LN, LQU and LQFI on the scattering angle $\theta$ for different hyperon-antihyperon pairs ($ \Lambda\bar{\Lambda}$, $\Xi^{0}\bar{\Xi^{0}}$, $\Xi^{-}\bar{\Xi^{+}}$, $\Sigma^{+}\bar{\Sigma^{-}}$), produced by electron-positron annihilation via charmonium decay. Without a dephasing effect, we show that, LN, LQU, and LQFI vanish at $\varphi=\pm\pi$ and are symmetric around $\varphi=\pi/2$. Furthermore, LN reaches its maximum when the scattering angle equals $\varphi=\pi/2$, while LQU and LQFI reach their maxima at different angles. We also explore the LN, LQU, and LQFI for different $\text{Y}\bar{\text{Y}}$ pairs subjected to three distinct types of decoherence channels. Specifically, we show that amplitude damping (AD) and phase damping (PD) lead to a decrease in LN, LQU, and LQFI with an increasing decoherence parameter $s$. In contrast, the phase flip (PF) channel exhibits symmetric behavior around $s=1/2$. Besides, we realize for teleportation, optimal fidelity for different hyperon-antihyperon pairs ($ \Lambda\bar{\Lambda}$, $\Xi^{0}\bar{\Xi^{0}}$, $\Xi^{-}\bar{\Xi^{+}}$, $\Sigma^{+}\bar{\Sigma^{-}}$). We discuss the influence of noisy channels, specifically (AD, PF and  PD), on the fidelity of quantum teleportation and on quantum correlations that can exist even beyond entanglement. Furthermore, the results show that the fidelity remains above the classical limit of $2/3$ in all three channels, even as the noise increases. This is a significant finding because it shows that not all quantum noise is detrimental. These results can have promising applications in quantum information and particle physics.\\

keywords: Teleportation; Fidelity; entanglement; LQU; LQFI; Disintegration; Hyperon-antihyperon pairs; Noisy channels.

\end{abstract}

\maketitle

\section{Introduction}    \label{sec:1}

In 1935, Albert Einstein, Boris Podolsky, and Nathan Rosen (collectively known as EPR) introduced a thought experiment that fundamentally questioned the completeness of quantum mechanics \cite{ref1}. This thought experiment, now referred to as the EPR paradox, illustrated the phenomenon of non-local correlations between quantum systems. EPR posited that a measurement performed on one particle could instantaneously influence the state of another particle, regardless of the distance separating them. In response to the challenges posed by the EPR paradox, Erwin Schrödinger introduced the concepts of entanglement and quantum steering. These ideas provided a profound framework for understanding the non-local correlations inherent in quantum mechanics \cite{ref2}. Additionally, Bell's theorem provided a crucial proof that local hidden variable theories are insufficient to describe all non-local correlations found between spatially separated quantum systems during local measurements \cite{ref3}. In 1981, Alain Aspect and his collaborators performed groundbreaking experiments using entangled photons to test Bell's inequalities \cite{Aspect81a,Aspect81b}. Their results strongly supported the predictions of quantum mechanics and demonstrated violations, thereby confirming the non-local correlations predicted by quantum entanglement. Entanglement, a purely quantum phenomenon with no classical equivalent, establishes correlations between parts of a multipartite quantum system. This unique resource is essential for numerous tasks in quantum information science \cite{ref4}.

Beyond entanglement, other forms of quantum correlations, such as non-locality without entanglement, have demonstrated utility in quantum technologies~\cite{X1,X2}. Both theoretical~\cite{X3} and experimental studies~\cite{X4} indicate that certain separable states can surpass classical states in specific computational tasks, underscoring the importance of a comprehensive investigation into various quantum correlations. Recent research has established a hierarchy of robustness against thermal noise: geometric quantum discord is more resilient than entanglement, which, in turn, is more robust than steering~\citep{J1,A1}. This makes geometric quantum discord a particularly valuable resource in noisy quantum environments. Furthermore, quantum coherence, exhibiting greater resistance to thermal noise, may offer protection for other quantum resources. A comprehensive analysis of these diverse correlations, even in gravitational analogs, underscores the inherent complexity of quantum phenomena across a broad range of physical contexts~\citep{J2}.

The concept of LQU is widely employed to quantify the quantum correlations inherent in a bipartite quantum state, serving as a discord-like measure \cite{L3}. Initially, LQU was defined for bipartite quantum systems, with an explicit formula available only in two dimensions. LQU's definition is often linked to Wigner-Yanase skew information, which quantifies the minimum uncertainty in measuring a single local observable on a quantum state \citep{X5}. Furthermore, the connection between LQU and QFI has also been explored in non-Markovian environments \cite{ref21}. Recent studies have investigated LQU under various decoherence models, with initial results reported for three-qubit systems \cite{ref21,ref23}. Importantly, in contrast to quantum discord, LQU avoids the necessity of a complex optimization procedure over measurements.

Quantum teleportation is of paramount importance in applications of quantum entanglement. By utilizing local operations, classical communication, and a shared correlated resource state, quantum information can be transmitted between two spatially separated agents, Alice and Bob. A pure, maximally entangled, two-qubit Bell state serves as the resource in the original standard teleportation protocol~\cite{ref8}. Subsequent investigations broadened the scope of the scheme by discovering that resource states beneficial for teleportation need not be maximally entangled or pure~\cite{ref9}. The link between these states and Bell inequality violations~\cite{reef10,reef11}, as well as discord-like correlations~\cite{reef12}, has since been examined.

Due to the interaction between Alice and Bob's entangled pair and its environment, which naturally induces state mixing, mixed resource states are often more realistic than pure ones. Research on noisy quantum teleportation schemes has been extensive~\cite{reef14,reef20}. Approaches include utilizing the Lindblad formalism to examine teleportation fidelity as a function of decoherence rates~\cite{reef13}, and determining the optimal Bell resource state in the presence of local Pauli noises~\cite{ref8}. Furthermore, recent advances include the development of methods to protect teleportation against certain decoherence channels~\cite{reef15,reef16}.

Studies have investigated the effect of different noisy channels on teleportation fidelity, with these channels typically modeling the interaction of Alice's and Bob's particles with local environments. Specifically, research has focused on dissipative interactions using Kraus operators for an amplitude damping channel~\cite{reef17,reef18}, as well as other common qubit noise channels such as bit flip, phase flip, depolarizing~\cite{reef19}, and phase damping~\cite{reef14}. Moreover, the teleportation protocol has also been explored in higher-dimensional systems under noisy channels~\cite{reef21}. Furthermore, from a theoretical and experimental perspective, these investigations reveal a crucial insight~\cite{reef22}: tailored channels (operating on suitable, initially pure resource states) can significantly mitigate the adverse effects of noise on teleportation fidelity, outperforming other noisy channels.

Extensions of the original teleportation protocol now encompass multi-party resource states with multipartite entanglement. This has enabled the development of strategies that utilize such entanglement to teleport multiple qubit states~\cite{reef23,reef24,reef25,reefe26,reefe27}. Research has also addressed multi-directional teleportation, allowing quantum information transfer between several agents~\cite{reefe28,reefe29,reefe30,reefe31}, including its behavior in noisy environments~\cite{reefe32,reefe33,reefe34,reefe35}.

While advancements have occurred in assessing teleportation success through noisy channels (whether acting on bipartite or multipartite resource states), the correlation between teleportation fidelity and the multipartite entanglement generated between the resource qubits and the environment warrants further investigation. This analysis aims to identify the processes (defined by the entanglement they induce) that optimize teleportation success. It would also potentially attribute fidelity improvements to the assistance of the generated multipartite entanglement. Moreover, studies in~\cite{reefe36,reefe37} have established a relationship between teleportation fidelity and the tripartite entanglement that emerges from the local interaction of a single resource qubit with a two-level environment.

Quantum entanglement has been studied using low-energy protons \cite{ref26}, while \cite{ref27} explores its potential generation at colliders by analyzing hadronic final states. Besides, quantum entanglement has also been investigated at smaller length scales and higher energies \cite{ref28}. In high-energy physics, diverse experimental approaches have been proposed to investigate quantum entanglement, including those based on neutral kaon physics 
\cite{ref29, ref30, ref31, ref32}, positronium \cite{ref33}, flavor oscillations in neutral B mesons \cite{ref34}, charmonium decays \cite{ref35}, and neutrino oscillations \cite{ref36}. Recent studies at the LHC have confirmed entanglement in top quark pair production \cite{ref39}, with \cite{ref40} additionally demonstrating the experimental feasibility of observing Bell inequality violations in this system. In this context, the investigation of quantum entanglement has been proposed across various areas of high-energy physics. These include top quark production \cite{ref41,ref42,ref43,ref44,ref45}, hyperon production \cite{ref46}, and the production of gauge bosons, both from Higgs boson decay \cite{ref47,ref48,ref49,ref50} and through direct production mechanisms \cite{ref49,ref50}.%

Recently, a novel method has emerged for studying strange baryon decays, centered on the investigation of hyperon-antihyperon pairs generated from $J/\psi$ resonances in electron-positron collisions \cite{ref51}. The angular distribution of these processes can be compactly described using real-valued matrices. These matrices represent both the initial spin-entangled state of the baryon-antibaryon pair and their subsequent weak two-body decay chains. The modular nature of these matrices allows for their rearrangement to model various decay scenarios, exemplified by the processes $e^+e^- \to \Lambda\bar{\Lambda}$, $e^+e^- \to \Xi\bar{\Xi}$, and analogous reactions \cite{ref51,ref52,ref53,ref54}. At the electron-positron collider, the BESIII experiment has performed extensive analyses of these angular distributions, employing multidimensional maximum-likelihood fits within this modular framework \cite{ref55,ref56}. The findings from these studies enable the exploration of new fundamental aspects of quantum mechanics, notably nonlocality and quantum entanglement \cite{ref57,ref58}. The study of quantum correlations in hyperon-antihyperon systems has garnered heightened interest \cite{ref59,ref60}, notably facilitated by recent technological advancements in the BESIII detector. These advancements facilitate more precise observations of entangled phenomena stemming from electron-positron collisions \cite{ref61,ref62}.

In this work, we investigate logarithmic negativity (LN), local quantum uncertainty (LQU), local quantum Fisher information (LQFI), and quantum teleportation through noisy channels in the process of $e^{+}e^{-} \to J/\psi \to \text{Y}\bar{\text{Y}}$ as studied at the BESIII experiment, where $\text{Y}$ and $\overline{\text{Y}}$ denote a spin-$1/2$ hyperon and its antihyperon ($\Lambda\bar{\Lambda}$, $\Xi^{0}\bar{\Xi}^{0}$, $\Xi^{-}\bar{\Xi}^{+}$, $\Sigma^{+}\bar{\Sigma}^{-}$). Utilizing the two-qubit density operator \cite{ref63,ref64}, we demonstrate that these quantum measures depend significantly on both the scattering angle $\varphi$, the amplitude $\theta$, and the parameters of dephasing channels $s$. We show that fidelity can exceed the classical limit of $2/3$ despite the noise in the considered channels, even as decoherence increases.

The remainder of this article is organized as follows. In Section \ref{sec:2}, we describe the physical model for the $\text{Y}\bar{\text{Y}}$ pair. Section \ref{sec:3} provides an overview of key quantum measures such as LN, LQU, and LQFI. The dynamics of LN, LQU, and LQFI under different dephasing noises are then discussed and compared in Section \ref{sec:4}. Our main results are presented in Section \ref{sec:5}. In Section \ref{sec:6}, we discuss the fidelity of quantum teleportation through different noisy channels. Finally, Section \ref{sec:7} offers a concluding summary.%

\section{Theoretical model}\label{sec:2}

Vector charmonia, including $J/\psi$ or $\psi(2S)$, exhibit decay modes that produce hyperon-antihyperon pairs, resulting in a massive two-qubit system, $\text{Y} \bar{\text{Y}}$, composed of two spin-$1/2$ particles. Due to momentum conservation in the center-of-mass frame, the hyperon and antihyperon are produced with momenta that are equal in magnitude but opposite in direction. One can use a two-qubit density operator to characterize their spin states \citep{ref63,ref64}
\begin{equation}
\begin{aligned}
\rho_{\text{Y}\bar{\text{Y}}} &= \frac{1}{4} \left[\mathds{1} \otimes \mathds{1} + \sum_{j} \mathbf{P}_j^{+} \left( \tau_j \otimes \mathds{1} \right) + \sum_{k} \mathbf{P}_k^{-} \left( \mathds{1} \otimes \tau_j \right) + \sum_{j,k} \mathbb{C}_{j,k} \left( \tau_j \otimes \tau_k \right) \right],
\end{aligned}
\label{eq:1}
\end{equation}

where $\tau = (\tau_1, \tau_2, \tau_3)$ represents the Pauli matrices, $\mathbf{P^{\pm}}$ denote the polarization vectors for the hyperon and antihyperon, and $\mathbb{C}_{j,k}$ specifies the elements of their correlation matrix. The aforementioned two-qubit density matrix admits a more concise representation, given by
\begin{equation}
\rho_{\text{Y}\bar{\text{Y}}} = \frac{1}{4}\sum_{\mu,\bar{\nu}=0}^{3}\Theta_{\mu\bar{\nu}}\tau^{\text{Y}}_{\mu} \otimes \tau^{\bar{\text{Y}}}_{\bar{\nu}}.
\end{equation}

We define $\Theta_{00} = 1$, $\Theta_{j0} = \mathbb{P}_j^{+}$, $\Theta_{0k} = \mathbb{P}_k^{-}$, and $\Theta_{jk} = \mathbb{C}_{j,k}$, with $\tau_0=\mathds{1}_{2\times 2}$. Using Pauli matrices $\tau^{\text{Y}}_{\mu}$($\tau^{\bar{\text{Y}}}_{\bar{\nu}}$) in the hyperon and antihyperon rest frames, respectively, we construct a $4\times 4$ real matrix, $\Theta_{\mu \bar{\nu}}$, representing their polarizations and spin correlations.

\begin{figure}[!h]
\begin{center}
\includegraphics[width=8.5cm,height=6cm]{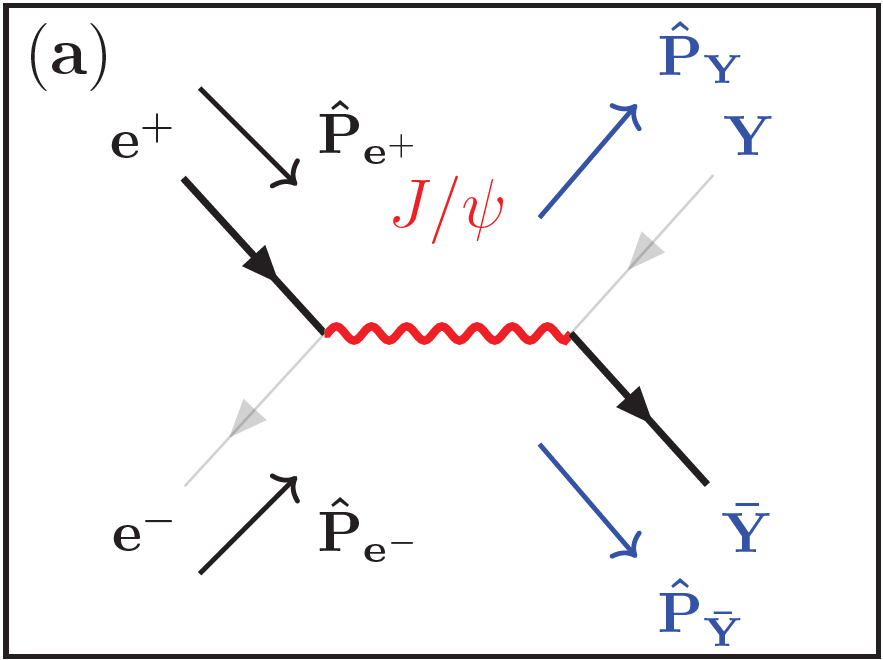}
\includegraphics[width=8.5cm,height=6cm]{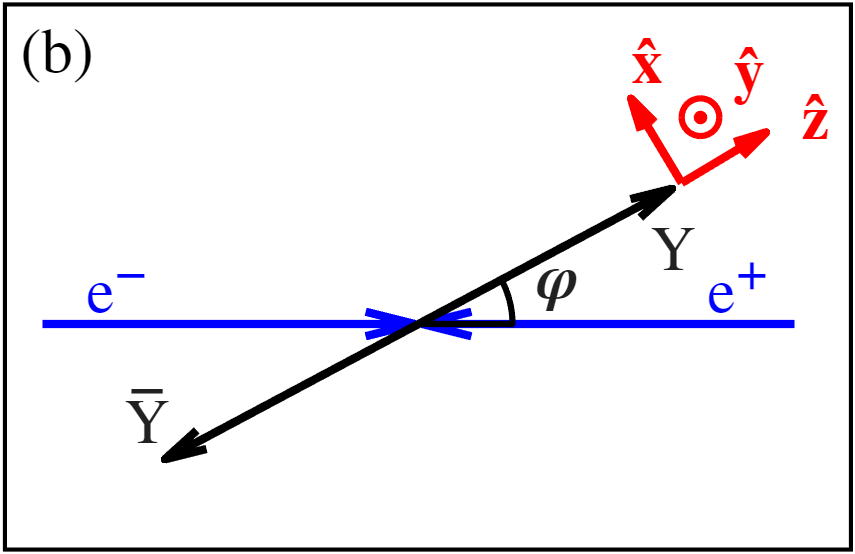}
\end{center}
\caption{(a): The Feynman diagram depicting the reaction $e^{+}e^{-}\to \text{Y}\bar{\text{Y}}$, (b): The coordinate system $\{\hat{x}, \hat{y}, \hat{z}\}$ is defined in the common rest frame of both the $\text{Y}$ and $\bar{\text{Y}}$ particles.}
\label{fig:e}
\end{figure}

One can define the helicity rest frame for the hyperon $\text{Y}$ as: 
$\mathbf{\hat{y}} = \frac{\mathbf{\hat{P}_e} \times \mathbf{\hat{P}}_\text{Y}}{|\mathbf{\hat{P}_e} \times \mathbf{\hat{P}}_\text{Y}|}, \quad \mathbf{\hat{z}} = \mathbf{\hat{P}}_\text{Y}, \quad \mathbf{\hat{x}} = \mathbf{\hat{y}} \times \mathbf{\hat{z}}. 
$
We adopt the rest frame of the antihyperon $\bar{\text{Y}}$, ensuring the coordinate axes are consistent with those defined for the hyperon $\left\lbrace \mathbf{\hat{x}},\mathbf{\hat{y}},\mathbf{\hat{z}}\right\rbrace$. The three axes we have chosen are the same as those in Ref. \cite{ref64}, which is shown in Fig. \ref{fig:e}(a).

The spin-correlation matrix $\Theta_{\mu \bar{\nu}}$ for the process $e^+ e^- \to \text{Y}\bar{\text{Y}}$ is, to lowest order, a function of two parameters: $\alpha_\psi$, where $-1\leq \alpha_\psi \leq 1$, and $\Delta\Phi$, where $ -\pi\leq \Delta\Phi \leq \pi$. The angle $\theta$, where $\cos(\theta) = \mathbf{\hat{P}_e} \cdot \mathbf{\hat{P}_\text{Y}}$, determines the elements of the matrix $\mathbb{C}_{j,k}$. Within the rest frames of $\text{Y}$ and $\bar{\text{Y}}$, the matrix $\Theta_{\mu\bar{\nu}}$ is expressed as
\begin{equation}
\Theta_{\mu\bar{\nu}} =
\begin{pmatrix}
\Theta_{0,0} & 0 & \Theta_{0,2} & 0 \\
0 & \Theta_{1,1} & 0 & \Theta_{1,3} \\
\Theta_{2,0} & 0 & \Theta_{2,2} & 0 \\
0 & \Theta_{3,1} & 0 & \Theta_{3,3}
\end{pmatrix}, \quad \text{where} \quad
\begin{aligned}
\Theta_{0,0}&= 1,  \Theta_{1,1}= \frac{\sin^{2}(\varphi)}{1+\alpha_{\psi}\cos^{2}(\varphi)},\\
\Theta_{2,2} &=\frac{-\alpha_{\psi}\sin^{2}(\varphi)}{1+\alpha_{\psi}\cos^{2}(\varphi)}, \Theta_{3,3} =\frac{\alpha_{\psi}+\cos^{2}\varphi}{1+\alpha_{\psi}\cos^{2}(\varphi)},\\
\Theta_{0,2} &=\Theta_{2,0}= \frac{\beta_{\psi}\sin(\varphi)\cos(\varphi)}{1+\alpha_{\psi}\cos^{2}(\varphi)}, \\
\Theta_{1,3} &=\Theta_{3,1}=  \frac{\gamma_{\psi}\sin(\varphi)\cos(\varphi)}{1+\alpha_{\psi}\cos^{2}(\varphi)},\\
\end{aligned}
\end{equation}
with the parameters $\beta_\psi$ and $\gamma_\psi$ are expressed in terms of $\alpha_\psi$ and $\Delta\Phi$ as
\begin{equation*}
\beta_\psi = \sqrt{1 - \alpha_\psi^2} \sin(\Delta\Phi), \quad \gamma_\psi = \sqrt{1 - \alpha_\psi^2} \cos(\Delta\Phi).
\end{equation*}
Defined within the rest frame of hyperon $\text{Y}$, with coordinates $(\mathbf{\hat{x}},\mathbf{\hat{y}},\mathbf{\hat{z}})$, the $\text{Y}$-hyperon polarization vector $\text{P}_{\text{Y}}$ is
\begin{equation}
\text{P}_{\text{Y}}= \frac{\beta_{\psi}\sin(\varphi)\cos(\varphi)}{1+\alpha_{\psi}\cos^{2}(\varphi)}. 
\end{equation}
The polarization vectors of the hyperon and antihyperon are equivalent in magnitude and share the same direction, expressed as $\text{P}_{\text{Y}}=\text{P}_{\bar{\text{Y}}}$. For simplicity, the two-qubit state can be transformed into an X-state by swapping the $\mathbf{\hat{y}}$ and $\mathbf{\hat{z}}$ axes. The resulting spin density matrix writes as
\begin{equation}
\rho_{\text{Y}\bar{\text{Y}}}^{\text{X}} = \frac{1}{4} \left( \mathbb{I}\otimes \mathbb{I} + \mathbb{A}\big(\tau_z \otimes \mathbb{I} + \mathbb{I} \otimes \tau_z\big) + \sum_{i=1}^{3} \mathbb{B}_{i} \tau_i \otimes \tau_i \right)
\end{equation}
which represents the symmetric X-state form for $\rho_{\text{Y}\bar{\text{Y}}}$. Thus, we denote this state as $\rho^{\text{X}}_{\text{Y}\bar{\text{Y}}}$. The matrix $\Theta_{\alpha\beta}$ for this state becomes
\begin{equation}
\Theta_{\mu\bar{\nu}} =
\begin{pmatrix}
1 & 0 & 0 & \mathbb{A} \\
0 & \mathbb{B}_1 & 0 & 0 \\
0 & 0 & \mathbb{B}_2 & 0 \\
\mathbb{A} & 0 & 0 & \mathbb{B}_3
\end{pmatrix}, \quad \text{where} \quad
\begin{aligned}
\mathbb{A} &= \frac{\beta_{\psi}\sin^{2}(\varphi)\cos(\varphi)}{1+\alpha_{\psi}\cos^{2}(\varphi)}, \\
\mathbb{B}_{1,2} &= \frac{1 + \alpha_{\psi} \pm \sqrt{(1 + \alpha_{\psi} \cos^2(2\varphi))^2 - \beta_{\psi}^2 \sin^2(2\varphi)}}{2(1+\alpha_{\psi}\cos^{2}(\varphi))}, \\
\mathbb{B}_3 &= \frac{-\alpha_{\psi}\sin^{2}(\varphi)}{1 + \alpha_{\psi}\cos^{2}(\varphi)},
\end{aligned}
\end{equation}

We note that $\mathbb{A} = \text{P}_{\text{Y}}=\text{P}_{\bar{\text{Y}}}$ and $\mathbb{B}_3 = \Theta_{2,2}$. Furthermore, $\mathbb{B}_1$ and $\mathbb{B}_2$ are derived from diagonalizing the block matrix of $\Theta_{i,j}$, where $i,j = x,y$ in $\Theta
_{\mu\bar{\nu}}$. We observe that the swapping of the $\mathbf{\hat{y}}$ and $\mathbf{\hat{z}}$ axes and the diagonalization of $\Theta_{i,j}$ can be obtained through a local unitary transformation
\begin{equation}
\rho^{\text{X}}_{\text{Y}\bar{\text{Y}}} = (U_{\text{Y}} \otimes U_{\bar{\text{Y}}})\rho_{\text{Y}\bar{\text{Y}}}(U_{\text{Y}} \otimes U_{\bar{\text{Y}}})^{\dagger},
\label{eq:8}
\end{equation}

where $\text{U}_{\text{Y}}$ and $\text{U}_{\bar{\text{Y}}}$ are unitary operators acting independently in the Hilbert spaces of $\text{Y}$ and $\bar{\text{Y}}$, respectively. When expressed in the $\tau_z$ basis, the spin density operator for the hyperon-antihyperon system can be rewritten by expanding equation (\ref{eq:8}) as 
\begin{equation}
\rho_{\text{Y}\bar{\text{Y}}}^{\text{X}} =
\begin{pmatrix}
\rho_{1,1}& 0 & 0 & \rho_{1,4} \\
0 & \rho_{2,2} & \rho_{2,3} & 0 \\
0 & \rho_{3,2} & \rho_{3,3}& 0 \\
\rho_{4,1} & 0 & 0 & \rho_{4,4}
\end{pmatrix}, \quad \text{where} \quad
\begin{aligned}
 \rho_{1,1}&= \frac{1}{4}\bigg(1+2\mathbb{A}+\mathbb{B}_{3}\bigg), \quad
\rho_{1,4} &= \rho_{4,1} = \frac{1}{4}\bigg(\mathbb{B}_{1}-\mathbb{B}_{2}\bigg), \\
\rho_{2,2} &= \rho_{3,3} = \frac{1}{4}\bigg(1-\mathbb{B}_{3}\bigg), \quad
\rho_{2,3} &= \rho_{3,2} = \frac{1}{4}\bigg(\mathbb{B}_{1}+\mathbb{B}_{2}\bigg), \\
\rho_{4,4} &= \frac{1}{4}\bigg(1-2\mathbb{A}+\mathbb{B}_{3}\bigg).
\end{aligned}
\label{eq:varrho}
\end{equation}

\section{Quantum correlation measures}\label{sec:3}

\subsection{Logarithmic negativity}
We employ logarithmic negativity, $L_N(\rho^{\text{X}}_{\text{Y}\bar{\text{Y}}})$, as proposed by Vidal and Werner \cite{EN1, EN2}, to quantify entanglement in bipartite mixed states \cite{EN3}. This measure is based on the absolute sum of the negative eigenvalues of the partially transposed density matrix, $\rho_{\text{Y}\bar{\text{Y}}}^{T_\text{Y}}$, with respect to subsystem $\text{Y}$ \cite{EN4}.
\begin{equation}
L_{N}(\rho^{\text{X}}_{\text{Y}\bar{\text{Y}}}) = \frac{||\rho_{\text{Y}\bar{\text{Y}}}^{T_\text{Y}}||_1 - 1}{2} = \sum_i \mu_i,
\label{eq:N1}
\end{equation}
where $\mu_i$ denote the negative eigenvalues of the partially transposed density matrix $\rho_{\text{Y}\bar{\text{Y}}}^{T_\text{Y}}$ (the notation $T_\text{Y}$ indicates that the partial transpose is performed with respect to subsystem $\text{Y}$). The trace norm is defined by
\begin{equation}
||\rho_{\text{Y}\bar{\text{Y}}}^{T_\text{Y}}||_1 = \text{Tr} \sqrt{\rho_{\text{Y}\bar{\text{Y}}}^{T_\text{Y}} \big(\rho_{\text{Y}\bar{\text{Y}}}^{T_\text{Y}}\big)^\dagger}.
\label{eq:N2}
\end{equation}
For a system composed of two qubits, the negativity writes as
\begin{equation}
L_{N}(\rho_{\text{Y}\bar{\text{Y}}}^{\text{X}}) = \max\{0, -2\mu_{\min}\}.
\label{eq:N3}
\end{equation}
For the state (\ref{eq:varrho}), the negative eigenvalues of its partially transposed density matrix in the computational basis are
\begin{equation}
\rho_{\text{Y}\bar{\text{Y}}}^{T_{\text{Y}}}=
\begin{pmatrix}
\rho_{1,1}& 0 & 0 & \rho_{2,3} \\
0 & \rho_{2,2} & \rho_{1,4} & 0 \\
0 & \rho_{1,4} & \rho_{2,2}& 0 \\
\rho_{2,3} & 0 & 0 & \rho_{4,4}
\end{pmatrix},
\label{eq:N4}
\end{equation}
and its eigenvalues are given by
\begin{equation}
\begin{aligned}
\e_{1,2}&=\frac{\big(\rho_{1,1}+\rho_{4,4}\big)\pm\sqrt{\big(\rho_{1,1}-\rho_{4,4}\big)^{2}+4|\rho_{2,3}|^{2}}}{2},\\
\e_{3,4}&=\frac{\big(\rho_{2,2}+\rho_{3,3}\big)\pm\sqrt{\big(\rho_{2,2}-\rho_{3,3}\big)^{2}+4|\rho_{1,4}|^{2}}}{2}.
\end{aligned}
\label{eq:N5}
\end{equation}
Equation (\ref{eq:N3}) describes the negativity, with
\[
\mu_{min}=\min\left\lbrace e_{1},e_{2},e_{3},e_{4}\right\rbrace.
\]

\subsection{Local quantum uncertainty} 

The preliminaries of LQU, a quantum correlation measure determined by the minimum of skew information, are presented in this section. Considering a bipartite quantum system with density matrix $\rho_{\text{Y}\bar{\text{Y}}}$, LQU is defined as \cite{L1}
\begin{equation}
LQU(\rho^{\text{X}}_{\text{Y}\bar{\text{Y}}})=\min_{K_{\text{Y}}}\mathcal{I}\Big(\rho^{\text{X}}_{\text{Y}\bar{\text{Y}}}, K_{\text{Y}}\otimes \mathbb{I}_{\bar{\text{Y}}}\Big).
\end{equation}

The quantity $\mathcal{I}\left(\rho^{\text{X}}_{\text{Y}\bar{\text{Y}}},\, K_{\text{Y}}\otimes \mathbb{I}_{\bar{\text{Y}}} \right) = -\frac{1}{2}\,\text{Tr}\left\lbrace\left[\sqrt{\rho^{\text{X}}_{\text{Y}\bar{\text{Y}}}},\, K_{\text{Y}}\otimes \mathbb{I}_{\bar{\text{Y}}}\right]^{2}\right\rbrace$ denotes the skew information, where $\left[\sqrt{\rho^{\text{X}}_{\text{Y}\bar{\text{Y}}}},\, K_{\text{Y}}\otimes \mathbb{I}_{\bar{\text{Y}}}\right]$ is the commutator, $\mathbb{I}_{\bar{\text{Y}}}$ represents the identity operator acting on the subsystem $\bar{\text{Y}}$, and $K_{\text{Y}}$ is a local Hermitian operator with a non-degenerate spectrum. When considering a bipartite $2\otimes d$ state with a single-qubit subsystem $\text{Y}$, LQU simplifies as follows 
\begin{equation}
LQU(\rho^{\text{X}}_{\text{Y}\bar{\text{Y}}})=1-\Gamma_{\max}(\mathcal{W}_{\text{Y}\bar{\text{Y}}}).
\label{eq:w2}
\end{equation}
The term $\Gamma_{\max}(\mathcal{W}_{\text{Y}\bar{\text{Y}}})$ corresponds to the largest eigenvalue of the symmetric $3\times3$ matrix $\mathcal{W}_{\text{Y}\bar{\text{Y}}}$. The elements of $\mathcal{W}_{\text{Y}\bar{\text{Y}}}$ are defined as
\begin{equation}
(\mathcal{W}_{\text{Y}\bar{\text{Y}}})_{ij}={\rm Tr}\left\lbrace \sqrt{\rho_{\text{Y}\bar{\text{Y}}}}(\tau_{\text{Y}_i}\otimes \mathbb{I}_{\bar{\text{Y}}})\sqrt{\rho_{\text{Y}\bar{\text{Y}}}}(\tau_{\text{Y}_j}\otimes \mathbb{I}_{\bar{\text{Y}}})\right\rbrace.
\label{eq:w3} 
\end{equation}
Here, $\tau_{\text{Y}_{i,j}}$ (where $i,j\in \left\lbrace x,y,z\right\rbrace $) denote the Pauli operators acting on subsystem $\text{Y}$. The eigenvalues of matrix (\ref{eq:w3}) for a thermal state (\ref{eq:varrho}) are given by
\begin{subequations}
\begin{align}
\mathcal{W}_{x,x} &= \big(\sqrt{\gamma_{1}}+\sqrt{\gamma_{2}}\big)\big(\sqrt{\gamma_{3}}+\sqrt{\gamma_{4}}\big)
+ \frac{\big(\rho_{2,2}-\rho_{3,3}\big)\big(\rho_{4,4}-\rho_{1,1}\big)+4|\rho_{1,4}\rho_{2,3}|}
{\left(\sqrt{\gamma_{1}}+\sqrt{\gamma_{2}}\right)\left(\sqrt{\gamma_{3}}+\sqrt{\gamma_{4}}\right)}, \\[6pt]
\mathcal{W}_{y,y} &= \big(\sqrt{\gamma_{1}}+\sqrt{\gamma_{2}}\big)\big(\sqrt{\gamma_{3}}+\sqrt{\gamma_{4}}\big)
- \frac{\big(\rho_{2,2}-\rho_{3,3}\big)\big(\rho_{4,4}-\rho_{1,1}\big)+4|\rho_{1,4}\rho_{2,3}|}
{\left(\sqrt{\gamma_{1}}+\sqrt{\gamma_{2}}\right)\left(\sqrt{\gamma_{3}}+\sqrt{\gamma_{4}}\right)}, \\[6pt]
\mathcal{W}_{z,z} &= \frac{\left(\sqrt{\gamma_{1}}+\sqrt{\gamma_{2}}\right)^{2}
+ \left(\sqrt{\gamma_{3}}+\sqrt{\gamma_{4}}\right)^{2}}{2}
+ \frac{\left(\rho_{4,4} - \rho_{1,1}\right)^{2} - 4|\rho_{1,4}|^{2}}
{2\left(\sqrt{\gamma_{1}}+\sqrt{\gamma_{2}}\right)^{2}}
- \frac{\big(\rho_{2,2}-\rho_{3,3}\big)^{2}-4|\rho_{2,3}|^{2}}
{2\left(\sqrt{\gamma_{3}}+\sqrt{\gamma_{4}}\right)^{2}}.
\end{align}
\end{subequations}

Here, $\gamma_i$ ($i=1,2,3,4$) are the eigenvalues of the density matrix in Eq. (\ref{eq:varrho}), satisfying $\gamma_i \geq 0$ and the normalization condition $\sum_i \gamma_i = 1$. These eigenvalues are given by
\begin{equation}
\begin{aligned}
\gamma_{1,2}&= \frac{1}{2}\Big(\rho_{1,1} + \rho_{4,4}) \pm \sqrt{(\rho_{1,1} - \rho_{4,4})^2 + 4|\rho_{1,4}|^2}\Big),\\
\gamma_{3,4}&= \frac{\big(\rho_{2,2}+\rho_{3,3}\big)\pm\sqrt{\big(\rho_{2,2}-\rho_{3,3}\big)^{2}+4|\rho_{2,3}|^{2}}}{2}.
\end{aligned}
\label{eq:VP}
\end{equation}

When $\mathcal{W}_{x,x}\geq \mathcal{W}_{y,y}$, an explicit expression for LQU (\ref{eq:w2}) is given by

\begin{equation}
LQU(\rho^{\text{X}}_{\text{Y}\bar{\text{Y}}}) = 1 - \max\{\mathcal{W}_{x,x}, \mathcal{W}_{z,z}\}.
\label{eq:WLQU}
\end{equation}

\subsection{Local quantum Fisher information}

Quantum Fisher Information (QFI) is a fundamental tool in quantum metrology, designed to enhance measurement precision in quantum systems \cite{f1,f2,f3}. It quantifies a quantum system's sensitivity to parameter variations, thereby defining the ultimate limit for parameter estimation accuracy \cite{f4,f5}. Specifically, Fisher information measures the distinguishability of a quantum state from its infinitesimally perturbed counterparts based on measurement outcomes, encoding the extractable information about the parameter \cite{f8}. The interplay between QFI ($\mathcal{F}$) and Local Quantum Fisher Information (LQFI) is critical for efficient parameter estimation, enabling the design of optimal measurement strategies adapted to subsystem properties. Here, we denote this measure by $LQFI$, representing the optimal LQFI with the measurement operator $H_{\text{Y}}$ acting on the subspace of party $\text{Y}$ in the bipartite system $\text{Y}\bar{\text{Y}}$.
\begin{equation}
LQFI(\rho^{\text{X}}_{\text{Y}\bar{\text{Y}}})=\min_{H_{\text{Y}}}\mathcal{F}\big(\rho^{\text{X}}_{\text{Y}\bar{\text{Y}}},H_{\text{Y}}\big).
\end{equation}
For the particular case where subsystem $\text{B}$ constitutes a qubit, the LQFI is given by
\begin{equation}
LQFI(\rho^{\text{X}}_{\text{Y}\bar{\text{Y}}})=1-\lambda^{\mathcal{M}}_{\max},
\label{eq:LQFI}
\end{equation}
$\lambda^{\mathcal{M}}_{\max}$ refers to the maximum eigenvalue of the $3\times3$ symmetric matrix $\mathcal{M}$, the elements of which are given by
\begin{equation}
\mathcal{M_{\alpha\beta}}=\sum_{i\neq j}\frac{2\gamma_{i}\gamma_{j}}{\gamma_{i}+\gamma_{j}}\bra{\psi_{i}}\tau_{\alpha}\otimes\mathbb{I}\ket{\psi_{j}}\bra{\psi_{j}}\tau_{\beta}\otimes\mathbb{I}\ket{\psi_{i}},
\label{eq:m}
\end{equation}
where $\gamma_i$ and $\ket{\psi_i}$ denote the eigenvalues and eigenvectors of the density matrix $\rho^{\text{X}}_{\text{Y}\bar{\text{Y}}}$, with the eigenvalues explicitly given in Eq. (\ref{eq:VP}). In addition, $\tau_\alpha$ and $\tau_\beta$ denote the standard Pauli matrices, with indices $\alpha, \beta \in \{x, y, z\}$. To derive the explicit expression for LQFI requires first computing the elements of matrix $\mathcal{M}$. Moreover, one can readily confirm from (\ref{eq:m}) that the off-diagonal elements are zero and the diagonal elements are written as
\begin{equation*}
\begin{aligned}
\mathcal{M}_{x,x} &= \frac{64\mathcal{M}_{1}\mathcal{M}_{2}}{\mathcal{M}_{3}}, \quad
\mathcal{M}_{y,y} &= \frac{64\mathcal{M}_{4}\mathcal{M}_{2}}{\mathcal{M}_{3}}, \quad \text{and} \quad
\mathcal{M}_{z,z} &= 1 - \Bigg(\frac{|\rho_{1,4}|}{\rho_{1,1}+\rho_{4,4}} + \frac{|\rho_{2,3}|}{\rho_{2,2}+\rho_{3,3}}\Bigg),
\end{aligned}
\end{equation*}
where 
\begin{equation*}
\begin{aligned}
\mathcal{M}_{1} &=\rho_{1,1}\rho_{3,3}+\rho_{2,2}\rho_{4,4}+\gamma_{1}\gamma_{2}+\gamma_{3}\gamma_{4}+2|\rho_{1,4}\rho_{2,3}|,\\
\mathcal{M}_{2} &= \big(\rho_{1,1}+\rho_{4,4}\big)\gamma_{3}\gamma_{4}+\big(\rho_{2,2}+\rho_{3,3}\big)\gamma_{1}\gamma_{2},\\
\mathcal{M}_{3} &= \Big[1-(\gamma_{1}-\gamma_{2})^{2}-(\gamma_{3}-\gamma_{4})^{2}\Big]^{2}-4\big(\gamma_{1}-\gamma_{2}\big)^{2}\big(\gamma_{3}-\gamma_{4}\big)^{2},\\
\mathcal{M}_{4}&=\rho_{1,1}\rho_{3,3}+\rho_{2,2}\rho_{4,4}+\gamma_{1}\gamma_{2}+\gamma_{3}\gamma_{4}-2|\rho_{1,4}\rho_{2,3}|.
\end{aligned}
\end{equation*}
We show that $\mathcal{M}_{x,x} \geq \mathcal{M}_{y,y}$, which implies that
\begin{equation}
LQFI(\rho^{\text{X}}_{\text{Y}\bar{\text{Y}}})=1-\max\left\lbrace \mathcal{M}_{x,x},\mathcal{M}_{z,z}\right\rbrace.
\label{eq:F}
\end{equation}
\section{Noisy channels}\label{sec:4}

Interacting with their surroundings leads to decoherence in quantum systems, resulting in lost quantum coherence and the decay of fragile quantum correlations. To model decoherence in quantum systems, decohering channels such as amplitude damping (AD), phase flip (PF), and phase damping (PD) are commonly employed \citep{L4}. These channels represent noise or perturbations that impact the quantum state, causing it to lose its purity and become mixed or probabilistic. Given an initial bipartite quantum state $\rho_{\text{Y}\bar{\text{Y}}}$, the final quantum state under the action of decohering channels can be rigorously derived via the Kraus operator formalism as
\begin{equation}
\hat{\varepsilon}(\rho^{\text{X}}_{\text{Y}\bar{\text{Y}}})=\sum_{kl}\hat{K}_{kl} \rho^{\text{X}}_{\text{Y}\bar{\text{Y}}} \hat{K}_{kl}^{\dagger}.
\label{eq:epsilon}
\end{equation}

Denoted as $\hat{K}_{kl} = \hat{K}_k \otimes \hat{K}_l$, the Kraus operators in this context correspond to one-qubit quantum channels represented by $\hat{K}_k$ and $\hat{K}_l$. It is crucial that these operators satisfy the closure condition $\sum_{kl} \hat{K}_{kl}^\dagger \hat{K}_{kl} = \mathbb{I}$, a fundamental requirement for valid quantum operations.

\subsection{Amplitude damping channel}

The AD decoherence channel effectively models spontaneous emission in quantum systems by describing energy dissipation. For a single-qubit AD channel, the description is formulated utilizing the following Kraus operators:
\begin{equation}
\hat{K}_1 = {\rm diag}\bigg(1,\sqrt{1-s}\bigg)
\quad, \quad
\hat{K}_2 =
\begin{pmatrix}
0 & \sqrt{s} \\
0 & 0
\end{pmatrix}.
\label{eq:K}
\end{equation}
The decoherence parameter, $s=1-{\rm Exp}[-v t]$, with $s\in[0,1]$ and $v$ representing the decay rate, characterizes the AD channel. Substituting Eqs. (\ref{eq:varrho}) and (\ref{eq:K}) into Eq. (\ref{eq:epsilon}), the resulting density matrix for the system under the AD effect is expressed as

\begin{equation}
\hat{\rho}^{\text{AD}}_{\text{Y}\bar{\text{Y}}} =
\begin{pmatrix}
\eta_{1,1}& 0 & 0 & \eta_{1,4}  \\
0 & \eta_{2,2} & \eta_{2,3} & 0 \\
0 & \eta_{2,3} & \eta_{3,3}& 0 \\
\eta_{1,4}  & 0 & 0 & \eta_{4,4}
\end{pmatrix}, \quad \text{where} \quad
\begin{aligned}
\eta_{1,1} &= \rho_{1,1} + s\left(2\rho_{2,2} + s\rho_{4,4}\right), \\
\eta_{1,4} &= (1 - s)\rho_{1,4}, \\
\eta_{2,2} &=\eta_{3,3}= -(s - 1)\left(\rho_{2,2} + s\rho_{4,4}\right), \\
\eta_{2,3} &= (1 - s)\rho_{2,3}, \\
\eta_{4,4} &= (s - 1)^2 \rho_{4,4}.
\end{aligned}
\label{eq:AD}
\end{equation}

Utilizing Equation (\ref{eq:N3}), the entanglement negativity can be writes as
\begin{equation}
L_{N}(\hat{\rho}_{\text{Y}\bar{\text{Y}}}^{\text{AD}}) = \max\{0, -2\mu^{\text{AD}}_{\min}\},
\end{equation}
where 
\begin{equation*}
\mu^{\text{AD}}_{\min}=\min\left\lbrace e^{\text{AD}}_{1},e^{\text{AD}}_{2},e^{\text{AD}}_{3},e^{\text{AD}}_{4}\right\rbrace, 
\end{equation*}

and $e^{\text{AD}}_i$ ($i=1,2,3,4$) denote the eigenvalues of the transpose of matrix (\ref{eq:AD})
\begin{equation*}
\begin{aligned}
e^{\text{AD}}_{1,2}&=\frac{\big(\eta_{1,1}+\eta_{4,4}\big)\pm\sqrt{\big(\eta_{1,1}-\eta_{4,4}\big)^{2}+4|\eta_{2,3}|^{2}}}{2},\\
e^{\text{AD}}_{3,4}&=\frac{\big(\eta_{2,2}+\eta_{3,3}\big)\pm\sqrt{\big(\eta_{2,2}-\eta_{3,3}\big)^{2}+4|\eta_{1,4}|^{2}}}{2}.
\end{aligned}
\label{eq:N5}
\end{equation*}

Furthermore, an explicit expression for LQU (\ref{eq:WLQU}) can be obtained as
\begin{equation}
LQU(\hat{\rho}^{\text{AD}}_{\text{Y}\bar{\text{Y}}}) = 1 - \max \left\{\mathcal{W}^{\text{AD}}_{x,x}, \mathcal{W}^{\text{AD}}_{z,z} \right\},
\end{equation}
where
\begin{equation*}
\begin{aligned}
\mathcal{W}^{\text{AD}}_{x,x} &= \left( \sqrt{\mathcal{A}_1} + \sqrt{\mathcal{A}_2} \right) \left( \sqrt{\mathcal{A}_3} + \sqrt{\mathcal{A}_4} \right) + \frac{\big(\eta_{2,2}-\eta_{3,3}\big)\big(\eta_{4,4}-\eta_{1,1}\big) +4\left| \eta_{1,4} \eta_{2,3} \right|}{\left( \sqrt{\mathcal{A}_1} + \sqrt{\mathcal{A}_2} \right) \left( \sqrt{\mathcal{A}_3} + \sqrt{\mathcal{A}_4} \right)}, \\
\mathcal{W}^{\text{AD}}_{z,z} &= \frac{\left( \sqrt{\mathcal{A}_1} + \sqrt{\mathcal{A}_2} \right)^2 + \left( \sqrt{\mathcal{A}_3} + \sqrt{\mathcal{A}_4} \right)^2}{2} + \frac{\left( \eta_{4,4} - \eta_{1,1} \right)^2 - 4 \left| \eta_{1,4} \right|^2}{2\left( \sqrt{\mathcal{A}_1} + \sqrt{\mathcal{A}_2} \right)^2}+ \frac{\big(\eta_{2,2}-\eta_{3,3}\big)^{2}-4\left| \eta_{2,3} \right|^2}{2\left( \sqrt{\mathcal{A}_3} + \sqrt{\mathcal{A}_4} \right)^2} \Bigg].
\end{aligned}
\end{equation*}
The quantities $\mathcal{A}_{i}$ denote the eigenvalues of the matrix defined in Eq.  (\ref{eq:AD}), and are explicitly given by
\begin{equation*}
\begin{aligned}
\mathcal{A}_{1,2} &= \frac{\eta_{1,1} + \eta_{4,4} \pm \sqrt{(\eta_{1,1} - \eta_{4,4})^2 + 4 \left| \eta_{1,4} \right|^2}}{2}, \\
\mathcal{A}_{3,4} &= \frac{\big(\eta_{2,2}+\eta_{3,3}\big)\pm\sqrt{\big(\eta_{2,2}-\eta_{3,3}\big)^{2}+4|\eta_{2,3}|^{2}}}{2}.
\end{aligned}
\end{equation*}

The LQFI for the thermal state $\hat{\rho}^{\text{AD}}_{\text{Y}\bar{\text{Y}}}$ is defined as
\begin{equation}
LQFI(\hat{\rho}^{\text{AD}}_{\text{Y}\bar{\text{Y}}})=1-\max\left\lbrace \mathcal{M}^{\text{AD}}_{x,x},\mathcal{M}^{\text{AD}}_{z,z}\right\rbrace 
\label{eq:FAD}
\end{equation}
where 

\begin{equation*}
\begin{aligned}
\mathcal{M}^{\text{AD}}_{x,x} &= \frac{64\mathcal{M}^{\text{AD}}_{1}\mathcal{M}^{\text{AD}}_{2}}{\mathcal{M}^{\text{AD}}_{3}}, 
\quad \text{and} \quad
\mathcal{M}^{\text{AD}}_{z,z} &= 1 - \Bigg(\frac{|\eta_{1,4}|}{\eta_{1,1}+\eta_{4,4}} + \frac{|\eta_{2,3}|}{\eta_{2,2}+\eta_{3,3}}\Bigg),
\end{aligned}
\end{equation*}

with
\begin{equation*}
\begin{aligned}
\mathcal{M}^{\text{AD}}_{1} &=\eta_{1,1}\eta_{3,3}+\eta_{2,2}\eta_{4,4}+\mathcal{A}_{1}\mathcal{A}_{2}+\mathcal{A}_{3}\mathcal{A}_{4}+2|\eta_{1,4}\eta_{2,3}|,\\
\mathcal{M}^{\text{AD}}_{2} &= \big(\eta_{1,1}+\eta_{4,4}\big)\mathcal{A}_{3}\mathcal{A}_{4}+\big(\eta_{2,2}+\eta_{3,3}\big)\mathcal{A}_{1}\mathcal{A}_{2},\\
\mathcal{M}^{\text{AD}}_{3} & = \Big[1-(\mathcal{A}_{1}-\mathcal{A}_{2})^{2}-(\mathcal{A}_{3}-\mathcal{A}_{4})^{2}\Big]^{2}-4\big(\mathcal{A}_{1}-\mathcal{A}_{2}\big)^{2}\big(\mathcal{A}_{3}-\mathcal{A}_{4}\big)^{2}.
\end{aligned}
\end{equation*}
\subsection{Phase flip channel}

This channel exhibits a dynamic process that occurs with a probability $s$. This dynamic, acting upon the initial state, induces a unitary transformation characterized by the Pauli operator $\tau_z$, thereby resulting in a phase flip (PF). The single-qubit PF decoherence channel is typically represented by the following set of Kraus operators:
\begin{equation}
\hat{K}_1 = {\rm diag}\bigg(\sqrt{s},\sqrt{s}\bigg)
\quad,\quad
\hat{K}_2 = {\rm diag}\bigg(\sqrt{1-s},-\sqrt{1-s}\bigg).
\label{eq:fp}
\end{equation}
The density matrix can be expressed as

\begin{equation}
\hat{\rho}^{\text{PF}}_{\text{Y}\bar{\text{Y}}}=
\begin{pmatrix}
\rho_{1,1}& 0 & 0 & \delta_{1,4}\\
0 & \rho_{2,2} & \delta_{2,3} & 0 \\
0 & \delta_{2,3} & \rho_{3,3} & 0 \\
\delta_{1,4} & 0 & 0 & \rho_{4,4}
\end{pmatrix}, \quad \text{where} \quad
\begin{aligned}
\delta_{1,4} &= \rho_{1,4}\Big(1 - 2s\Big)^2,\\
\delta_{2,3} &=\rho_{2,3}\Big(1 - 2s\Big)^2.
\end{aligned}
\label{eq:PF}
\end{equation}

By applying Equation (\ref{eq:N3}), we obtain the logarithmic negativity as
\begin{equation}
L_{N}(\hat{\rho}^{\text{PF}}_{\text{Y}\bar{\text{Y}}}) = \max\{0, -2\mu^{\text{PF}}_{\min}\},
\end{equation}
where 
\begin{equation*}
\mu^{\text{PF}}_{\min}=\min\left\lbrace e^{\text{PF}}_{1},e^{\text{PF}}_{2},e^{\text{PF}}_{3},e^{\text{PF}}_{4}\right\rbrace,
\end{equation*}

with \( e^{\text{PF}}_i \) (\(i=1,2,3,4\)) denote the eigenvalues of the transpose of matrix (\ref{eq:PF})
\begin{equation*}
\begin{aligned}
\e^{\text{PF}}_{1,2} &=\frac{\rho_{1,1}+\rho_{4,4}\pm \sqrt{\big(\rho_{1,1}-\rho_{4,4}\big)^{2}+4|\delta_{2,3}|}}{2}, \\
\e^{\text{PF}}_{3,4}&=\frac{\big(\rho_{2,2}+\rho_{3,3}\big)\pm\sqrt{\big(\rho_{2,2}-\rho_{3,3}\big)^{2}+4|\delta_{1,4}|^{2}}}{2}. 
\end{aligned}
\label{eq:VPPF}
\end{equation*}

The LQU for the thermal state $\hat{\rho}^{\text{PF}}_{\text{Y}\bar{\text{Y}}}$ is defined as
\begin{equation}
LQU(\hat{\rho}^{\text{PF}}_{\text{Y}\bar{\text{Y}}}) = 1 - \max\{\mathcal{W}^{\text{PF}}_{x,x}, \mathcal{W}^{\text{PF}}_{z,z}\}, 
\label{eq:lqu_definition}
\end{equation}
where
\begin{equation*}
\begin{aligned}
\mathcal{W}^{\text{PF}}_{x,x} &= \Big( \sqrt{\mathcal{B}_1} + \sqrt{\mathcal{B}_2} \Big) \Big( \sqrt{\mathcal{B}_3} + \sqrt{\mathcal{B}_4} \Big) + \frac{\Big( \rho_{2,2} - \rho_{3,3} \Big)\Big( \rho_{4,4} - \rho_{1,1}\Big)+4 \left| \delta_{1,4} \delta_{2,3} \right|}{\Big( \sqrt{\mathcal{B}_1} + \sqrt{\mathcal{B}_2} \Big) \Big( \sqrt{\mathcal{B}_3} + \sqrt{\mathcal{B}_4} \Big)}, \\
\mathcal{W}^{\text{PF}}_{z,z} &= \frac{\Big( \sqrt{\mathcal{B}_1} + \sqrt{\mathcal{B}_2} \Big)^2 + \Big( \sqrt{\mathcal{B}_3} + \sqrt{\mathcal{B}_4} \Big)^2}{2} + \frac{\Big( \rho_{4,4} - \rho_{1,1} \Big)^2 - 4 \left| \delta_{1,4} \right|^2}{2 \Big( \sqrt{\mathcal{B}_1} + \sqrt{\mathcal{B}_2} \Big)^2}+\frac{\Big( \rho_{2,2} - \rho_{3,3} \Big)^2- 4\left| \delta_{2,3} \right|^2}{\Big( \sqrt{\mathcal{B}_3} + \sqrt{\mathcal{B}_4} \Big)^2}.
\end{aligned}
\end{equation*}

The quantities $\mathcal{B}_{i}$ denote the eigenvalues of the matrix defined in Eq.  (\ref{eq:PF}), and are explicitly given by
\begin{equation*}
\begin{aligned}
\mathcal{B}_{1,2} &= \frac{\rho_{1,1} +\rho_{4,4} \pm \sqrt{(\rho_{1,1} - \rho_{4,4})^2 + 4\big|\delta_{1,4}\big|^2}}{2},\\
\mathcal{B}_{3,4} &= \frac{\big(\rho_{2,2}+\rho_{3,3}\big)\pm\sqrt{\big(\rho_{2,2}-\rho_{3,3}\big)^{2}+4|\delta_{2,3}|^{2}}}{2}.
\end{aligned}
\end{equation*}

The LQFI for the thermal state $\hat{\rho}^{\text{PF}}_{\text{Y}\bar{\text{Y}}}$ is defined as
\begin{equation}
LQFI(\hat{\rho}^{\text{PF}}_{\text{Y}\bar{\text{Y}}})=1-\max\left\lbrace \mathcal{M}^{\text{AD}}_{x,x},\mathcal{M}^{\text{PF}}_{z,z}\right\rbrace,
\label{eq:FPF}
\end{equation}
where 
\begin{equation*}
\begin{aligned}
\mathcal{M}^{\text{PF}}_{x,x} &= \frac{64\mathcal{M}^{\text{PF}}_{1}\mathcal{M}^{\text{PF}}_{2}}{\mathcal{M}^{\text{PF}}_{3}}, \quad  \text{and} \quad
\mathcal{M}^{\text{PF}}_{z,z} &= 1 - \Bigg(\frac{|\delta_{1,4}|}{\rho_{1,1}+\rho_{4,4}} + \frac{|\delta_{2,3}|}{\rho_{2,2}+\rho_{3,3}}\Bigg),
\end{aligned}
\end{equation*}
with
\begin{equation*}
\begin{aligned}
\mathcal{M}^{\text{PF}}_{1} &=\rho_{1,1}\rho_{3,3}+\rho_{4,4}\rho_{2,2}+\mathcal{B}_{1}\mathcal{B}_{2}+\mathcal{B}_{3}\mathcal{B}_{4}+2|\delta_{1,4}\delta_{2,3}|,\\
\mathcal{M}^{\text{PF}}_{2} &= \big(\rho_{1,1}+\rho_{4,4}\big)\mathcal{B}_{3}\mathcal{B}_{4}+\big(\rho_{2,2}+\rho_{3,3}\big)\mathcal{B}_{1}\mathcal{B}_{2},\\
\mathcal{M}^{\text{PF}}_{3} &= \Big[1-(\mathcal{B}_{1}-\mathcal{B}_{2})^{2}-(\mathcal{B}_{3}-\mathcal{B}_{4})^{2}\Big]^{2}-4\big(\mathcal{B}_{1}-\mathcal{B}_{2}\big)^{2}\big(\mathcal{B}_{3}-\mathcal{B}_{4}\big)^{2}.
\end{aligned}
\end{equation*}

\subsection{Phase Damping channel}

PD (Phase Damping) decoherence channels represent quantum noise that specifically degrades quantum phase information, leaving the system's energy unaffected. The Kraus operators for a single-qubit PD channel are explicitly written as
\begin{equation}
\hat{K}_1 = {\rm diag}\bigg(1,\sqrt{1-s}\bigg)
\quad,\quad
\hat{K}_2 = {\rm diag}\bigg(0,\sqrt{s}\bigg).
\label{eq:pdc}
\end{equation}
Substituting Eqs. (\ref{eq:varrho}) and (\ref{eq:pdc}) into Eq. (\ref{eq:epsilon}), the resulting density matrix for the system under the PD effect is expressed as
\begin{equation}
\hat{\rho}^{\text{PD}}_{\text{Y}\bar{\text{Y}}} =
\begin{pmatrix}
\rho_{1,1} & 0 & 0 & \kappa_{1,4} \\
0 & \rho_{2,2} & \kappa_{2,3} & 0 \\
0 & \kappa_{2,3} & \rho_{3,3} & 0 \\
\kappa_{1,4} & 0 & 0 & \rho_{4,4}
\end{pmatrix}, \quad \text{where} \quad
\begin{aligned}
\kappa_{1,4} &= \rho_{1,4}(1-s),\\
\kappa_{2,3} &=\rho_{2,3}(1-s).
\end{aligned}
\label{eq:PD}
\end{equation}

The logarithmic negativity, for the thermal state $\hat{\rho}^{\text{PD}}_{\text{Y}\bar{\text{Y}}}$ is defined as
\begin{equation}
L_{N}(\hat{\rho}^{\text{PD}}_{\text{Y}\bar{\text{Y}}}) = \max\{0, -2\mu^{\text{PD}}_{\min}\},
\end{equation}
where 
\begin{equation*}
\mu^{\text{PD}}_{\min}=\min\left\lbrace \e^{\text{PD}}_{1},\e^{\text{PD}}_{2},\e^{\text{PD}}_{3},\e^{\text{PD}}_{4}\right\rbrace,
\end{equation*}

and \( \e^{\text{PD}}_i \) (\(i=1,2,3,4\)) denote the eigenvalues of the transpose of matrix (\ref{eq:PD})
\begin{equation*}
\begin{aligned}
\e^{\text{PD}}_{1,2}&=\frac{\rho_{1,1}+\rho_{4,4}\pm \sqrt{\big(\rho_{1,1}-\rho_{4,4}\big)^{2}+4|\kappa_{2,3}|}}{2}, \\
\e^{\text{PD}}_{3,4}&=\frac{\big(\rho_{2,2}+\rho_{3,3}\big)\pm\sqrt{\big(\rho_{2,2}-\rho_{3,3}\big)^{2}+4|\kappa_{1,4}|^{2}}}{2}.
\end{aligned}
\label{eq:N5}
\end{equation*}

The LQU for the density matrix $\hat{\rho}^{\text{PD}}_{\text{Y}\bar{\text{Y}}}$ is given by
\begin{equation}
LQU(\hat{\rho}^{\text{PD}}_{\text{Y}\bar{\text{Y}}}) = 1 - \max\{\mathcal{W}^{\text{PD}}_{x,x}, \mathcal{W}^{\text{PD}}_{z,z}\},
\label{eq:lqu_pd}
\end{equation}
where
\begin{equation*}
\begin{aligned}
\mathcal{W}^{\text{PD}}_{x,x} &= \Big( \sqrt{\mathcal{C}_1} + \sqrt{\mathcal{C}_2} \Big) \Big( \sqrt{\mathcal{C}_3} + \sqrt{\mathcal{C}_4} \Big) + \frac{\Big( \rho_{2,2} - \rho_{3,3} \Big)\Big( \rho_{4,4} - \rho_{1,1}\Big)+4 \left| \kappa_{1,4}\kappa_{2,3} \right|}{\Big( \sqrt{\mathcal{C}_1} + \sqrt{\mathcal{C}_2} \Big) \Big( \sqrt{\mathcal{C}_3} + \sqrt{\mathcal{C}_4} \Big)}, \\
\mathcal{W}^{\text{PD}}_{z,z} &= \frac{\Big( \sqrt{\mathcal{C}_1} + \sqrt{\mathcal{C}_2} \Big)^2 + \Big( \sqrt{\mathcal{C}_3} + \sqrt{\mathcal{C}_4} \Big)^2}{2} + \frac{\Big(\rho_{4,4} - \rho_{1,1} \Big)^2 - 4 \left| \kappa_{1,4} \right|^2}{2\Big( \sqrt{\mathcal{C}_1} + \sqrt{\mathcal{C}_2} \Big)^2}+\frac{\Big(\rho_{2,2} - \rho_{3,3} \Big)^2 - 4\left|\kappa_{2,3} \right|^2}{\Big( \sqrt{\mathcal{C}_3} + \sqrt{\mathcal{C}_4} \Big)^2} \Bigg].
\end{aligned}
\end{equation*}

The quantities $\mathcal{C}_{i}$ denote the eigenvalues of the matrix defined in Eq.  (\ref{eq:PD}), and are explicitly given by
\begin{equation*}
\begin{aligned}
\mathcal{C}_{1,2} &= \frac{\rho_{1,1}+\rho_{4,4}\pm \sqrt{\big(\rho_{1,1}-\rho_{4,4}\big)^2 + 4\big|\kappa_{1,4}\big|^2}}{2}, \\
\mathcal{C}_{3,4} &= \frac{\big(\rho_{2,2}+\rho_{3,3}\big)\pm\sqrt{\big(\rho_{2,2}-\rho_{3,3}\big)^{2}+4|\kappa_{2,3}|^{2}}}{2}.
\end{aligned}
\end{equation*}

The LQFI for the thermal state $\hat{\rho}^{\text{PF}}_{\text{Y}\bar{\text{Y}}}$ is defined as
\begin{equation}
LQFI(\hat{\rho}^{\text{PD}}_{\text{Y}\bar{\text{Y}}})=1-\max\left\lbrace \mathcal{M}^{\text{PD}}_{x,x},\mathcal{M}^{\text{PD}}_{z,z}\right\rbrace,
\label{eq:FPD}
\end{equation}
where 
\begin{equation*}
\begin{aligned}
\mathcal{M}^{\text{PD}}_{x,x} &= \frac{64\mathcal{M}^{\text{PD}}_{1}\mathcal{M}^{\text{PD}}_{2}}{\mathcal{M}^{\text{PF}}_{3}}, \quad  \text{and} \quad
\mathcal{M}^{\text{PD}}_{z,z} &= 1 - \Bigg(\frac{|\kappa_{1,4}|}{\rho_{1,1}+\rho_{4,4}} + \frac{|\kappa_{2,3}|}{\rho_{2,2}+\rho_{3,3}}\Bigg),
\end{aligned}
\end{equation*}
with

\begin{equation*}
\begin{aligned}
\mathcal{M}^{\text{PD}}_{1} &=\rho_{1,1}\rho_{3,3}+\rho_{2,2}\rho_{4,4}+\mathcal{C}_{1}\mathcal{C}_{2}+\mathcal{C}_{3}\mathcal{C}_{4}+2|\kappa_{1,4}\kappa_{2,3}|,\\
\mathcal{M}^{\text{PD}}_{2} &= \big(\rho_{1,1}+\rho_{4,4}\big)\mathcal{C}_{3}\mathcal{C}_{4}+\big(\rho_{2,2}+\rho_{3,3}\big)\mathcal{C}_{1}\mathcal{C}_{2},\\
\mathcal{M}^{\text{PD}}_{3} &= \Big[1-(\mathcal{C}_{1}-\mathcal{C}_{2})^{2}-(\mathcal{C}_{3}-\mathcal{C}_{4})^{2}\Big]^{2}-4\big(\mathcal{C}_{1}-\mathcal{C}_{2}\big)^{2}\big(\mathcal{C}_{3}-\mathcal{C}_{4}\big)^{2}.
\end{aligned}
\end{equation*}
\section{Results and discussions}\label{sec:5}
We investigate the evolution of LN, LQU, and LQFI in the $e^{+}e^{-} \rightarrow \text{Y}\overline{\text{Y}}$ process within this section, focusing on its dependence on the scattering angle $\varphi$ and the dephasing parameter, characterized by $s$. In particle physics, the process $e^{+}e^{-} \rightarrow J/\psi \rightarrow \text{Y} \overline{\text{Y}}$ involves the production of a pair of ground-state octet hyperons. Understanding the parameters associated with this process is crucial for studying the properties and interactions of these hyperons. Thanks to the Beijing Spectrometer III (BESIII) experiment, the following table summarizes some key parameters relevant to this process

\begin{table}[H]
\centering
\caption{Parameters relevant to $e^{+}e^{-} \rightarrow J/\psi \rightarrow \text{Y} \overline{\text{Y}}$, where $\text{Y} \overline{\text{Y}}$ denotes a pair of ground-state octet hyperons.}
 \label{t1}
\begin{tabular}{c c c c c}
\hline
\hline

$\text{Y}\bar{\text{Y}}$ & $\Lambda\overline{\Lambda}$ & $\Sigma^{+}\overline{\Sigma}^{-}$ &  $\Xi^{-}\overline{\Xi}^{+}$ &  $\Xi^{0}\overline{\Xi}^{0}$\\

\hline

$\alpha_{\psi}$ & 0.475(4) & -0.508(7)  & 0.586(16)& 0.514(16) \\
\hline
$\Delta\Phi/rad$ & 0.752(8) & -0.270(15) & 1.213(49) & 1.168(26)\\
\hline
Ref & \cite{ref61,T2} & \cite{T3,T4} & \cite{ref56,T6} & \cite{ref63,ref64} \\
\hline
\hline
\end{tabular}
\label{tab1}
\end{table}

\subsection{Without dephasing effect}

\begin{figure}[!h]
    \includegraphics[scale=0.27]{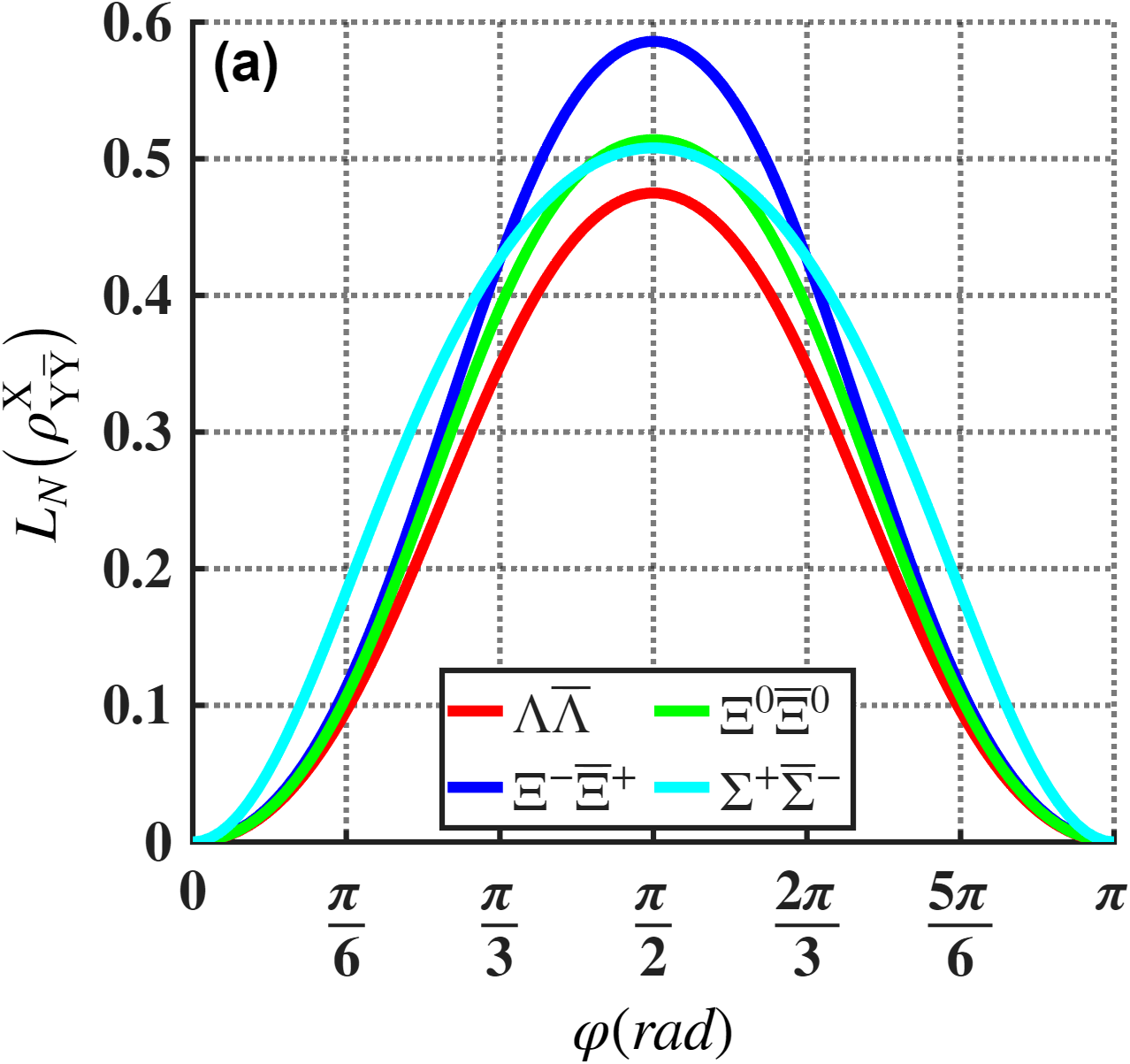}
    \includegraphics[scale=0.27]{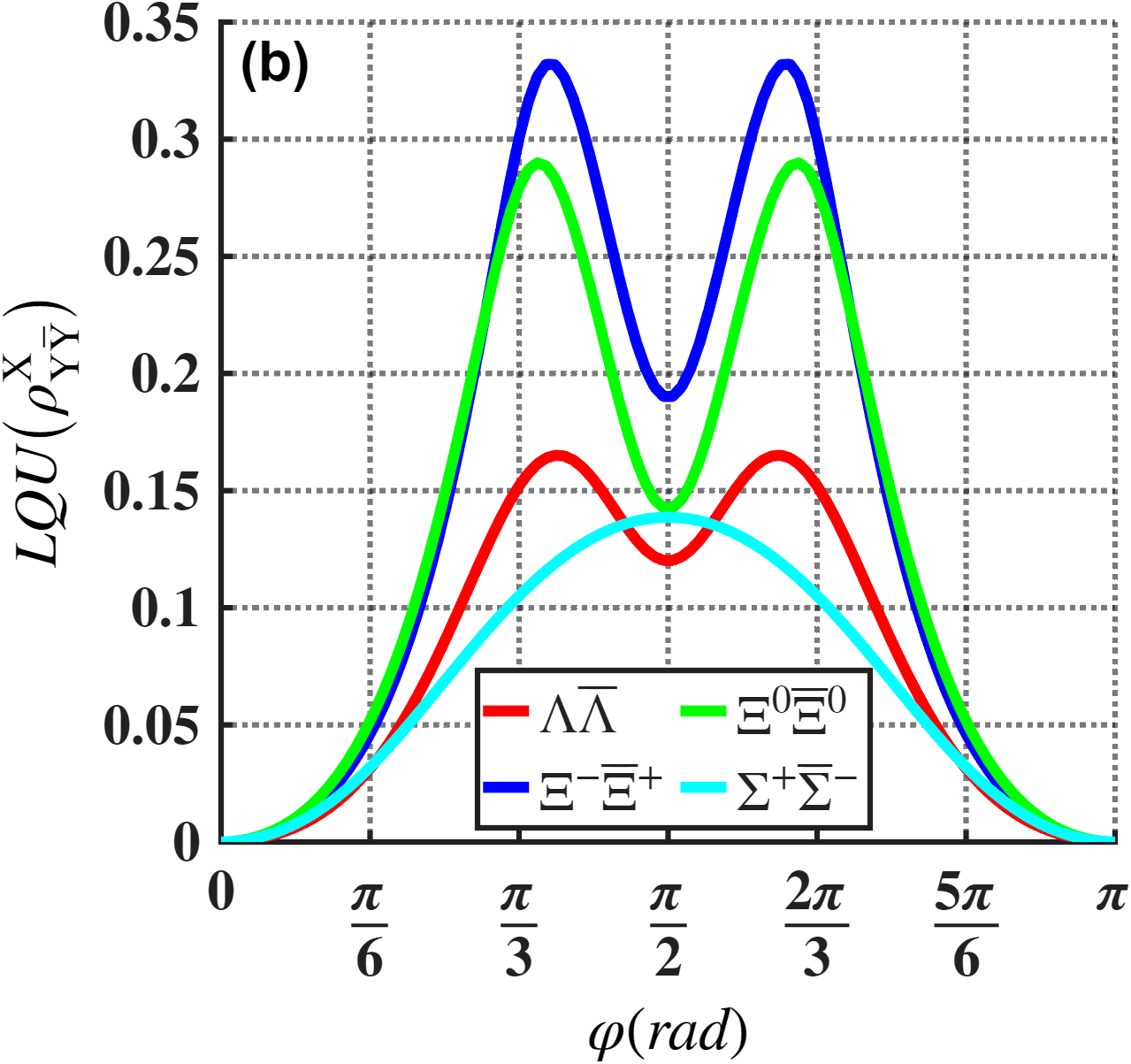}
    \includegraphics[scale=0.27]{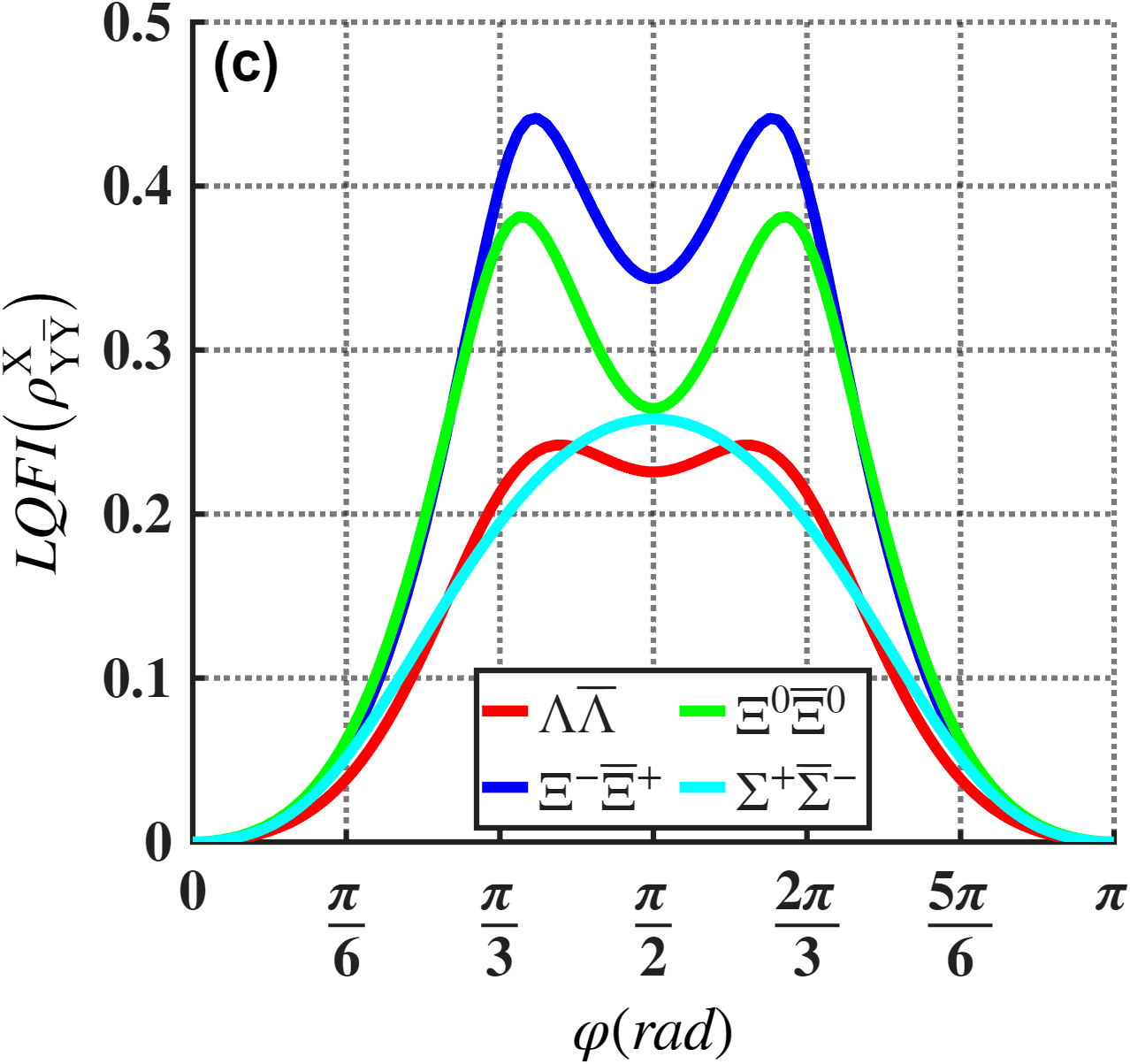}
\caption{Plot of the LN, LQU and LQFI as a function of the scattering angle $\varphi$ in $e^+e^- \to J/\psi \to \text{Y} \bar{\text{Y}}$ for various decay channels: $\Lambda\bar{\Lambda}$, $\Sigma^+\bar{\Sigma}^-$, $\Xi^0\bar{\Xi}^0$, and $\Xi^-\Xi^+$, taking into account the experimental parameters as in table \ref{t1}.}
\label{fig:L}
\end{figure}

Using the experimental parameters from Table~\ref{t1}, the foremost objective of this investigation, as depicted in Fig.~\ref{fig:L}(a-c), is to analyze the logarithmic negativity (LN), local quantum uncertainty (LQU), and local quantum Fisher information (LQFI) for four decay channels ($\Lambda\bar{\Lambda}$, $\Sigma^+\bar{\Sigma}^-$, $\Xi^0\bar{\Xi}^0$, and $\Xi^-\bar{\Xi}^+$) as functions of $\varphi$, in the absence of dephasing effects. This figure shows that the LN exhibits symmetry with respect to $\varphi=\pi/2$ within the range $\varphi\in[0,\pi]$. As shown in Fig.~\ref{fig:L}(a), as the scattering angle $\varphi$ increases from $\varphi=0$ towards $\pi/2$, the LN of the four hyperons exhibits a monotonic increase from $L_{N}=0$ at $\varphi=0$. Subsequently, LN monotonically diminishes after reaching its maximum value at $\varphi = \pi/2$ (see Table~\ref{tab:t2}), vanishing at $\varphi = \pi$, also displayed in Fig.~\ref{fig:L}(a). 

\begin{table}[h!]
\centering
\caption{The maximum logarithmic negativity $L_{Nmax}$ (as defined in Eq. (\ref{eq:N3})) and its corresponding angle $\varphi_{max}$ in the reaction $e^{+}e^{-} \rightarrow J/\psi \rightarrow \text{Y} \overline{\text{Y}}$.}
\begin{tabular}{c c c c c}
\hline
\hline
$\text{Y}\bar{\text{Y}}$ & $\Lambda\bar{\Lambda}$ & $\Sigma^{+}\bar{\Sigma}^{-}$ &  $\Xi^{-}\bar{\Xi}^{+}$ &  $\Xi^{0}\bar{\Xi}^{0}$  \\
\hline
$L_{Nmax}$ & 0.475 & 0.508  & 0.586  & 0.514   \\
\hline
$\varphi_{max}$ & $90^\circ$ & $90^\circ$ & $90^\circ$ & $90^\circ$\\
\hline
\hline  
\end{tabular}
\label{tab:t2}
\end{table}
When $\varphi=\pi/2$, the mixed state $\rho_{\text{Y}\bar{\text{Y}}}^{\text{X}}$, as described by matrix (\ref{eq:varrho}), assumes a particularly simple form and can be explicitly written as
\begin{equation}
\rho_{\text{Y}\bar{\text{Y}}}^{\text{X}} =\frac{1}{4}
\begin{pmatrix}
1 - \alpha_{\psi} & 0 & 0 & 1 + \alpha_{\psi} \\
0 & 1 + \alpha_{\psi} & 1 + \alpha_{\psi} & 0 \\
0 & 1 + \alpha_{\psi} & 1 + \alpha_{\psi} & 0 \\
1 + \alpha_{\psi} & 0 & 0 & 1 - \alpha_{\psi}
\end{pmatrix}
\label{eq:varrhoTh}
\end{equation}
Its eigenvalues are given by
\begin{equation}
\e_1 = \frac{1}{2}, \quad 
\e_2 = \frac{-\alpha_{\psi}}{2}, \quad
\e_3 = \frac{1}{2}(1 + \alpha_{\psi}), \quad 
\e_4 = 0.
\label{eq:epsilon_values}
\end{equation}
Notably, one of the eigenvalues becomes negative as $\alpha_{\psi} > 0$, signaling the presence of entanglement in the state. In particular, the negativity is, in this case, directly proportional to the parameter $\alpha_{\psi}$. It is given by
\begin{equation}
L_{N}(\rho^{\text{X}}_{\text{Y}\bar{\text{Y}}}) = \alpha_{\psi},
\end{equation}
Thus $L_{N}(\rho^{\text{X}}_{\text{Y}\bar{\text{Y}}})$ is independent of the relative phase $\Delta\phi$, highlighting that, for $\varphi = \pi/2$, the degree of entanglement is solely governed by $\alpha_{\psi}$. It is also noteworthy that the $\Xi^{-}\bar{\Xi}^{+}$ hyperon-antihyperon shows stronger quantum correlations as the scattering angle approaches $\pi/2$, highlighting the significance of this angle in high-energy phenomena. This behavior indicates a significant impact of the scattering angle on the entanglement between different hyperons. Furthermore, these observations underscore the intricate nature of particle interactions, particularly the species-dependent response of hyperons to angular variations. Such insights are essential for advancing our understanding of particle physics and the fundamental dynamics governing hyperon interactions.

\begin{table}[!h]
\centering
\caption{The maximum LQU values and their corresponding angles $\varphi_{\text{max}}$ observed in the reaction $e^{+}e^{-} \rightarrow J/\psi \rightarrow \text{Y} \bar{\text{Y}}$.}
\begin{tabular}{c c c c c}
\hline
\hline
$\text{Y}\bar{\text{Y}}$ & $\Lambda\overline{\Lambda}$ & $\Sigma^{+}\overline{\Sigma}^{-}$ &  $\Xi^{-}\overline{\Xi}^{+}$ &  $\Xi^{0}\overline{\Xi}^{0}$  \\
\hline
$LQU_{max}$ & $0.165$  & $0.138$  & $0.332$  & $0.290$   \\
\hline
$\varphi_{max}$ & $108^\circ$,$71.94^\circ$ & $90^\circ$ & $113,45^\circ$,$66.54^\circ$ & $115,72^\circ$,$64.27^\circ$\\
\hline
\hline  
\end{tabular}
\label{tab:t3}
\end{table}

We now examine how the scattering angle $\varphi$ influences LQU. In Fig.~\ref{fig:L}(b), we present LQU versus $\varphi$ for various decay channels ($\Lambda\bar{\Lambda}$, $\Sigma^+\bar{\Sigma}^-$, $\Xi^0\bar{\Xi}^0$, and $\Xi^-\Xi^+$). Symmetric around $\varphi=\frac{\pi}{2}$, the LQU of $\text{Y}\bar{\text{Y}}$ pairs is observed throughout the $\varphi$ range, excluding the collinear limits where $\varphi=0$ and $\varphi=\pi$. However, the maximum LQU for $\Lambda\bar{\Lambda}$, $\Xi^0\bar{\Xi}^0$, and $\Xi^-\Xi^+$ pairs is not necessarily at $\varphi=\pi/2$, contrasting with the behavior observed for entanglement negativity (LN). The maximum can instead be observed at different angles, which are listed in Table~\ref{tab:t3}. An exception is $\Sigma^+\bar{\Sigma}^-$, which retains its maximum at $\varphi=\pi/2$, as displayed in Fig.~\ref{fig:L}(b). This result highlights that even when particles are perfectly aligned at certain angles, quantum fluctuations significantly impact the behavior of the $\text{Y}\bar{\text{Y}}$ pairs.

In Fig.~\ref{fig:L}(c), we present the evolution of the Local quantum Fisher Information (LQFI) as a function of $\varphi$ without a dephasing effect. It can be observed that the evolution of LQFI and LQU (see Fig.~\ref{fig:L}(b)) exhibit approximately similar behavior versus $\varphi$ for all four decay channels ($\Lambda\bar{\Lambda}$, $\Sigma^+\bar{\Sigma}^-$, $\Xi^0\bar{\Xi}^0$, and $\Xi^-\Xi^+$) (See Table \ref{tab:t4}).

\begin{table}[!h]
\centering
\caption{LQFI maxima and their respective angles $\varphi_{\text{max}}$ as observed in $e^{+}e^{-} \rightarrow J/\psi \rightarrow \text{Y} \overline{\text{Y}}$ production.}
\begin{tabular}{c c c c c}
\hline
\hline
$\text{Y}\bar{\text{Y}}$ & $\Lambda\overline{\Lambda}$ & $\Sigma^{+}\overline{\Sigma}^{-}$ &  $\Xi^{-}\overline{\Xi}^{+}$ &  $\Xi^{0}\overline{\Xi}^{0}$  \\
\hline
$LQFI_{max}$ & $0.241$  & $1.57$  & $0.441$  & $0.381$   \\
\hline
$\varphi_{max}$ & $108^\circ$,$71.94^\circ$ & $90^\circ$ & $113.45^\circ$,$66.54^\circ$ & $115.72^\circ$,$64.27^\circ$\\
\hline
\hline  
\end{tabular}
\label{tab:t4}
\end{table}
\subsection{Under dephasing effect}

This subsection investigates the three-dimensional evolution of LN, LQU, and LQFI in $e^{+}e^{-}\to \Lambda\bar{\Lambda}$. We will examine their dependence on the parameters $s$ and $\varphi$ across different decoherence channels, including Amplitude Damping (AD), Phase Damping (PD), and Phase Flip (PF). The results obtained for this hyperon under these channels also apply to the hyperons $\Xi^{-}$, $\Xi^{0} $, and $\Sigma^{+}$.

%\textcolor{red}{Based respectively on Eqs.~(\ref{eq:N3}), (\ref{eq:w2}), and (\ref{eq:LQFI}), we investigate the evolution of the logarithmic negativity (LN), the local quantum uncertainty (LQU), and the local quantum Fisher information (LQFI) under the three types of decohering channels, using the final expressions corresponding to the AD (\ref{eq:AD}), PF (\ref{eq:PF}), and PD (\ref{eq:PD}) channels}.

\begin{figure}[!h]
\includegraphics[scale=0.3]{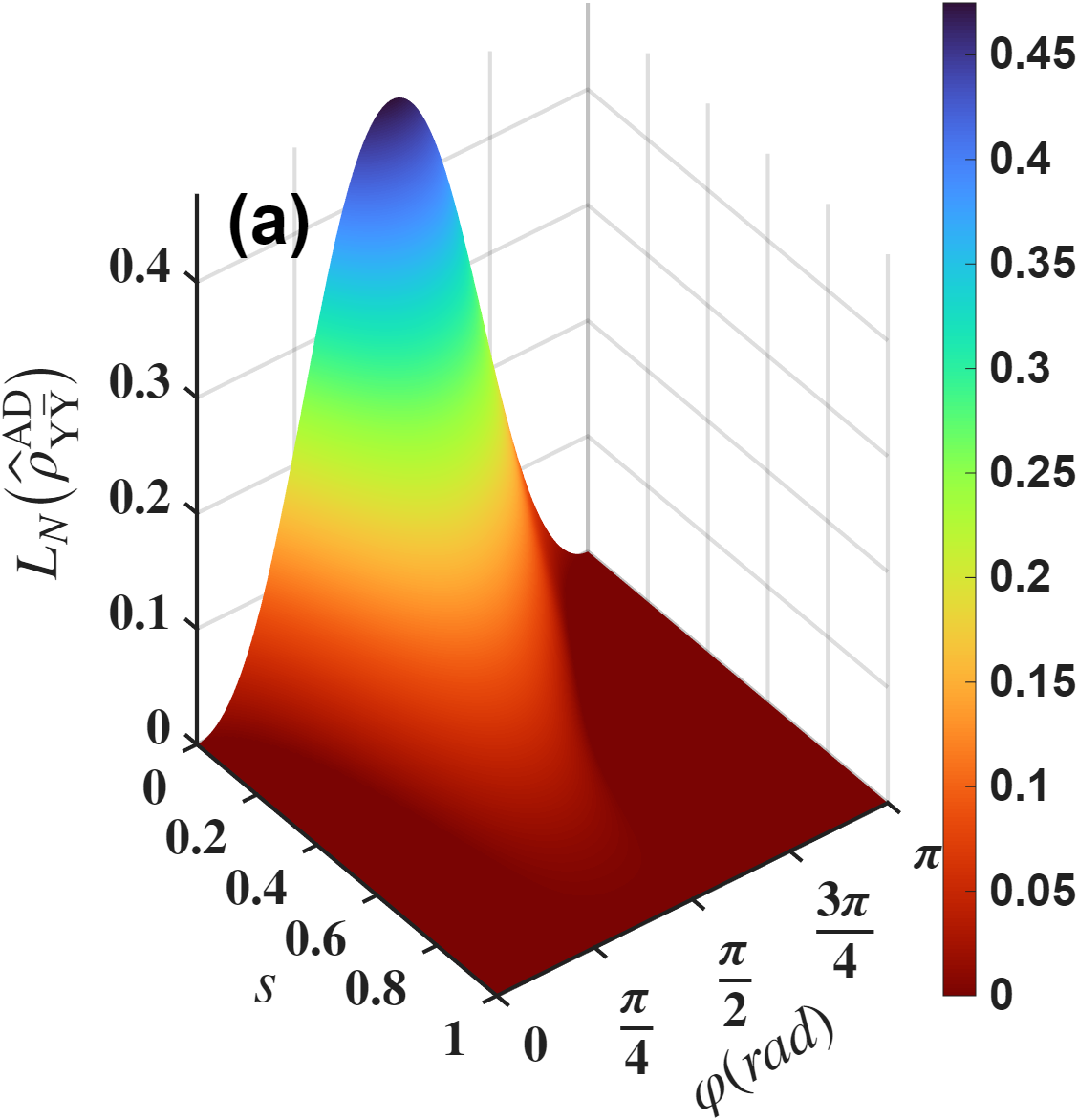}
\includegraphics[scale=0.3]{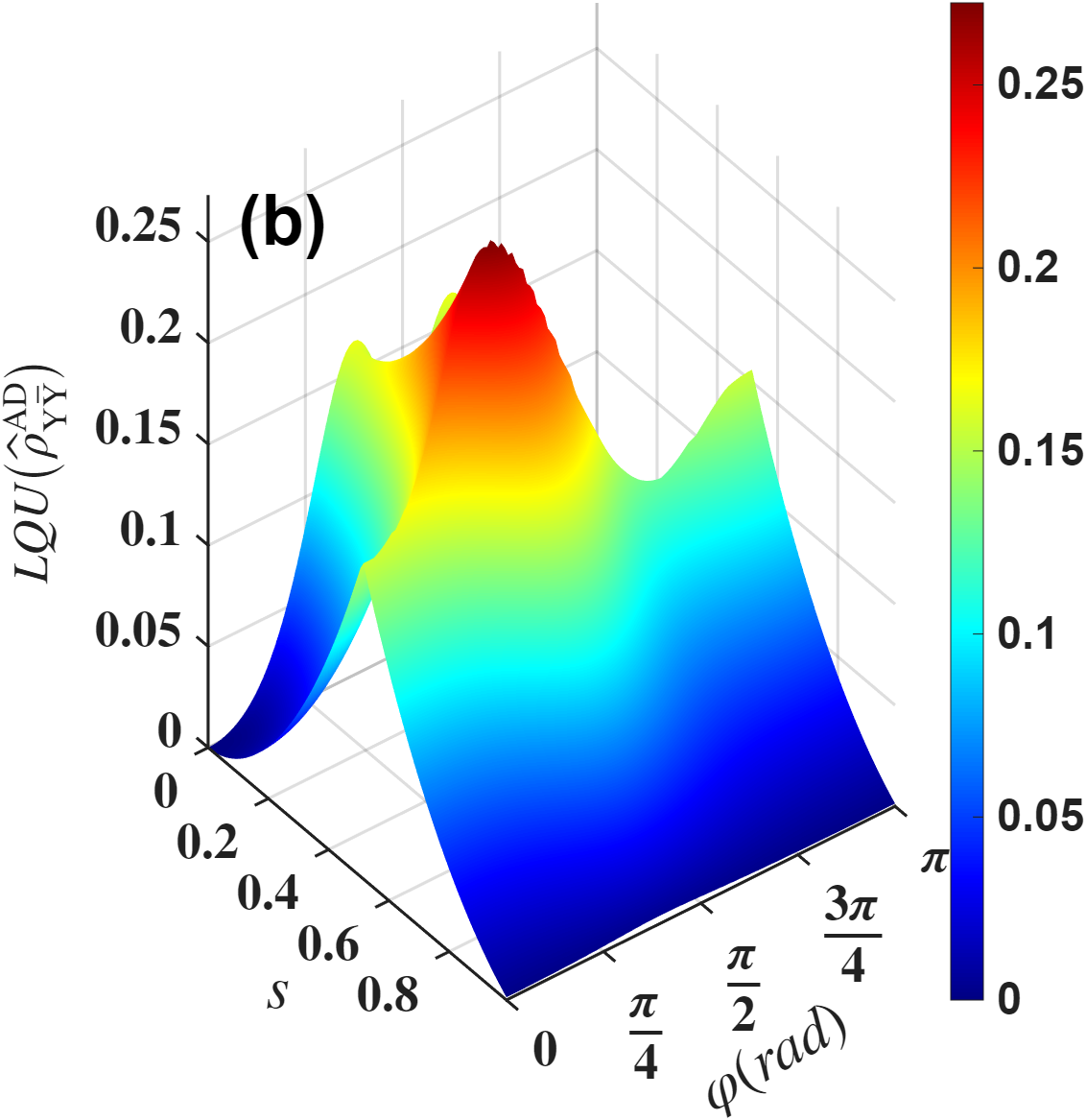}
\includegraphics[scale=0.3]{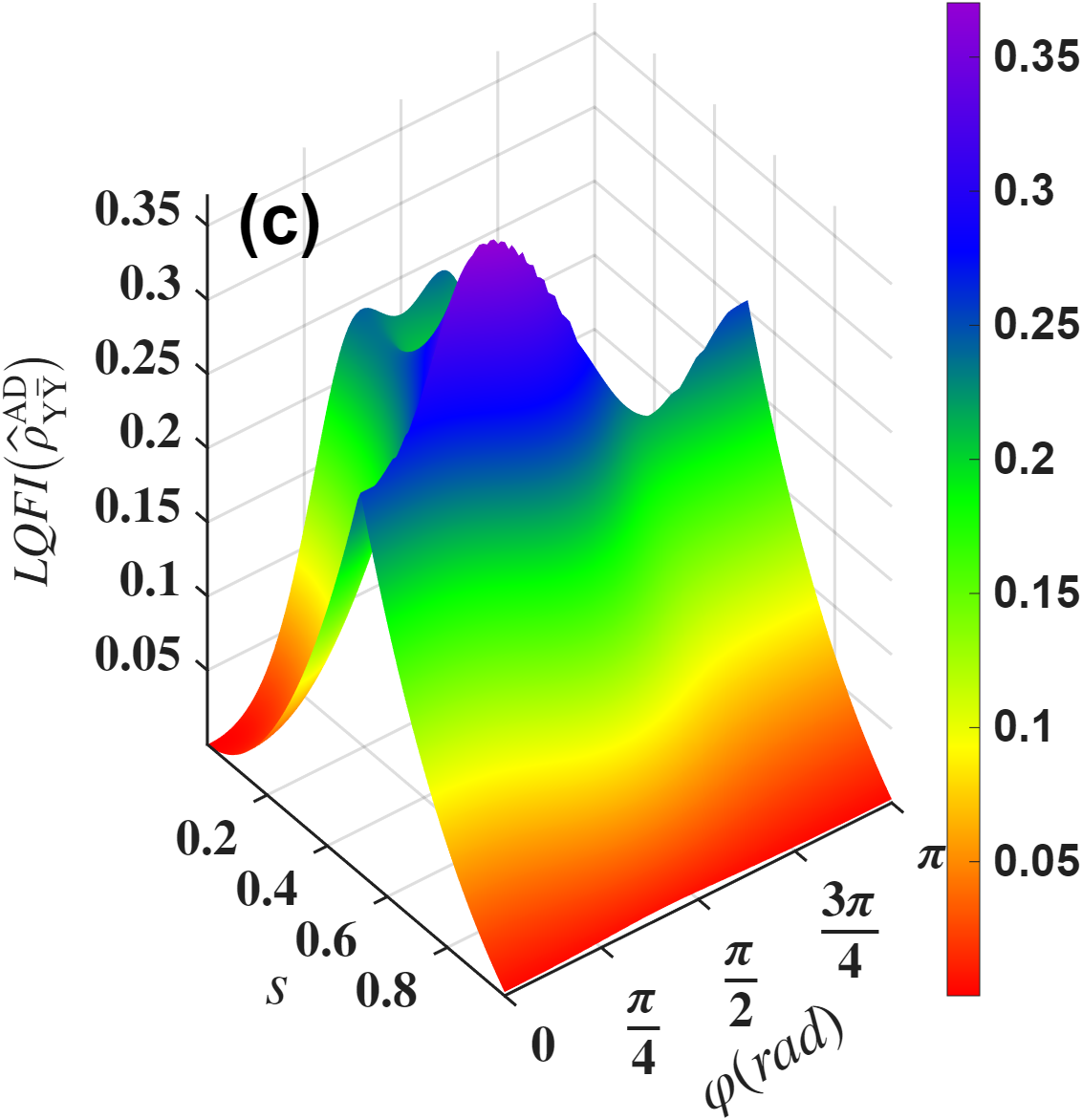}
\caption{Plot of the LN (a), LQU (b) and LQFI (c) versus $\text{s}$ and the scattering angle $\varphi$ in $e^+e^- \to J/\psi \to \Lambda \bar{\Lambda}$, for AD channel. Using the parameters value setting in the Table~\ref{t1}.}
\label{fig:NAD}
\end{figure}

Figure~\ref{fig:NAD}(a) explores the three-dimensional evolution of LN under the AD channel, highlighting the persistence of quantum correlations in the presence of decoherence. LN reaches its maximum at $s=0$ and $\varphi=\pi/2$. As $s$ increases, the AD channel's effect leads to a monotonically decreasing LN, as shown in Fig.~\ref{fig:NAD}(a), ultimately causing its complete vanishing. This behavior underscores the significant impact of decoherence on quantum systems and the challenges it poses for Quantum Information Processing. Figures \ref{fig:NAD}(b) and \ref{fig:NAD}(c) show the evolution of the LQU and the LQFI under the Amplitude Damping (AD) channel. Both LQU and LQFI initially increase, reaching a maximum value, and then decrease as the parameter $s$ increases.

\begin{figure}[!h]
\includegraphics[scale=0.3]{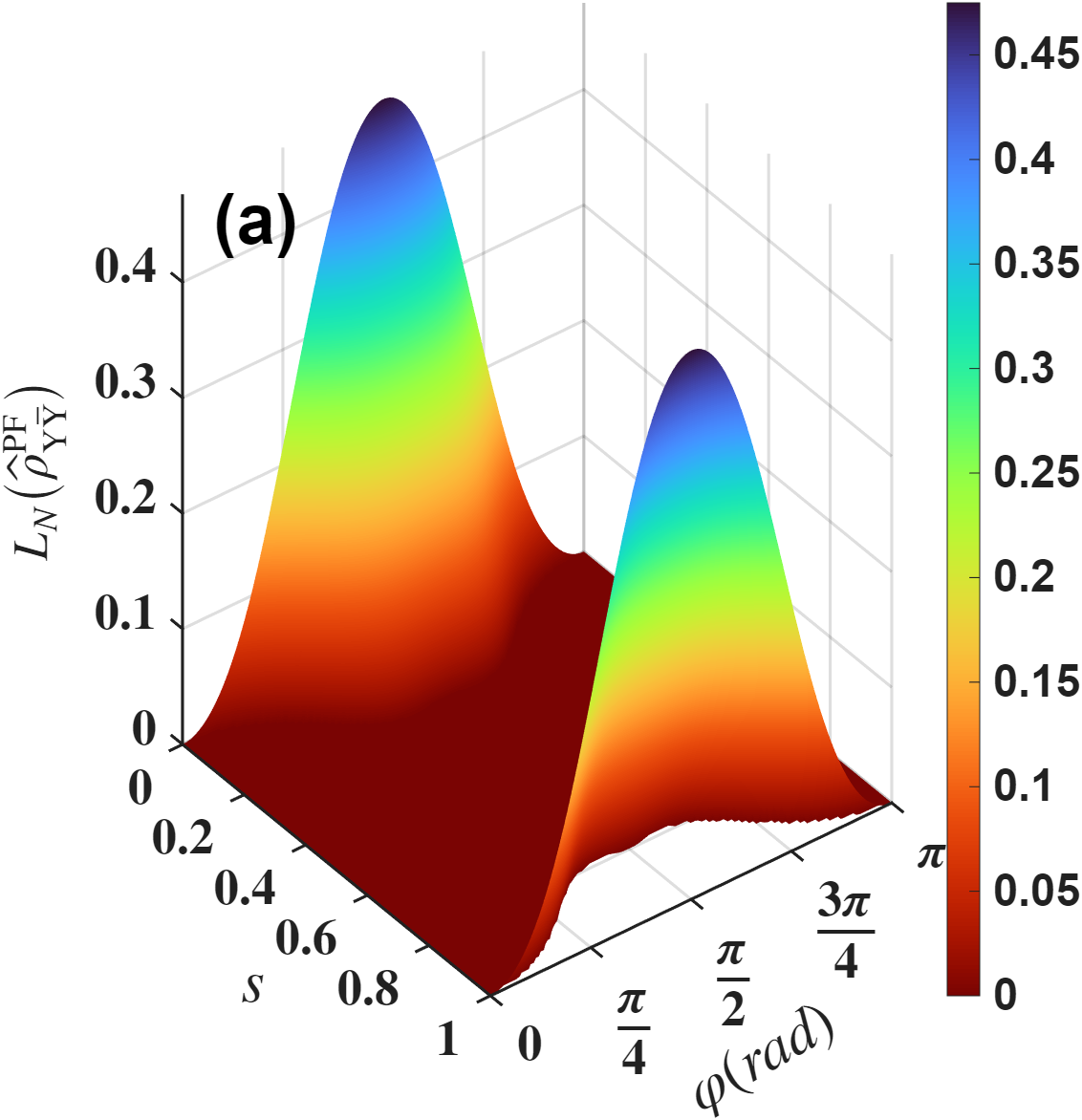}
\includegraphics[scale=0.3]{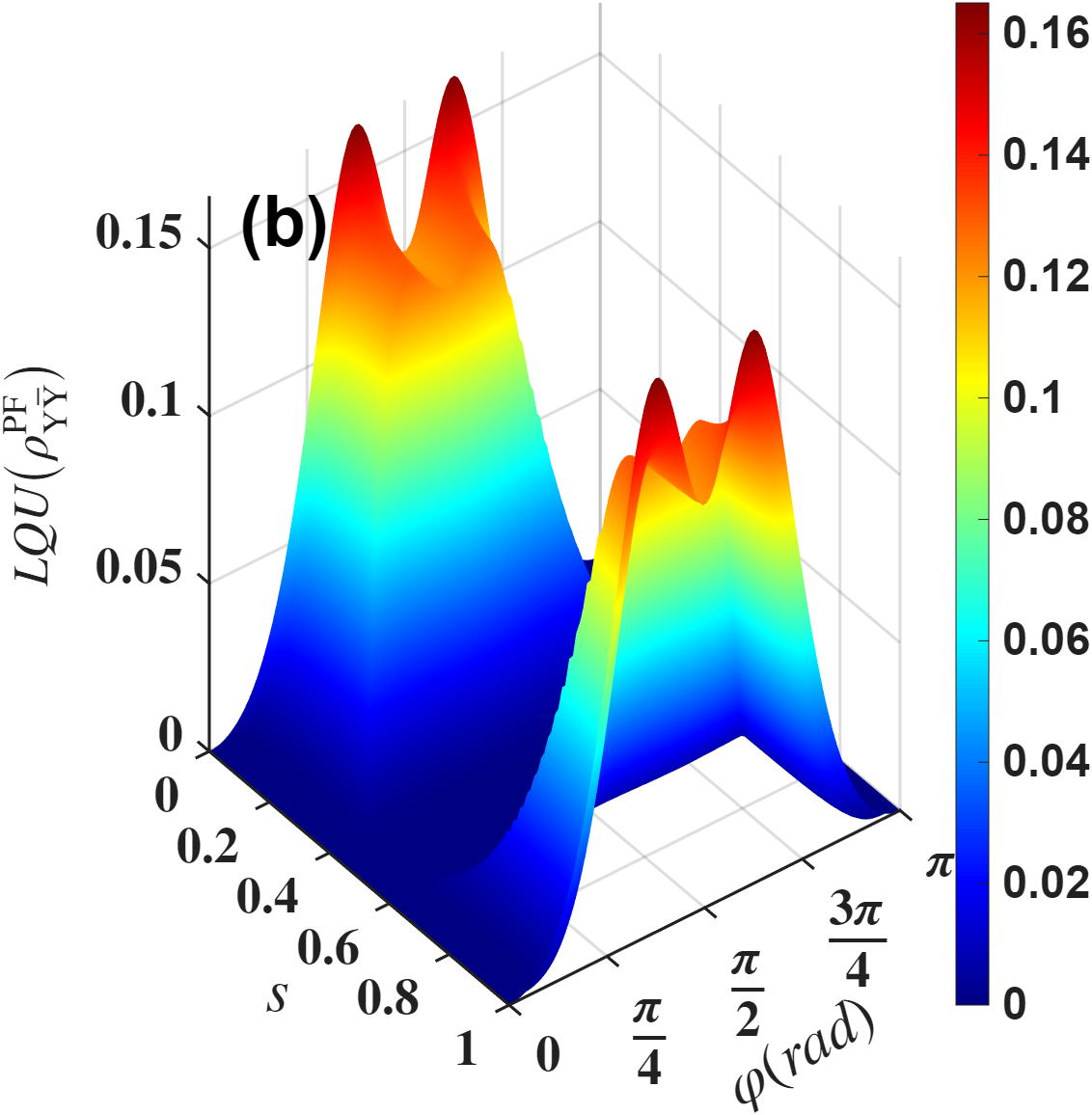}
\includegraphics[scale=0.3]{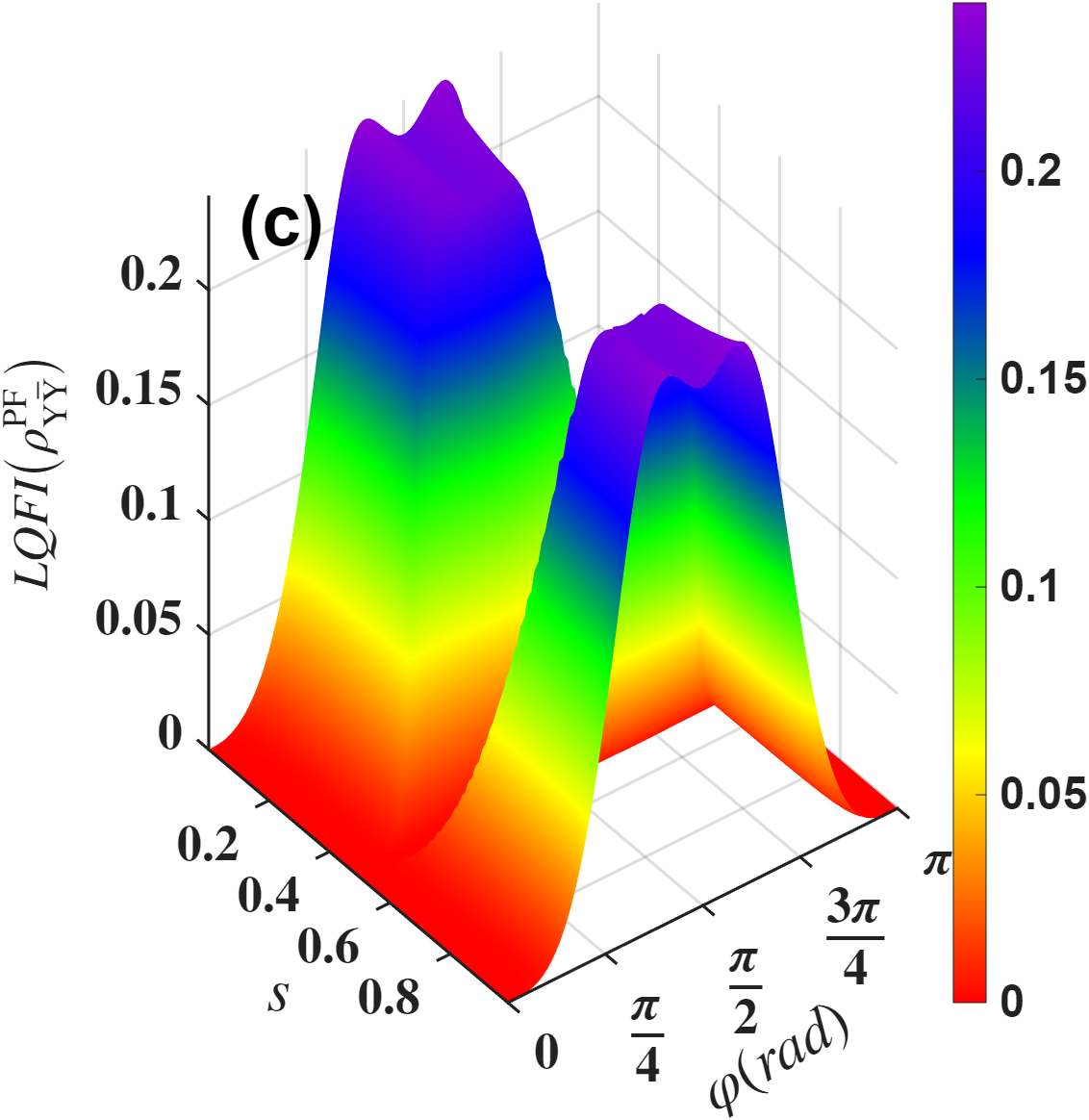}
\caption{Plot of the LN (a), LQU (b) and LQFI (c) versus $\text{s}$ and the scattering angle $\varphi$ in $e^+e^- \to J/\psi \to \Lambda \bar{\Lambda}$, for PF channel. Using the parameters value setting in the Table~\ref{t1}.}
\label{fig:NPF}
\end{figure}

\begin{figure}[!h]
\includegraphics[scale=0.3]{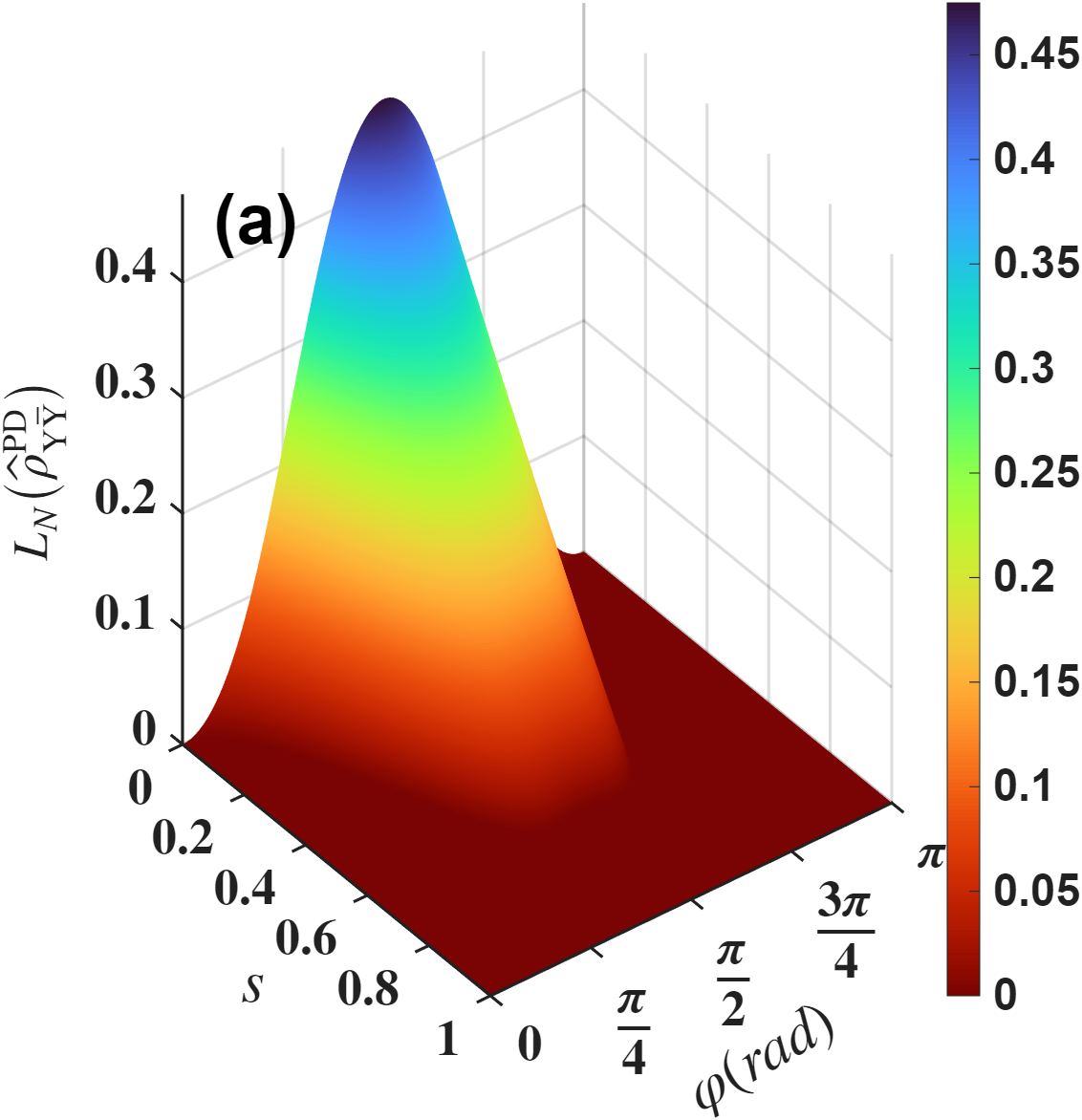}
\includegraphics[scale=0.3]{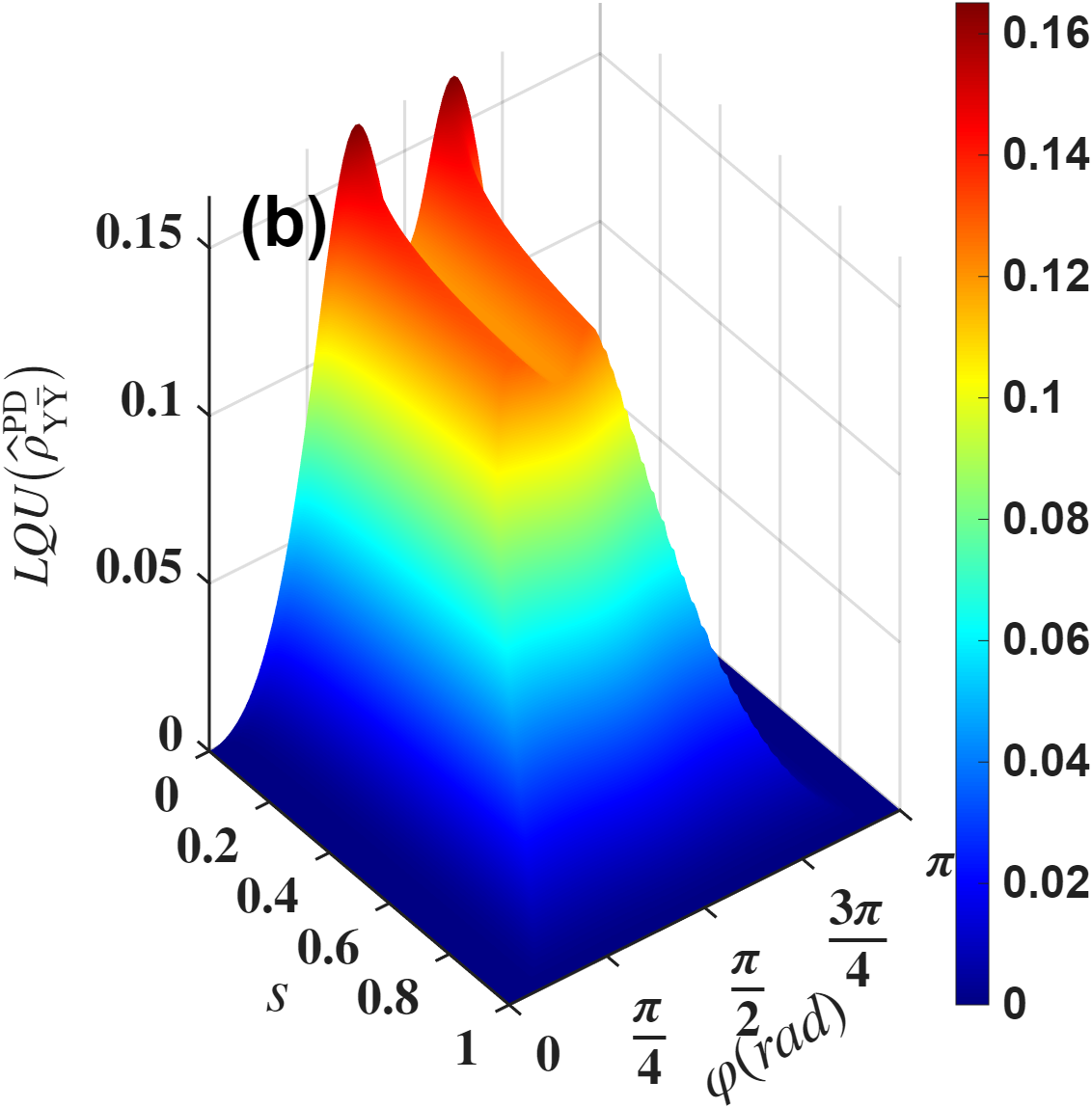}
\includegraphics[scale=0.3]{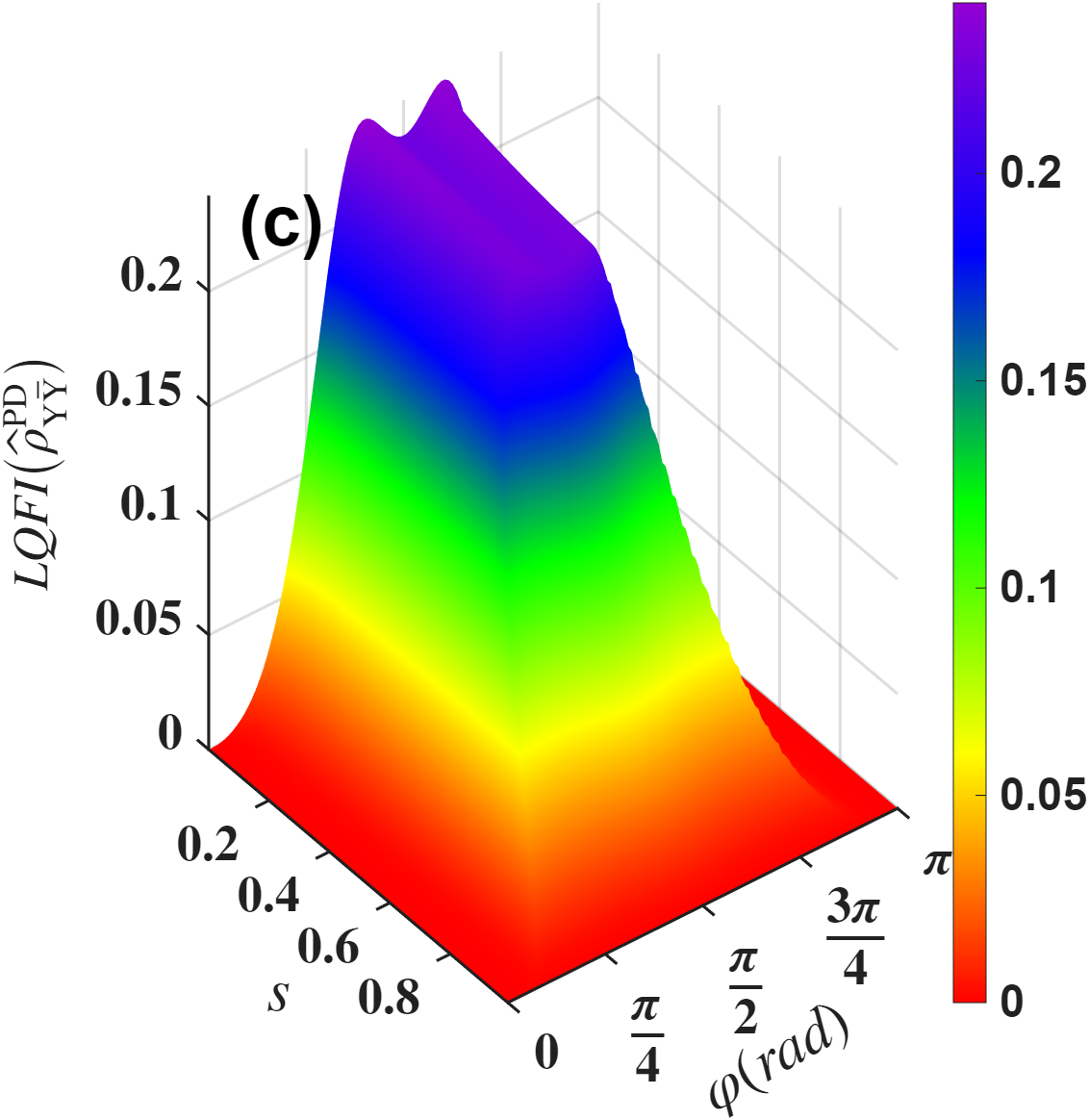}
\caption{Plot of the LN (a), LQU (b) and LQFI (c) versus $\text{s}$ and the scattering angle $\varphi$ in $e^+e^- \to J/\psi \to \Lambda \bar{\Lambda}$, for PD channel. Using the parameters value setting in the Table~\ref{t1}.}
\label{fig:NPD}
\end{figure}

\begin{figure}[!h]
\begin{center}
\includegraphics[scale=0.33]{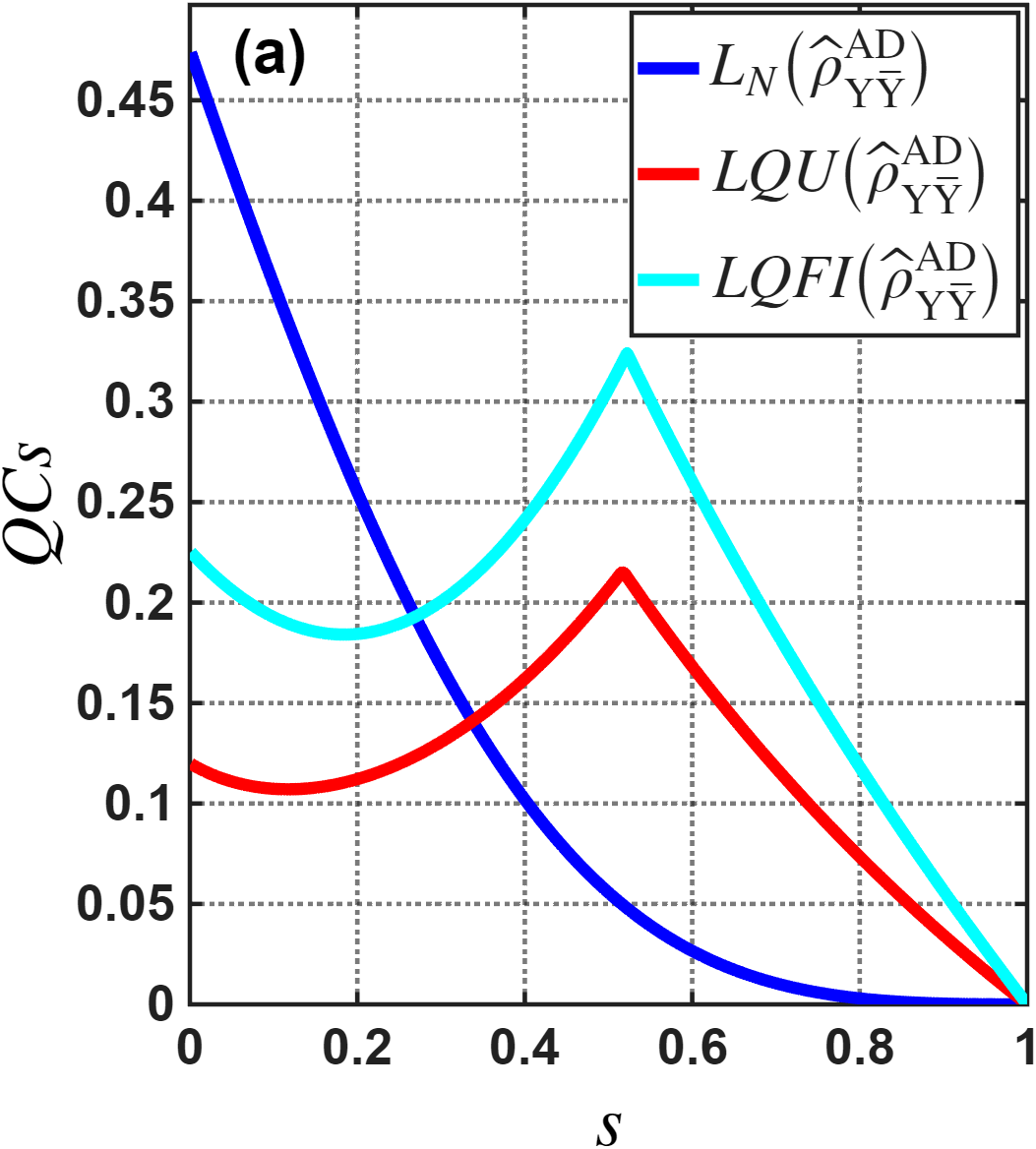}
\includegraphics[scale=0.33]{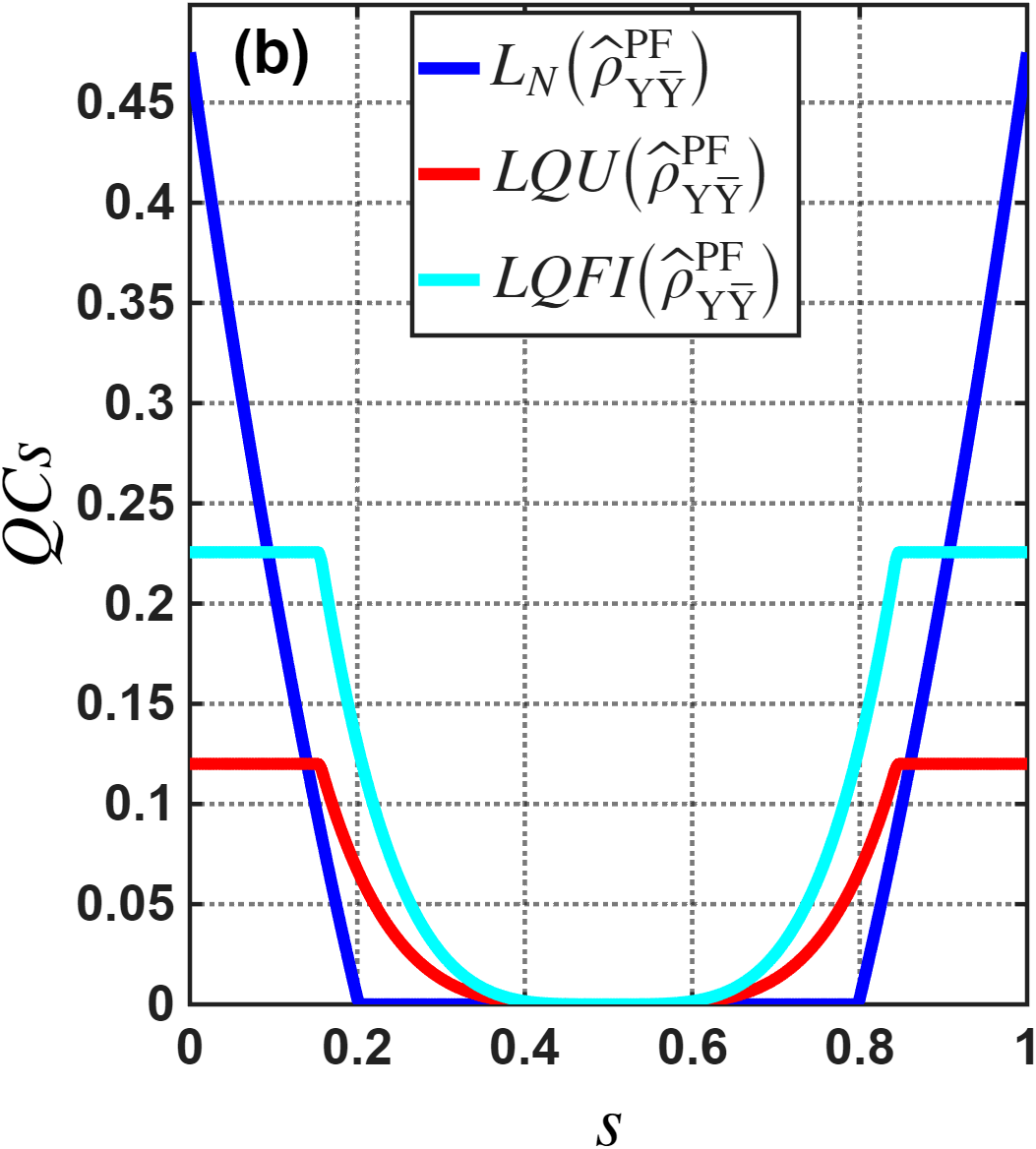}
\includegraphics[scale=0.33]{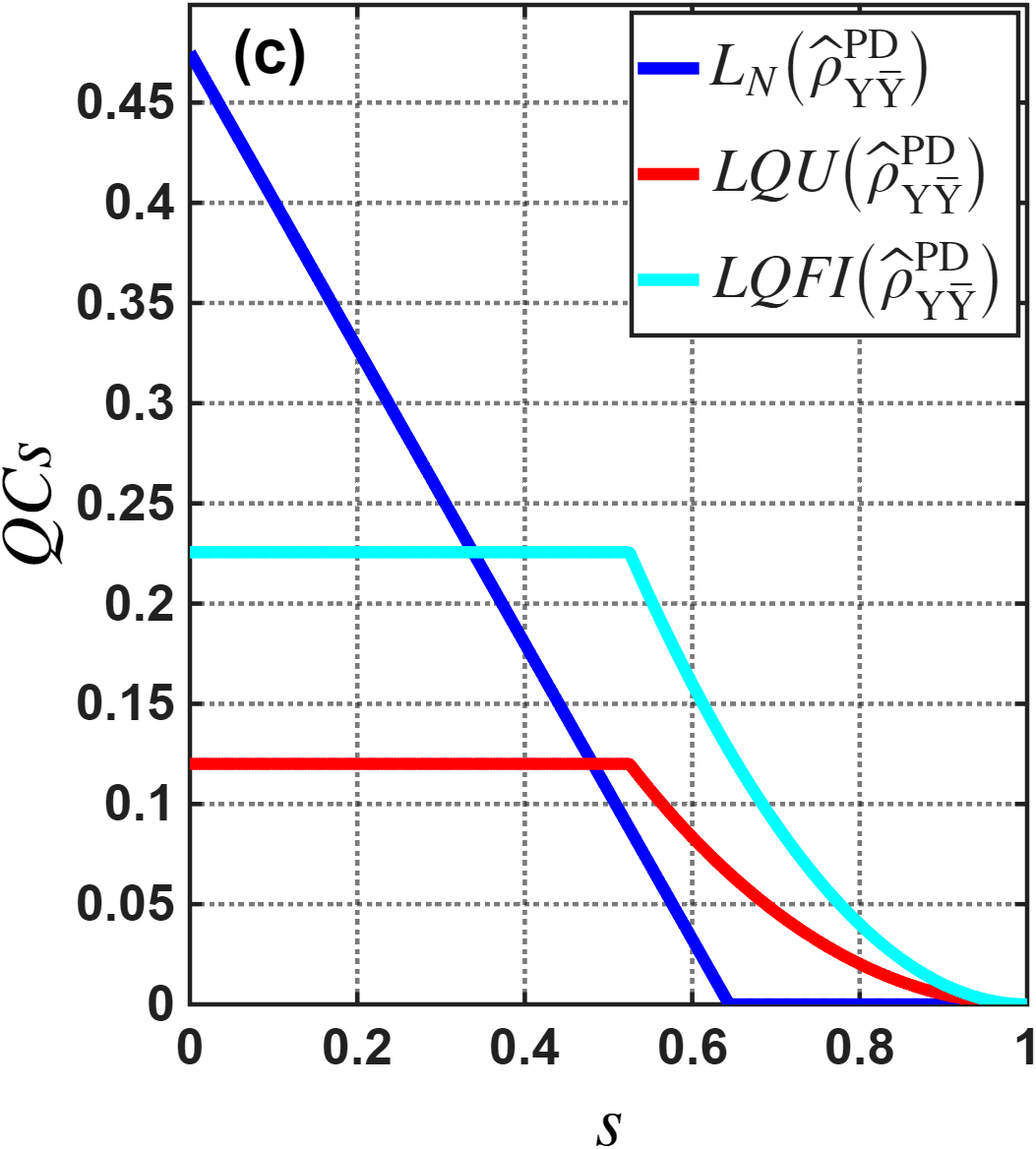}
\end{center}
\caption{Plot of LN, LQFI, and LQU versus the decoherence parameter $s$ with $\theta=\pi/2$ for hyperon pair $\text{Y}=\Lambda$, under Amplitude Damping (AD), Phase Flip (PF), and Phase Damping  (PD) channels. Taking into account the experimental parameters as in table \ref{t1}}
\label{fig:Dep}
\end{figure}

%%%%%%%%%%%%%%%%%%%%%%%%%%%%%%
The evolution of LN, as illustrated in Fig.~\ref{fig:NPF}(a), reveals essential information about the interaction between the scattering angle and the decoherence parameter $\text{s}$ under the PF channel. Notably, the LN reaches its maximum at a scattering angle of $\varphi=\pi/2$ and for the decoherence values $\text{s}=0$ and $\text{s}=1$. A detailed analysis of the evolution of LN as a function of $\text{s}$ highlights several distinct regimes. In the range $\text{s} \in [0, 0.2]$, the negativity decreases monotonically, as shown in Fig.~\ref{fig:Dep}, indicating a loss of entanglement as decoherence begins to affect the system. This trend continues until $\text{s}$ reaches $0.2$, where the LN becomes zero, suggesting complete decoherence. For $\text{s} \in [0.2,0.8]$, the LN remains zero, indicating the total absence of entanglement. However, when $\text{s}$ exceeds $0.8$, the LN reappears and increases monotonically until $\text{s}=1$. This behavior highlights the nonlinear dynamics of entanglement under the effect of decoherence and underscores the existence of regimes where entanglement can be partially restored despite environmental influence. The Phase Flip (PF) channel causes LQU and LQFI to decrease until they vanish around $s=0.5$, only to return to their previous maximum values thereafter, as shown in Figures \ref{fig:NPF}(b) and \ref{fig:NPF}(c).

%%%%%%%%%%%%%%%%%%%%%%%%%%%%%%%%%%%

The phase damping (PD) channel exhibits a behavior similar to that of the amplitude damping (AD) channel, particularly in terms of its impact on LN. In Fig. \ref{fig:NPD}(a) and Fig. \ref{fig:Dep}, LN reaches its maximum when the decoherence parameter is zero and $\varphi =\pi/2$, demonstrating the resilience of quantum correlations against environmental interactions. However, as the decoherence parameter $\text{s}$ increases, the PD channel induces a linear decrease in LN, ultimately leading to its complete disappearance. This highlights the fundamental role of quantum decoherence in reducing quantum correlations and underscores the necessity of strategies to preserve these correlations in quantum information processing. The Phase Damping (PD) channel exhibits a distinct pattern: LQU and LQFI remain relatively stable for low values of $s$ but decline more rapidly than in the Amplitude Damping (AD) channel as $\text{s}$ increases, as illustrated in Figures \ref{fig:NPD}(b) and \ref{fig:NPD}(c). Overall, while the general trends of LQU and LQFI are similar across these channels, their specific dynamics underscore the importance of understanding each distinct decoherence mechanism in quantum information processing.

In Fig. \ref{fig:Dep}(a-c), we plot LN, LQU and LQFI as a function of the decoherence parameter $s$ for the $e^{+}e^{-}\to \Lambda\bar{\Lambda}$ process under AD, PF and PD channels. As shown in Fig.~\ref{fig:Dep}(a) and Fig.~\ref{fig:Dep}(c), LN decreases monotonically with increasing $\text{s}$, vanishing as $s \to 1$. It can be seen that under the amplitude damping (AD) channel, the local quantum uncertainty (LQU) and local quantum Fisher information (LQFI) decrease after reaching their maximum at $s=0.5$, as depicted in Fig~\ref{fig:Dep}(a). Besides, the Fig.~\ref{fig:Dep}(b) indicates that, under the phase flip (PF) channel, LQU and LQFI remain resistant to decoherence for small values of $s$. They then decrease until $s=0.5$ before increasing again to recover their initial values, subsequently remaining robust against decoherence as $s \to 1$. Moreover, Fig.~\ref{fig:Dep}(c) illustrates that, for the phase damping (PD) channel, both quantities remain resistant to decoherence effects up to $s=0.5$ and then gradually vanish as $s \to 1$. Moreover, LQU and LQFI exhibit a similar evolution, both becoming zero at $s=1$. LQU and LQFI can also remain positive even when LN is zero, as depicted in Fig.~\ref{fig:Dep}(a-c). Therefore, LQU and LQFI are more robust than entanglement under the effect of decoherence. This indicates that LQU and LQFI quantify quantum correlations even in the absence of entanglement. Additionally, we notice that LQU is always bounded by LQFI, as shown in Fig.~\ref{fig:Dep}(a-c).

Let's now discuss the evolution of LN, LQU, and LQFI under the three decoherence channels (AD, PF, and PD). We remark that LN, LQU, and LQFI exhibit symmetric behavior under PF channel. It can be seen that LN is more robust under PF than AD and PD for $s\in[0.8,1]$. However, LN is zero under PF for $s\in[0.2,0.8]$. This indicates greater robustness of AD and PD compared to PF for a specific region of $s$. We also note that LQU and LQFI are more robust under PF than AD and PD channels, particularly in the interval $s \in [0.6,1]$ as shown in Fig.~\ref{fig:Dep}(b), and in the region $s \in [0,0.5]$, as illustrated in Fig.~\ref{fig:Dep}(c). 

\section{Quantum teleportation}\label{sec:6}

Quantum teleportation enables the transfer of an unknown quantum state using shared entanglement. In this study, we examine quantum teleportation using mixed entangled states as a resource, acting as a generalized depolarization channel \cite{Te1,Te2,Te3}. We analyze the influence of the diffusion angles $\varphi$, highlighting their key role in optimizing quantum teleportation protocols. Let us assume the input state is an arbitrary unknown two-qubit pure state $\ket{\psi_{\text{in}}}$, such as
\begin{equation}
\ket{\psi_{\text{in}}} = \cos\left(\frac{\theta}{2}\right)\ket{01} + e^{i\phi} \sin\left(\frac{\theta}{2}\right)\ket{10},
\end{equation}
where, $\theta \in [0, \pi]$ describes all possible states with varying amplitudes, while $\phi \in [0, 2\pi]$ represents their phase. The corresponding density matrix writes as

\begin{align}
\hat{\varrho}_{\text{in}} &= \ket{\psi_{\text{in}}}\bra{\psi_{\text{in}}} \notag \\
&=
\begin{pmatrix}
0 & 0 & 0 & 0 \\
0 & \sin^2\!\left(\tfrac{\theta}{2}\right) & \tfrac{1}{2} e^{i\phi} \sin\theta & 0 \\
0 & \tfrac{1}{2} e^{-i\phi} \sin\theta & \cos^2\!\left(\tfrac{\theta}{2}\right) & 0 \\
0 & 0 & 0 & 0
\end{pmatrix}.
\label{eq:in}
\end{align}

A quantum channel is mathematically defined as a completely positive and trace-preserving (CPTP) map, essential for transforming an input density operator into an output density operator~\cite{Te4}. In the context of quantum teleportation, this transformation is crucial as it enables the transmission of quantum information without the physical transport of the state itself~\cite{Te5}.

When considering mixed channels subject to noise, the output state can be represented by a mixed density operator. The process involves applying joint measurements and local unitary transformations to the input state $\hat{\varrho}_{in}$ to obtain the replicated output state $\hat{\varrho}_{out}$, hence
\begin{equation}
\hat{\varrho}^{\text{X}}_{out} = \sum_{i,j} p_{ij} \left( \tau_i \otimes \tau_j \right) \hat{\varrho}_{\text{in}} \left( \tau_i \otimes \tau_j \right),
\label{eq:out}
\end{equation}
Where $\tau_i$ ($i=0,x,y,z$) denote the identity matrix $\mathbb{I}_2$ and the three Pauli matrices. The probabilities are given by $p_{ij}=\text{Tr}[E_i \rho^{\text{X}}_{\text{Y}\bar{\text{Y}}}] \, \text{Tr}[E_j \rho^{\text{X}}_{\text{Y}\bar{\text{Y}}}]$ with $\sum p_{ij}=1$. The measurement operators correspond to the Bell states
\[
E_0 = \ket{\chi^-}\bra{\chi^-}, \quad
E_1 = \ket{\psi^-}\bra{\psi^-}, 
\]
\[
E_2 = \ket{\psi^+}\bra{\psi^+},\quad
E_3 = \ket{\chi^+}\bra{\chi^+},
\]
with
\[
\ket{\chi^\pm} = \frac{1}{\sqrt{2}}(\ket{01} \pm \ket{10}), \quad
\ket{\psi^\pm} = \frac{1}{\sqrt{2}}(\ket{00} \pm \ket{11}).
\]
Thereafter, by applying the local unitary transformation to the input state (Eq.~(\ref{eq:out})), the teleported output state at Bob’s lab remains an X-shaped matrix

\begin{equation}
\hat{\varrho}^{\text{X}}_{out}=
\begin{pmatrix}
\hat{\varrho}^{\text{X}}_{1,1} & 0 & 0 & \hat{\varrho}^{\text{X}}_{1,4} \\
0 & \hat{\varrho}^{\text{X}}_{2,2} & \hat{\varrho}^{\text{X}}_{2,3} & 0 \\
0 & \hat{\varrho}^{\text{X}}_{2,3} & \hat{\varrho}^{\text{X}}_{3,3}& 0 \\
\hat{\varrho}^{\text{X}}_{1,4} & 0 & 0 & \hat{\varrho}^{\text{X}}_{1,1}
\end{pmatrix}, \quad \text{where} \quad
\begin{aligned}
\hat{\varrho}^{\text{X}}_{1,1} &= 2 \rho_{2,2} (\rho_{1,1} + \rho_{4,4}), \\
\hat{\varrho}^{\text{X}}_{2,2} &= (\rho_{1,1} + \rho_{4,4})^2 \cos^2\left(\frac{\theta}{2}\right) + 4 \rho_{2,2}^2 \sin^2\left(\frac{\theta}{2}\right), \\
\hat{\varrho}^{\text{X}}_{3,3} &= (\rho_{1,1} + \rho_{4,4})^2 \sin^2\left(\frac{\theta}{2}\right) + 4 \rho_{2,2}^2 \cos^2\left(\frac{\theta}{2}\right), \\
\hat{\varrho}^{\text{X}}_{2,3} &= \hat{\varrho}_{3,2} = 2\left(\rho_{2,3}^2 e^{-i\phi} + \rho_{1,4}^2 e^{i\phi} \right) \sin\theta, \\
\hat{\varrho}^{\text{X}}_{1,4} &= 4 \rho_{2,3} \rho_{1,4} \sin\theta \cos\phi.
\end{aligned}
\label{eq:OF}
\end{equation}

When a hyperon--antihyperon input state $\text{Y}\bar{\text{Y}}$ is teleported through the amplitude-damping channel $\rho^{\text{AD}}_{\text{Y}\bar{\text{Y}}}$, the output replica state can be determined using Eqs.~(\ref{eq:AD}) and (\ref{eq:out})
\begin{equation}
\hat{\varrho}^{\text{AD}}_{out}=
\begin{pmatrix}
\hat{\varrho}^{\text{AD}}_{1,1} & 0 & 0 & \hat{\varrho}^{\text{AD}}_{1,4} \\
0 & \hat{\varrho}^{\text{AD}}_{2,2} & \hat{\varrho}^{\text{AD}}_{2,3} & 0 \\
0 & \hat{\varrho}^{\text{AD}}_{2,3} & \hat{\varrho}^{\text{AD}}_{3,3}& 0 \\
\hat{\varrho}^{\text{AD}}_{1,4} & 0 & 0 & \hat{\varrho}^{\text{AD}}_{1,1}
\end{pmatrix}, \quad \text{where} \quad
\begin{aligned}
\hat{\varrho}^{\text{AD}}_{1,1} &= 2 \eta_{2,2} (\eta_{1,1} + \eta_{4,4}), \\
\hat{\varrho}^{\text{AD}}_{2,2} &= (\eta_{1,1} + \eta_{4,4})^2 \cos^2\left(\frac{\theta}{2}\right) + 4 \eta_{2,2}^2 \sin^2\left(\frac{\theta}{2}\right), \\
\hat{\varrho}^{\text{AD}}_{3,3} &= (\eta_{1,1} + \eta_{4,4})^2 \sin^2\left(\frac{\theta}{2}\right) + 4 \eta_{2,2}^2 \cos^2\left(\frac{\theta}{2}\right), \\
\hat{\varrho}^{\text{AD}}_{2,3} &= \hat{\eta}_{3,2} = 2\left(\eta_{2,3}^2 e^{-i\phi} + \eta_{1,4}^2 e^{i\phi} \right) \sin\theta, \\
\hat{\varrho}^{\text{AD}}_{1,4} &= 4 \eta_{2,3} \eta_{1,4} \sin\theta \cos\phi.
\end{aligned}
\label{eq:OFAD}
\end{equation}

Similarly, when the input state $\text{Y}\bar{\text{Y}}$ is teleported through the phase-flip channel $\rho^{\text{PF}}_{\text{Y}\bar{\text{Y}}}$, the output replica state can be expressed using Eqs.~(\ref{eq:PF}) and (\ref{eq:out})
\begin{equation}
\hat{\varrho}^{\text{PF}}_{out}=
\begin{pmatrix}
\hat{\varrho}^{\text{X}}_{1,1} & 0 & 0 & \hat{\varrho}^{\text{PF}}_{1,4} \\
0 & \hat{\varrho}^{\text{X}}_{2,2} & \hat{\varrho}^{\text{PF}}_{2,3} & 0 \\
0 & \hat{\varrho}^{\text{PF}}_{2,3} & \hat{\varrho}^{\text{X}}_{3,3}& 0 \\
\hat{\varrho}^{\text{PF}}_{1,4} & 0 & 0 & \hat{\varrho}^{\text{X}}_{1,1}
\end{pmatrix}, \quad \text{where} \quad
\begin{aligned}
\hat{\varrho}^{\text{PF}}_{2,3} &= \hat{\varrho}^{\text{PF}}_{3,2} = 2\left(\delta_{2,3}^2 e^{-i\phi} + \delta_{1,4}^2 e^{i\phi} \right) \sin\theta, \\
\hat{\varrho}^{\text{PF}}_{1,4} &= 4 \delta_{2,3} \delta_{1,4} \sin\theta \cos\phi.
\end{aligned}
\label{eq:OFPF}
\end{equation}

Finally, for the pure dephasing channel $\rho^{\text{PD}}_{\text{Y}\bar{\text{Y}}}$, the teleported replica state is given by using Eqs.~(\ref{eq:PD}) and (\ref{eq:out}).
\begin{equation}
\hat{\varrho}^{\text{PD}}_{out}=
\begin{pmatrix}
\hat{\varrho}^{\text{X}}_{1,1} & 0 & 0 & \hat{\varrho}^{\text{PF}}_{1,4} \\
0 & \hat{\varrho}^{\text{X}}_{2,2} & \hat{\varrho}^{\text{PF}}_{2,3} & 0 \\
0 & \hat{\varrho}^{\text{PF}}_{2,3} & \hat{\varrho}^{\text{X}}_{3,3}& 0 \\
\hat{\varrho}^{\text{PF}}_{1,4} & 0 & 0 & \hat{\varrho}^{\text{X}}_{1,1}
\end{pmatrix}, \quad \text{where} \quad
\begin{aligned}
\hat{\varrho}^{\text{PF}}_{2,3} &= \hat{\varrho}^{\text{PF}}_{3,2} = 2\left(\kappa_{2,3}^2 e^{-i\phi} + \kappa_{1,4}^2 e^{i\phi} \right) \sin\theta, \\
\hat{\varrho}^{\text{PF}}_{1,4} &= 4 \kappa_{2,3} \kappa_{1,4} \sin\theta \cos\phi.
\end{aligned}
\label{eq:OFPD}
\end{equation}

\subsection{Fidelity}

The hyperon-antihyperon spin density operator plays a central role in analyzing the quantum properties of the $\text{Y}\bar{\text{Y}}$ system, particularly for assessing the fidelity of quantum teleportation. It enables a deep understanding of the spin dynamics of hyperons and antihyperons, which is essential for reliable quantum communication. To this end, we study the decays of $\Lambda\overline{\Lambda}$, $\Sigma^{+}\bar{\Sigma}^{-}$, $\Xi^{-}\bar{\Xi}^{+}$, and $\Xi^{0}\bar{\Xi}^{0}$ to highlight the physical effects and quantum phenomena specific to this system (See table \ref{tab1}). These hyperons play a key role in understanding the strong interaction, thus contributing to the overall analysis of the system’s interactions. Investigating these decays sheds light on the limitations and potential applications of quantum teleportation, aiming to improve the fidelity of quantum information transfer.

The quality of the teleportation process is often characterized by studying the fidelity between the input state $\hat{\varrho}_{in}$ and the output state $\hat{\varrho}_{out}$. For pure input states, fidelity can be effectively applied as an indicator of a quantum channel's teleportation capabilities \citep{Te6,Te7}. The fidelity writes as \citep{L7}
\begin{equation}
F(\hat{\varrho}_{in},\hat{\varrho}^{\text{X}}_{out})=\bra{\psi_{in}}\hat{\varrho}^{\text{X}}_{out}\ket{\psi_{in}}=\Big\lbrace\text{Tr}\Big( \sqrt{\sqrt{\hat{\varrho}_{in}}\hat{\varrho}^{\text{X}}_{out}\sqrt{\hat{\varrho}_{in}}}\Big)\Big\rbrace^{2},
\label{eq:50}
\end{equation}
a straightforward calculation, we find that

\begin{equation}
\begin{aligned}
F (\hat{\varrho}_{in},\hat{\varrho}^{\text{X}}_{out})&= \hat{\varrho}^{\text{X}}_{2,2}\sin^{2}\Big(\frac{\theta}{2}\Big)+\hat{\varrho}^{\text{X}}_{3,3}\cos^{2}\Big(\frac{\theta}{2}\Big)+\Re\text{e}\Big(\hat{\varrho}^{\text{X}}_{2,3}\e^{-i\phi}\Big)
\end{aligned}
\label{eq:F}
\end{equation}

Orthogonality between the input and output states leads to a near-zero fidelity, implying that the quantum information is fully degraded during transmission, and consequently, the teleportation process is unsuccessful. Fidelity approaching unity indicates a near-perfect transmission where the output state closely mirrors the input state. However, a fidelity in the range $0<F<1$ suggests that the quantum information undergoes partial degradation during the transmission process.

To evaluate the quality of quantum state transfer in our system, it's essential to study the evolution of the fidelity between the input state $\hat{\varrho}_{\text{in}}$ and the output state $\hat{\varrho}^{\text{X}}_{out}$. Fidelity is a key measure that quantifies how close the teleported state remains to the initial state. In our study, we particularly focus on the influence of the system's geometric parameters: the scattering angle $\varphi$ and the amplitude angle $\theta$. These parameters modulate the nature of the states involved in the teleportation. Understanding this dependence allows us to optimize quantum teleportation protocols by considering the physical and experimental conditions related to angular interactions.

Initially, the fidelity threshold $F = 2/3$, recognized in the literature as the theoretical boundary between classical and quantum regimes for quantum teleportation systems. The classical regime is characterized by a fidelity $F\leq 2/3$, while the quantum regime emerges when $F$ exceeds this threshold, i.e., $F\geq 2/3$.
\begin{figure}[!h]
\begin{center}
\includegraphics[scale=0.37]{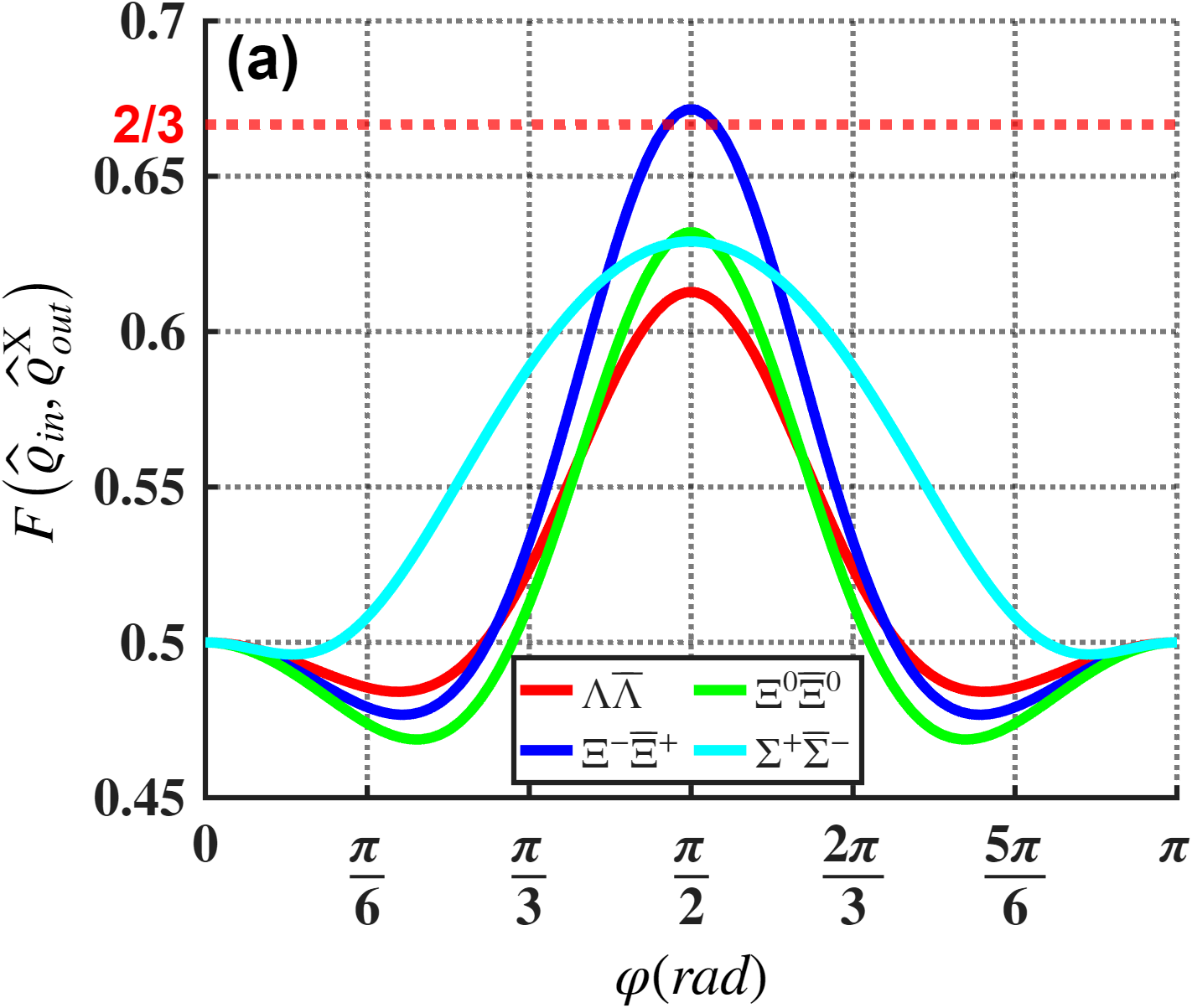}
\includegraphics[scale=0.37]{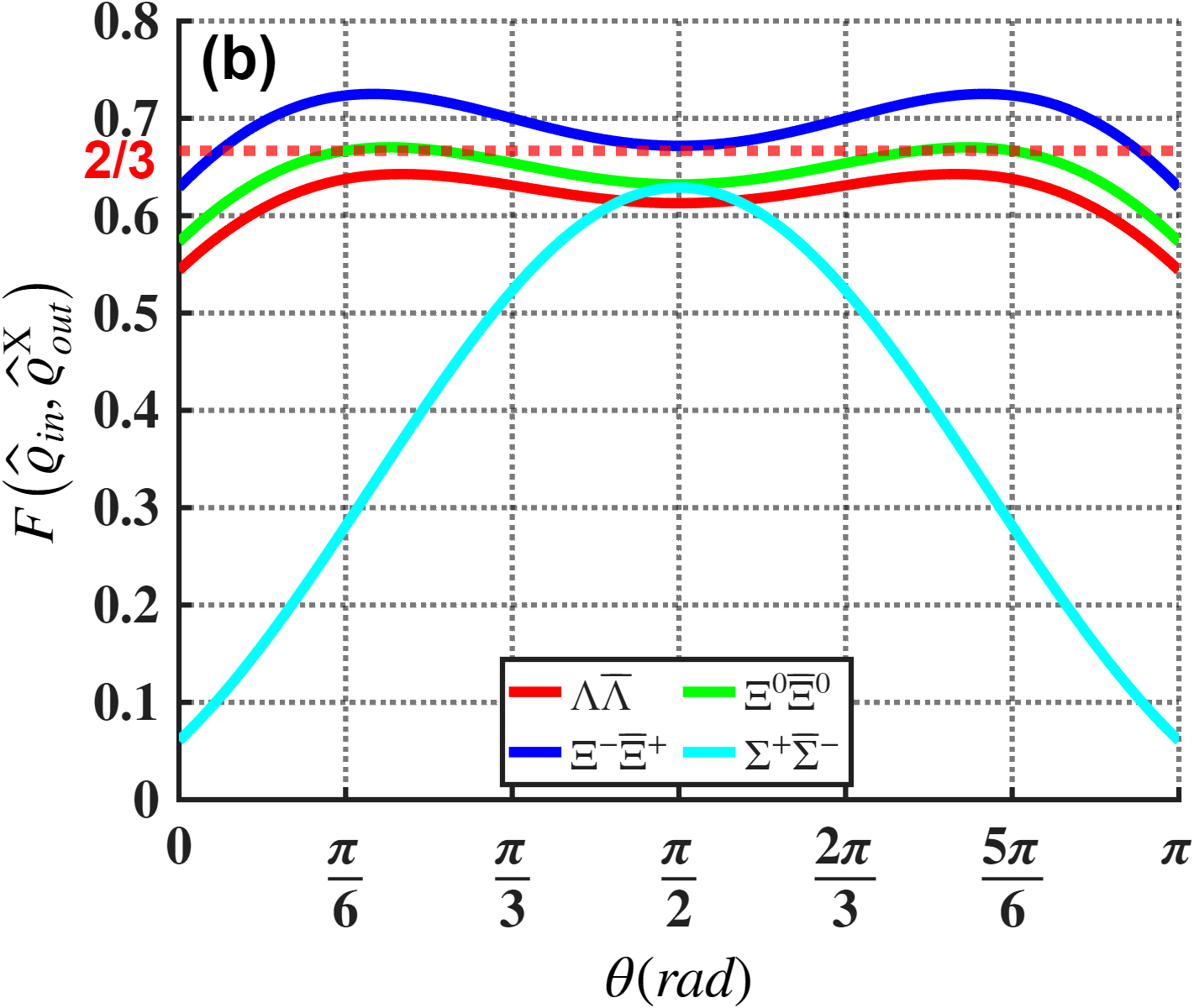}
\end{center}
\caption{Plot of the fidelity $F$ as a function of: (a) the scattering angle $\varphi$ (with $\phi=0$ and $\theta=\pi/2$), (b) the amplitude angle $\theta$ (with $\phi=0$ and $\varphi=\pi/2$) and the phase $\phi$ (with $\varphi=\theta=\pi/2$), in $\text{e}^+\text{e}^- \to \text{Y} \bar{\text{Y}}$ processes with $\text{Y} = \Lambda, \Sigma^+, \Xi^-$, and $\Xi^0$, utilizing the experimental parameters listed in Table \ref{tab1}.}
\label{fig:FI7}
\end{figure}

The examination of the results presented in Fig.~\ref{fig:FI7}(a) reveals the notable effect of the scattering angle $\varphi$ on the fidelity of the quantum teleportation protocol, with the amplitude angle $\theta$ fixed at $\pi/2$. The fidelity $F$ starts at a non-zero value, decreases with increasing $\varphi$, and then increases proportionally with $\varphi$ for various hyperon--antihyperon pairs $\text{Y}\bar{\text{Y}}$. For the pairs $\Lambda\bar{\Lambda}$, $\Sigma^{+}\bar{\Sigma}^{-}$, and $\Xi^{0}\bar{\Xi}^{0}$, the fidelity peaks at $\varphi = \pi/2$, remaining in the classical regime ($F < 2/3$). In contrast, the $\Xi^{-}\bar{\Xi}^{+}$ pair surpasses the threshold $F = 2/3$, indicating quantum behavior. Furthermore, a pronounced symmetry around $\varphi = \pi/2$ is observed, suggesting equivalent system behavior for values of $\varphi$ symmetric about this point.

The analysis of Fig.~\ref{fig:FI7}(b) underscores the significant impact of the amplitude angle $\theta$ on the fidelity of the protocol, with the scattering angle $\varphi$ fixed at $\pi/2$. The fidelity varies noticeably across different hyperon--antihyperon pairs, highlighting the protocol's sensitivity to the parameters of the system under study.A fidelity threshold of $F = 2/3$ is established to differentiate between classical and quantum regimes in the analysis of hyperon--antihyperon pairs. For the $\Sigma^{+}\bar{\Sigma}^{-}$ system, the fidelity increases steadily, reaching its maximum at $\theta = \pi/2$, while remaining in the classical regime ($F < 2/3$). In contrast, the $\Lambda\bar{\Lambda}$, $\Xi^{-}\bar{\Xi}^{+}$, and $\Xi^{0}\bar{\Xi}^{0}$ systems exhibit symmetry around $\theta = \pi/2$, with fidelity values fluctuating around the $F = 2/3$ threshold in certain configurations. An asymmetry in the process is also observed near this angle. The classical regime dominates for the $\Lambda\bar{\Lambda}$ and $\Sigma^{+}\bar{\Sigma}^{-}$ pairs, where the fidelity generally remains below or close to $F = 2/3$. In contrast, the quantum regime prevails for the $\Xi^{0}\bar{\Xi}^{0}$ and $\Xi^{-}\bar{\Xi}^{+}$ pairs, where the fidelity frequently exceeds this threshold. The quantum regime is particularly pronounced for the $\Xi^{-}\bar{\Xi}^{+}$ pair across the entire range of $\theta$, except at the extrema, where the fidelity may approach or exceed $F = 2/3$.

\begin{figure}[!h]
\begin{center}
\begin{minipage}{0.5\linewidth}
\includegraphics[scale=0.3]{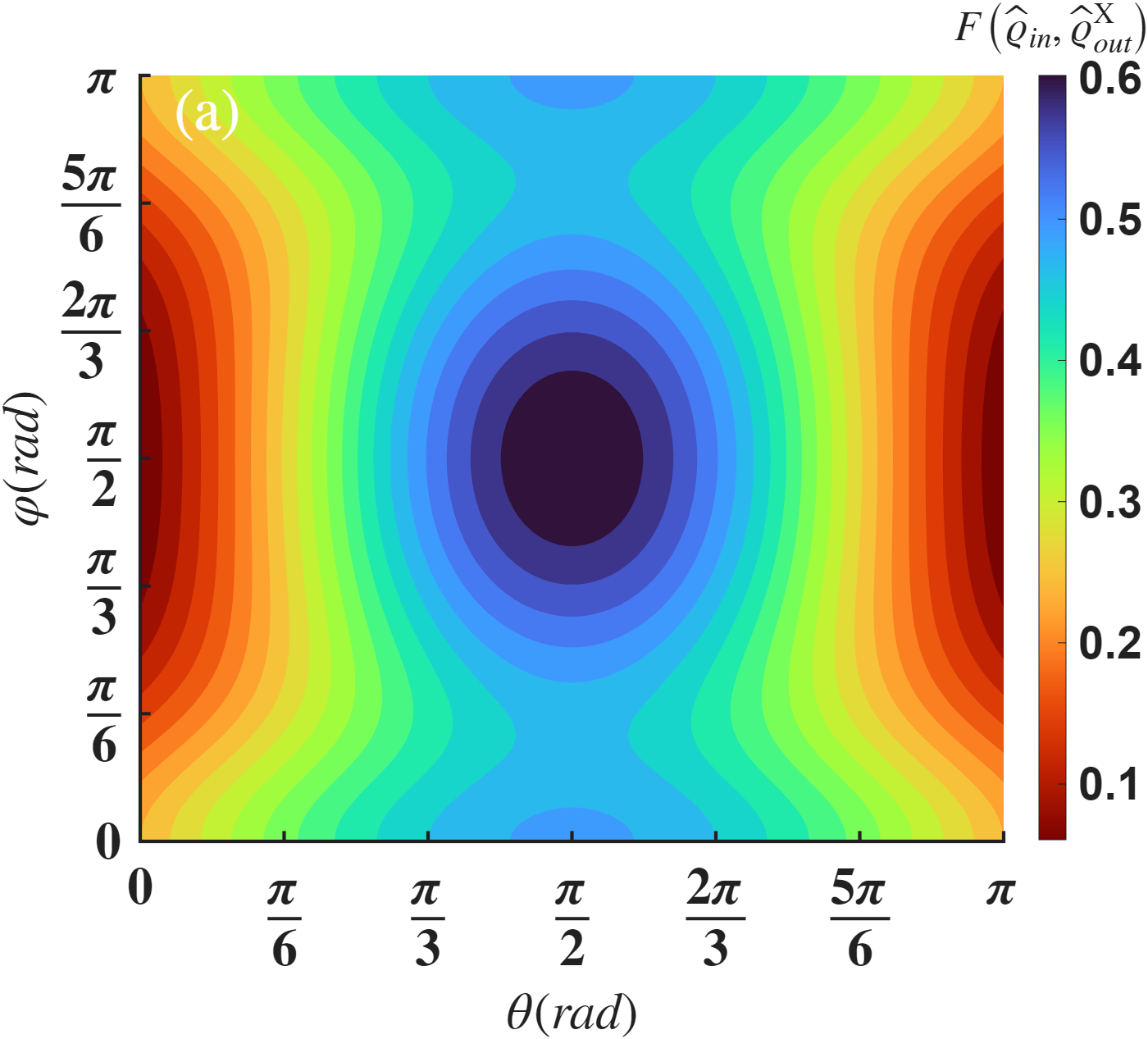}
\end{minipage}\hfill
\begin{minipage}{0.5\linewidth}
\includegraphics[scale=0.3]{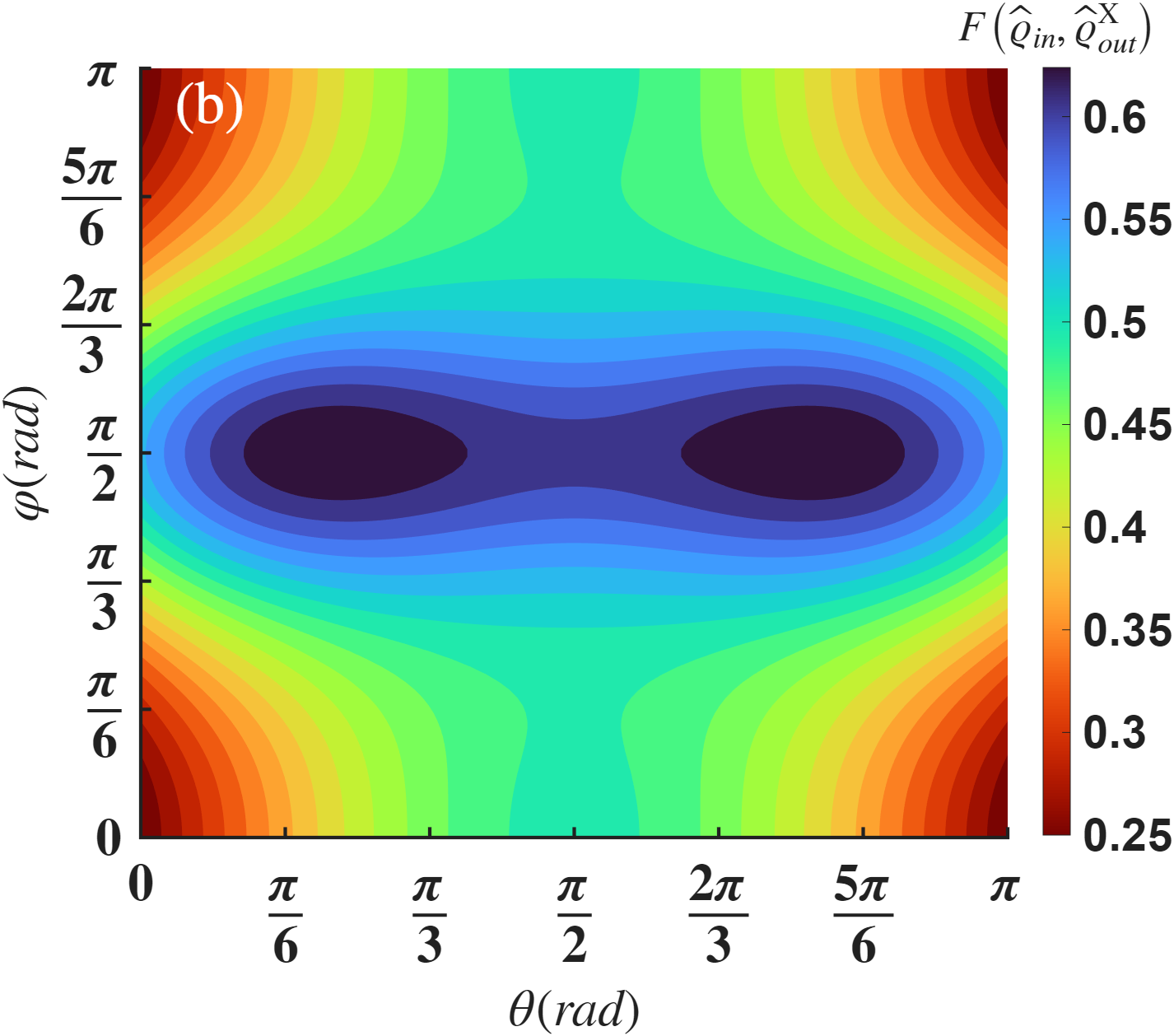}
\end{minipage}
\begin{minipage}{0.5\linewidth}
\includegraphics[scale=0.3]{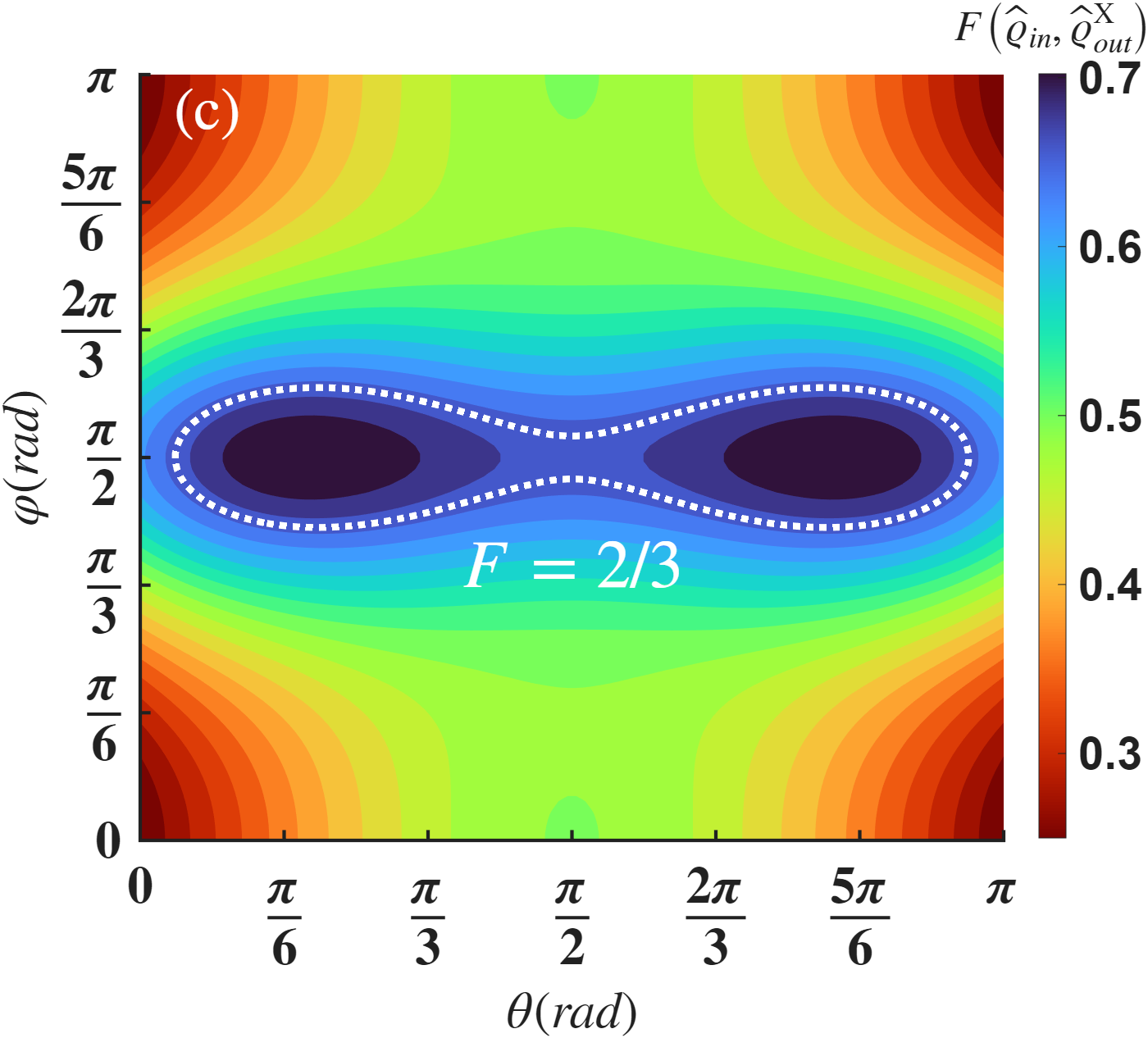}
\end{minipage}\hfill
\begin{minipage}{0.5\linewidth}
\includegraphics[scale=0.3]{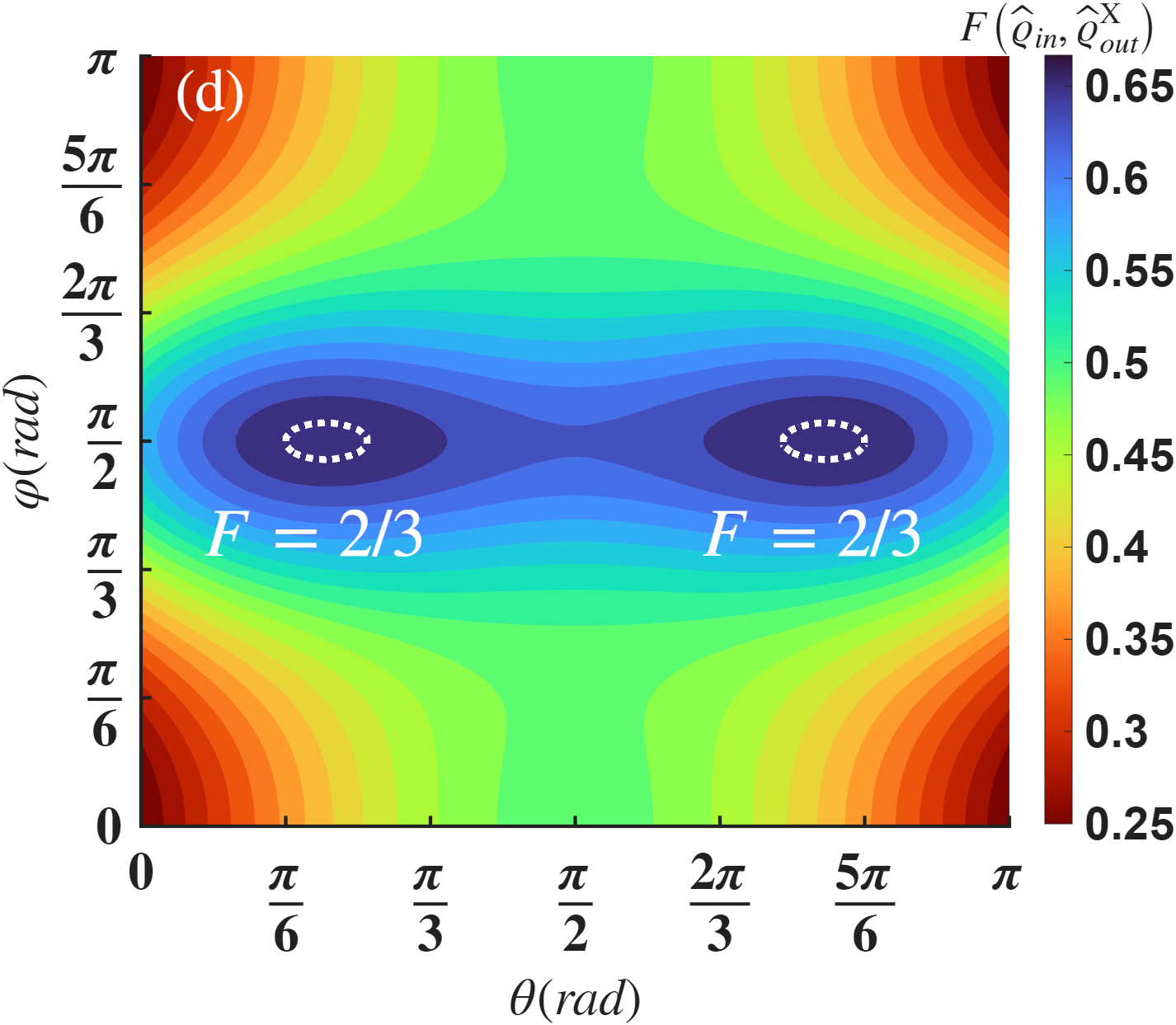}
\end{minipage}
\caption{Plot of the Fidelity $F$ as a function of the scattering angle $\varphi$ and the amplitude angle $\theta$ in $e^{+}e^{-} \to \text{Y}\bar{\text{Y}}$ processes, with $\phi=0$, for various decay channels: (a) $\Sigma^{+}\bar{\Sigma}^{-}$, (b) $\Xi^{-}\bar{\Xi^{+}}$, (c) $\Xi^{-}\bar{\Xi}^{+}$, and (d) $\Xi^{0}\bar{\Xi}^{0}$. The white dashed line at $F=2/3$ represents the critical value that separates the classical and quantum regimes. Experimental parameters are taken from Table \ref{t1}.}
\label{fig:FWd}
\end{center}
\end{figure}

The study illustrated in Fig.~\ref{fig:FWd} focuses on the critical transitions between quantum and classical regimes by jointly optimizing the amplitude angle \(\theta\) and the scattering angle \(\varphi\). This study builds on the observations from Fig.~\ref{fig:L}(b--c), which highlight the significance of the system parameters $\{\theta, \varphi\}$. The results presented in Fig.~\ref{fig:FWd} indicate that simultaneous optimization of these parameters significantly enhances the teleportation fidelity, with \(\theta\) playing a central role in extending the quantum regime and ensuring effective quantum teleportation. More precisely, the optimal configuration occurs at \(\theta = \pi/2\), which maximizes the extent of the quantum regime. This adjustment is crucial, as relying solely on \(\varphi\) is insufficient to induce quantum behavior; suboptimal parameter settings lead to conventional (classical) channel behavior, as shown in Fig.~\ref{fig:FWd}(c-d). These results highlight the importance of parameter optimization in quantum teleportation, where enhanced fidelity is essential to advancing quantum communication and information processing.

%\subsection{Under dephasing effect}

Let us study the evolution of the fidelity for our system under three types of decoherence channels as a function of the scattering angle \(\varphi\), the angle $\theta$, and the decoherence parameter \(s\), using the final formulas for the AD~(\ref{eq:OFAD}), PF~(\ref{eq:OFPF}), and PD~(\ref{eq:OFPD}) channels. 

In this study, we investigate the behavior of the $\Xi^{-}\bar{\Xi^{+}}$ hyperon under the action of amplitude damping (AD), phase damping (PD), and depolarizing (PF) channels. The results obtained for this baryon under these channels also apply to the baryons $\Lambda$, $\Xi^{0} $, and $\Sigma^{+}$.

The expression in Eq.~\eqref{eq:F} provides the general form of the teleportation fidelity. In the following, we specify this result for the case where the shared entangled state undergoes different types of noisy quantum channels.\\
Based on Eqs.~(\ref{eq:50}) and (\ref{eq:OFAD}), the fidelity for the amplitude damping (AD) channel can be written as
\begin{equation}
\begin{aligned}
F (\hat{\varrho}_{in},\hat{\varrho}^{\text{AD}}_{out})&= \hat{\varrho}^{\text{AD}}_{2,2}\sin^{2}\Big(\frac{\theta}{2}\Big)+\hat{\varrho}^{\text{AD}}_{3,3}\cos^{2}\Big(\frac{\theta}{2}\Big)+\Re\text{e}\Big(\hat{\varrho}^{\text{AD}}_{2,3}\e^{-i\phi}\Big)
\end{aligned}
\label{eq:FAD}
\end{equation}

Using a similar method, and with the help of Eqs.~(\ref{eq:50}) and (\ref{eq:OFPF}), the fidelity under the phase flip (PF) channel can be expressed as
\begin{equation}
\begin{aligned}
F (\hat{\varrho}_{in},\hat{\varrho}^{\text{PF}}_{out})&= \hat{\varrho}^{\text{X}}_{2,2}\sin^{2}\Big(\frac{\theta}{2}\Big)+\hat{\varrho}^{\text{X}}_{3,3}\cos^{2}\Big(\frac{\theta}{2}\Big)+\Re\text{e} \Big(\hat{\varrho}^{\text{PF}}_{2,3}\e^{-i\phi}\Big)
\end{aligned}
\label{eq:FPF}
\end{equation}

Based on a similar approach, and with the help of Eqs.~(\ref{eq:50}) and (\ref{eq:OFPD}), the fidelity for the phase damping (PD) channel is expressed as
\begin{equation}
\begin{aligned}
F (\hat{\varrho}_{in},\hat{\varrho}^{\text{PF}}_{out})&= \hat{\varrho}^{\text{X}}_{2,2}\sin^{2}\Big(\frac{\theta}{2}\Big)+\hat{\varrho}^{\text{X}}_{3,3}\cos^{2}\Big(\frac{\theta}{2}\Big)+\Re\text{e}\Big(\hat{\varrho}^{\text{PD}}_{2,3}\e^{-i\phi}\Big)
\end{aligned}
\label{eq:FPD}
\end{equation}

\begin{figure}[!h]
\begin{center}
\includegraphics[scale=0.26]{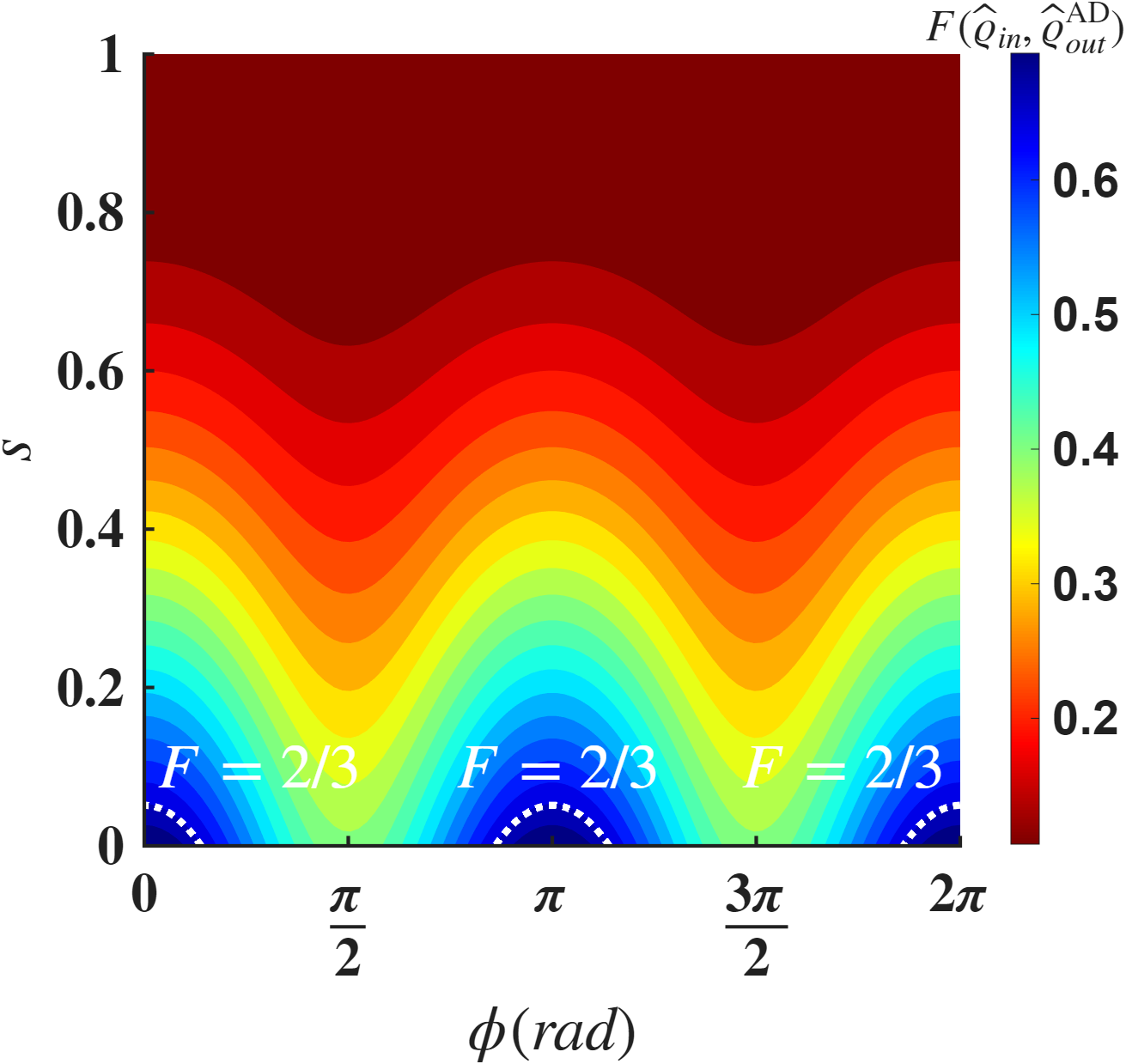}
\includegraphics[scale=0.26]{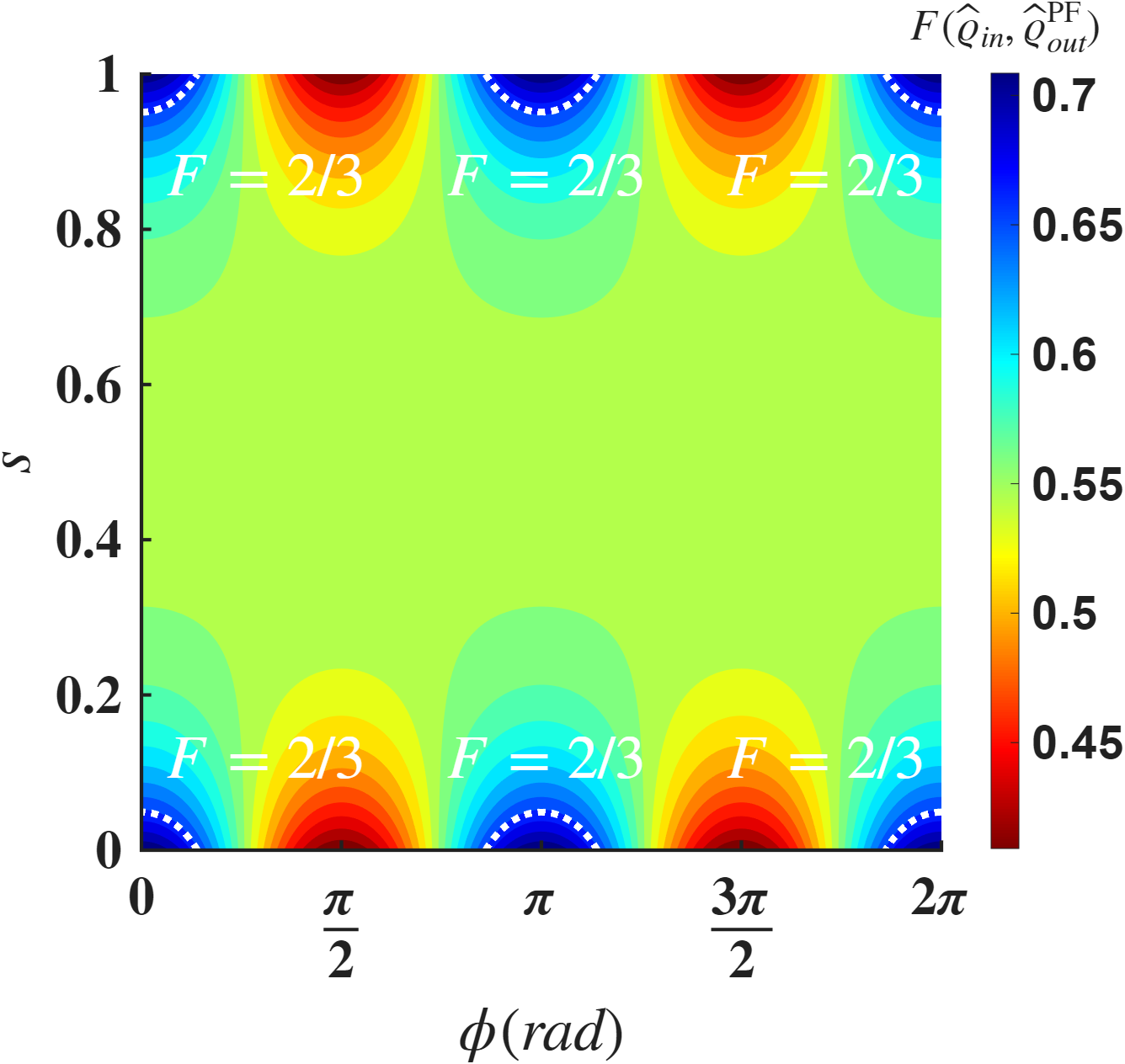}
\includegraphics[scale=0.26]{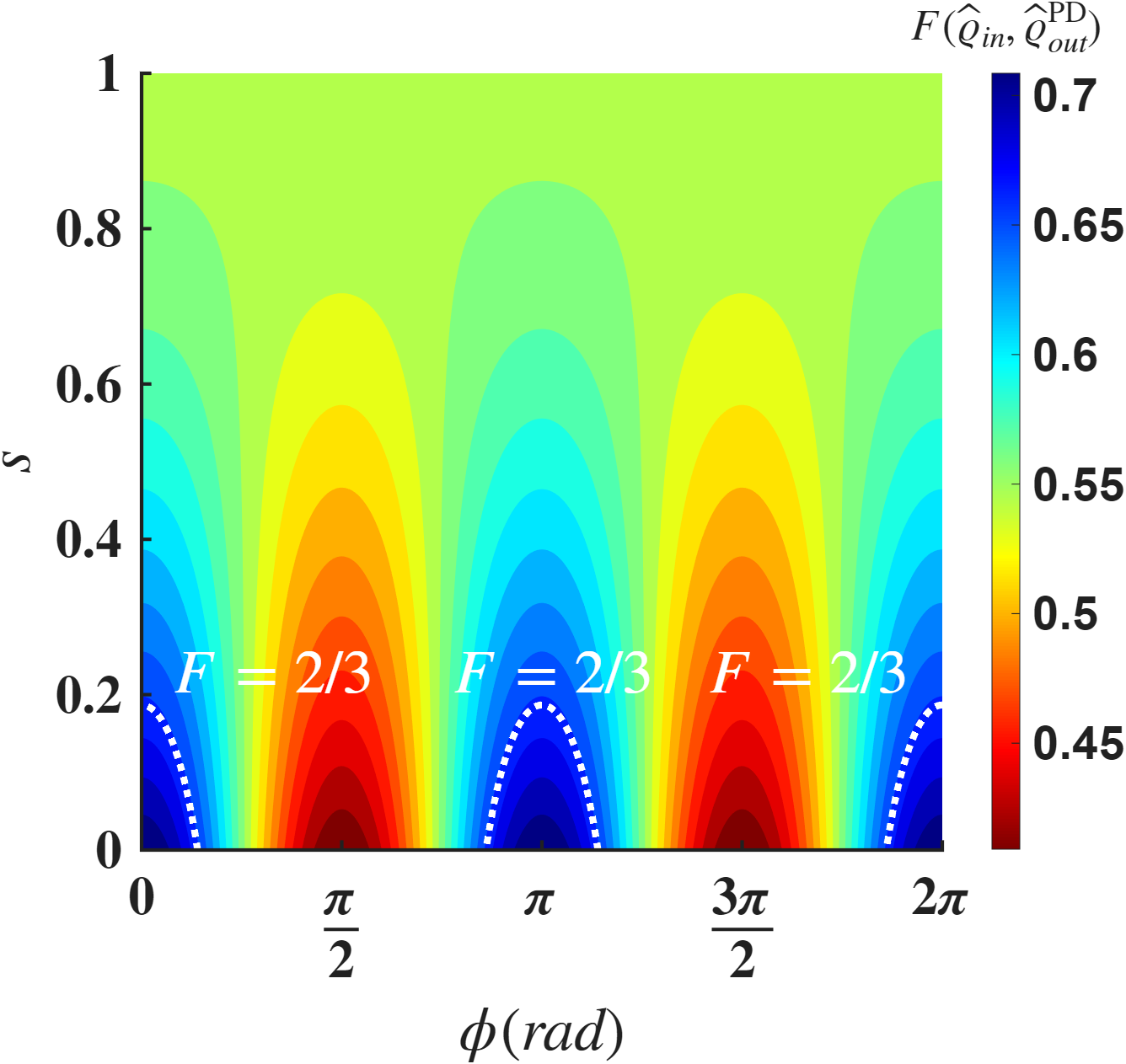}
\put(-485,150){{\bfseries (a)}}
\put(-315,150){{\bfseries (b)}}
\put(-145,150){{\bfseries (c)}}
\end{center}
\caption{Plot of the Fidelity $F$ as a function of the phase angle $\phi$ and the decoherence parameter $s$ in $e^{+}e^{-} \to \Xi^{-}\bar{\Xi}^{+}$ processes, with $\theta=\pi/6$ and $\varphi=\pi/2$, for various noisy channels: (a) AD channel, (b) PF channel, (c) PD channel. The white dashed line at $F=2/3$ represents the critical value that separates the classical and quantum regimes. Experimental parameters are taken from Table \ref{t1}.}
\label{fig:FAD}
\end{figure}
\begin{figure}[!h]
\begin{center}
\includegraphics[scale=0.26]{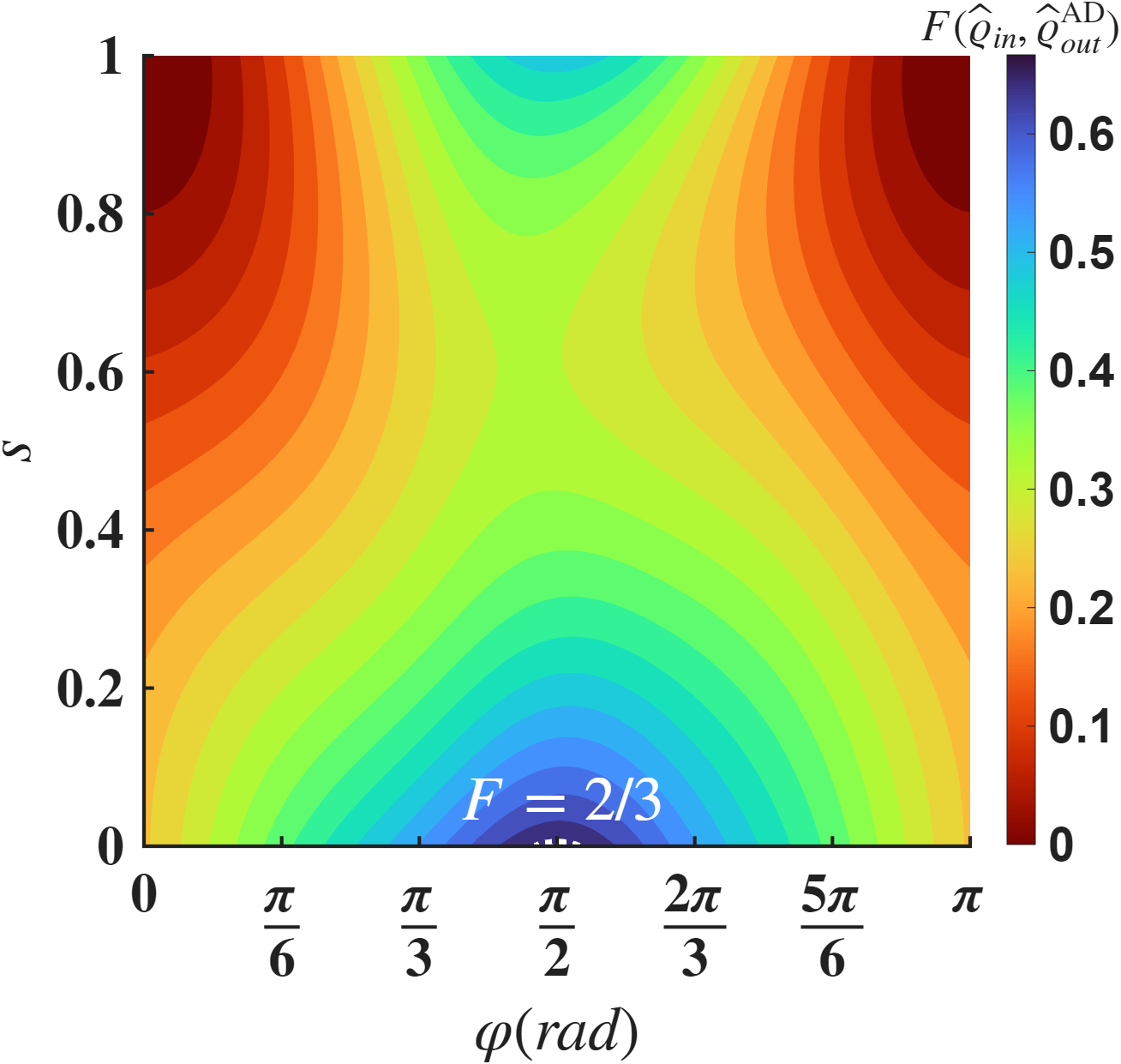}
\includegraphics[scale=0.26]{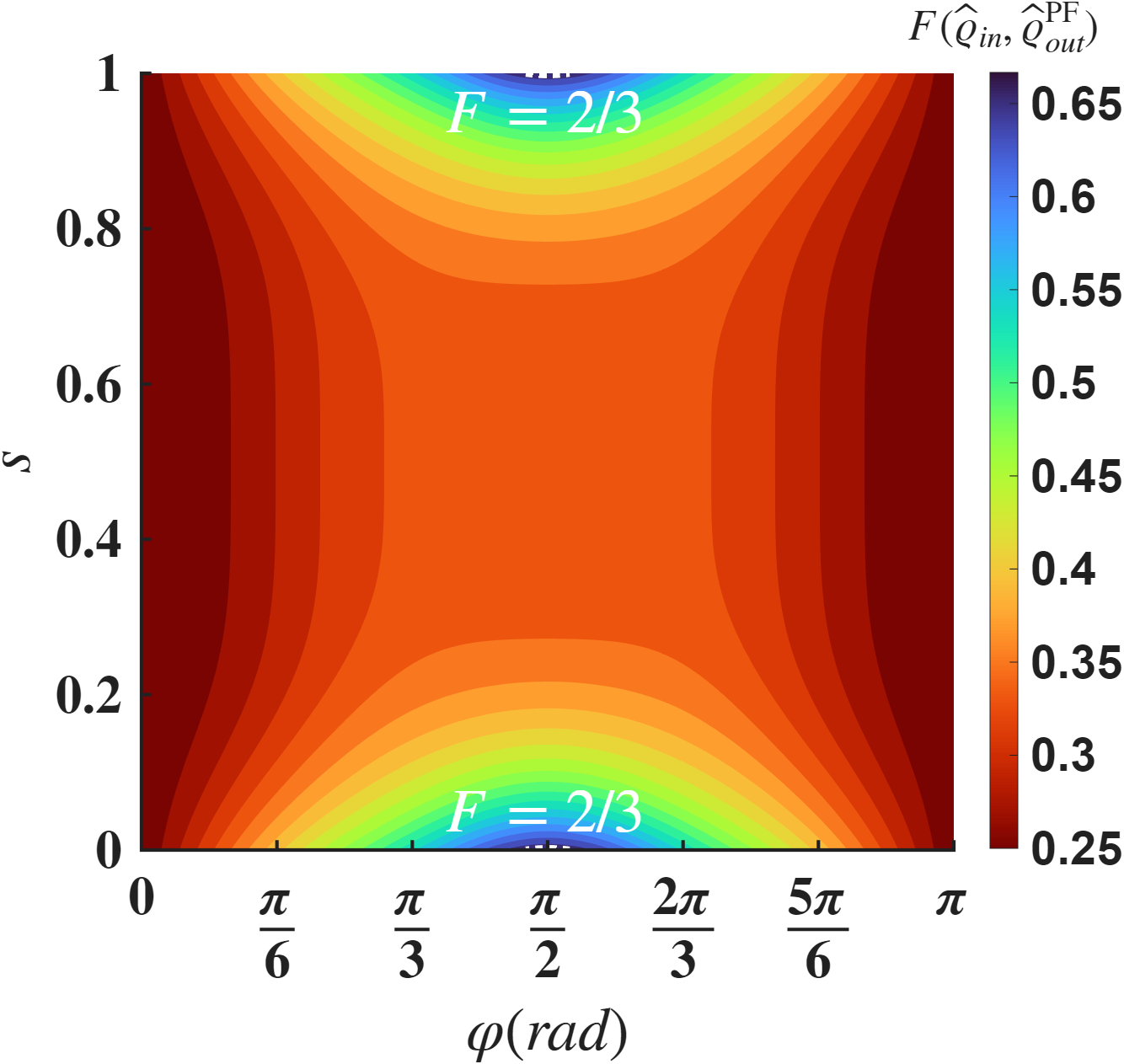}
\includegraphics[scale=0.26]{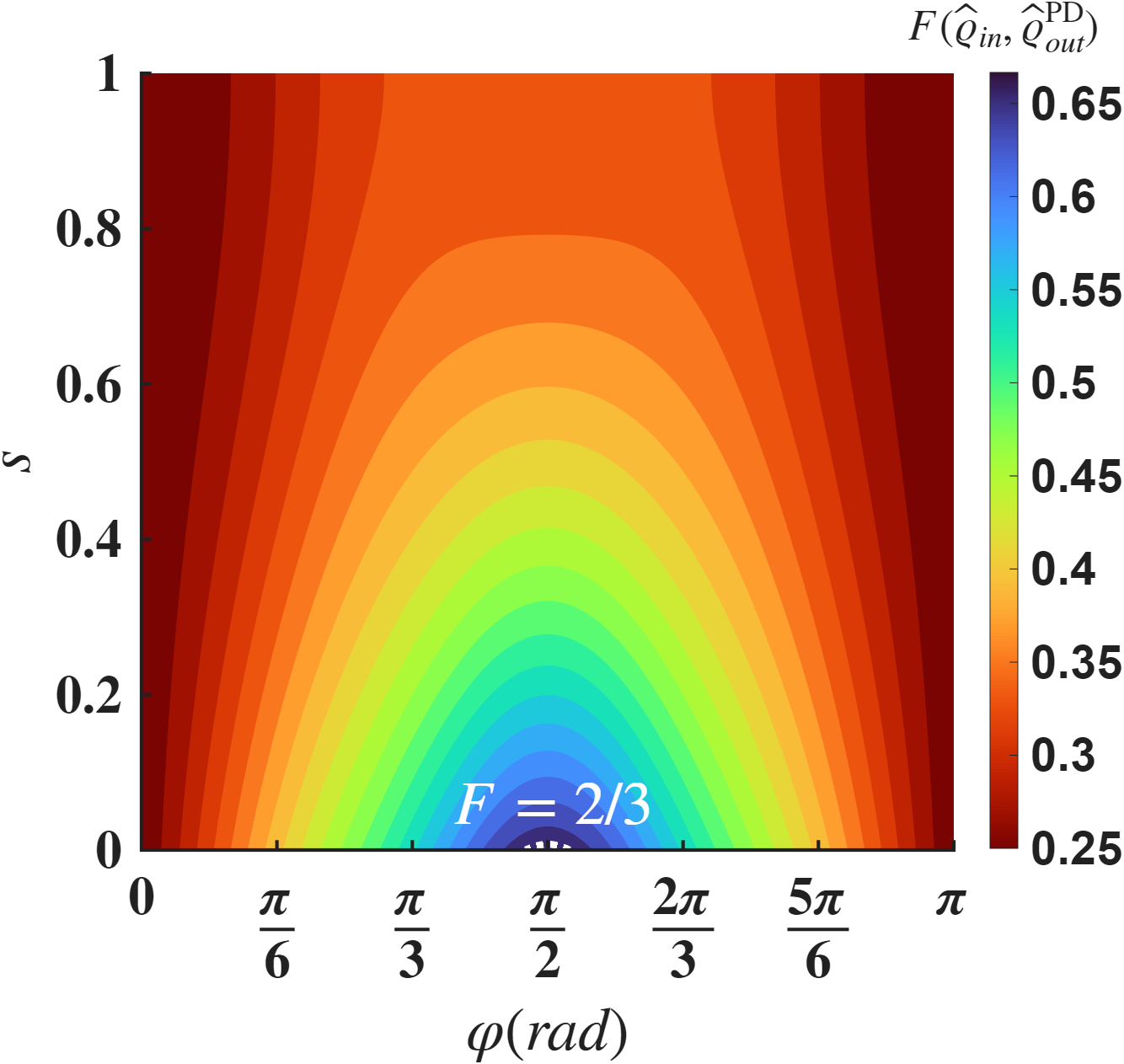}
\put(-485,150){{\bf(a)}}
\put(-315,150){{\bf (b)}}
\put(-145,150){{\bf(c)}}
\end{center}
\caption{Plot of the Fidelity $F$ as a function of the scattering angle $\varphi$ and the the decoherence parameter $s$ in $e^{+}e^{-} \to  \Xi^{-}\bar{\Xi}^{+}$ processes, with $\phi=0$ and $\theta=\pi/6$, for various noisy channels: (a) AD channel, (b) PF channel, (c) PD channel. The white dashed line at $F=2/3$ represents the critical value beyond which the quantum regime begins. Experimental parameters are taken from Table \ref{t1}.}
\label{fig:FPF}
\end{figure}
\begin{figure}[!h]
\begin{center}
\includegraphics[scale=0.26]{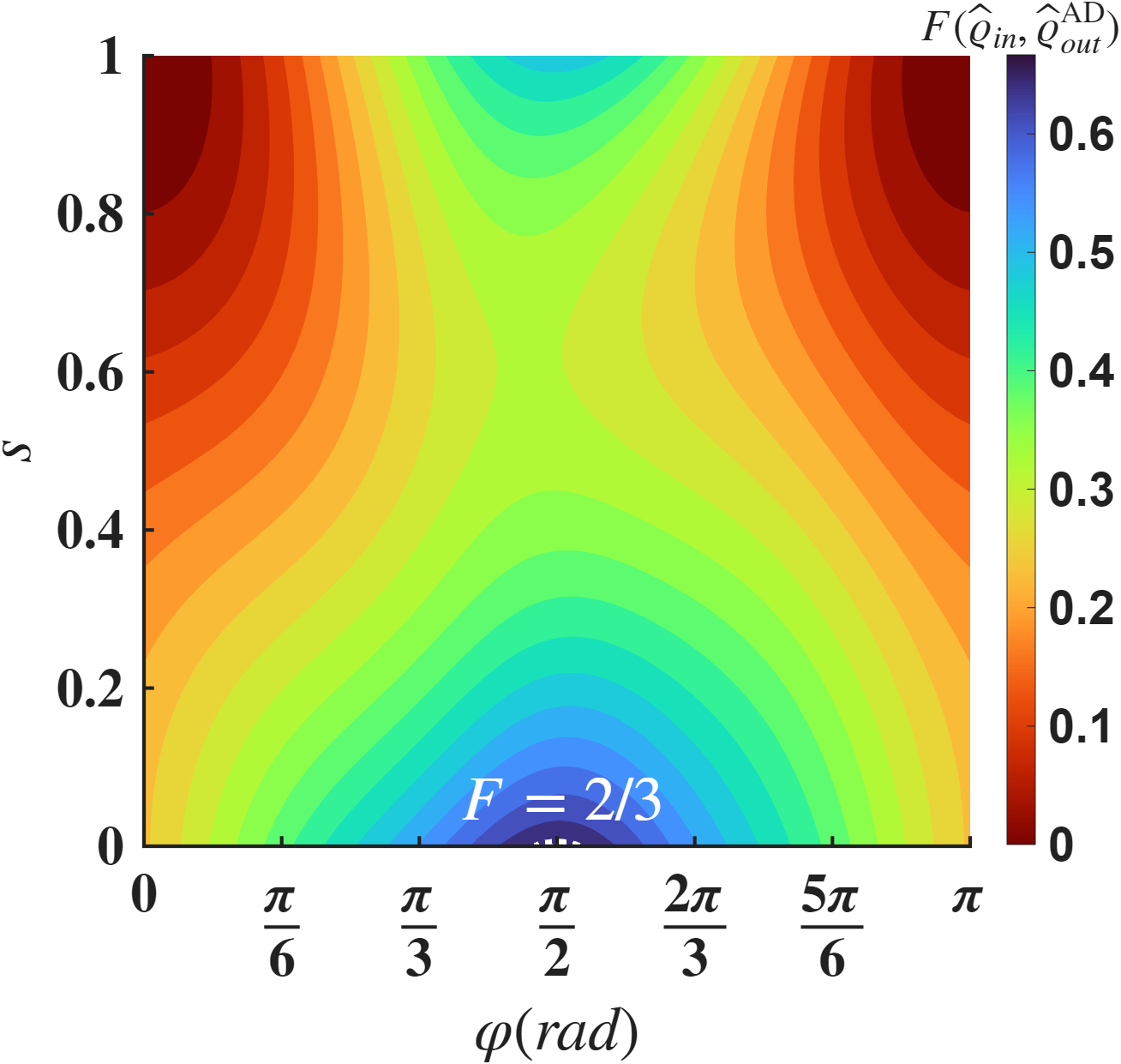}
\includegraphics[scale=0.26]{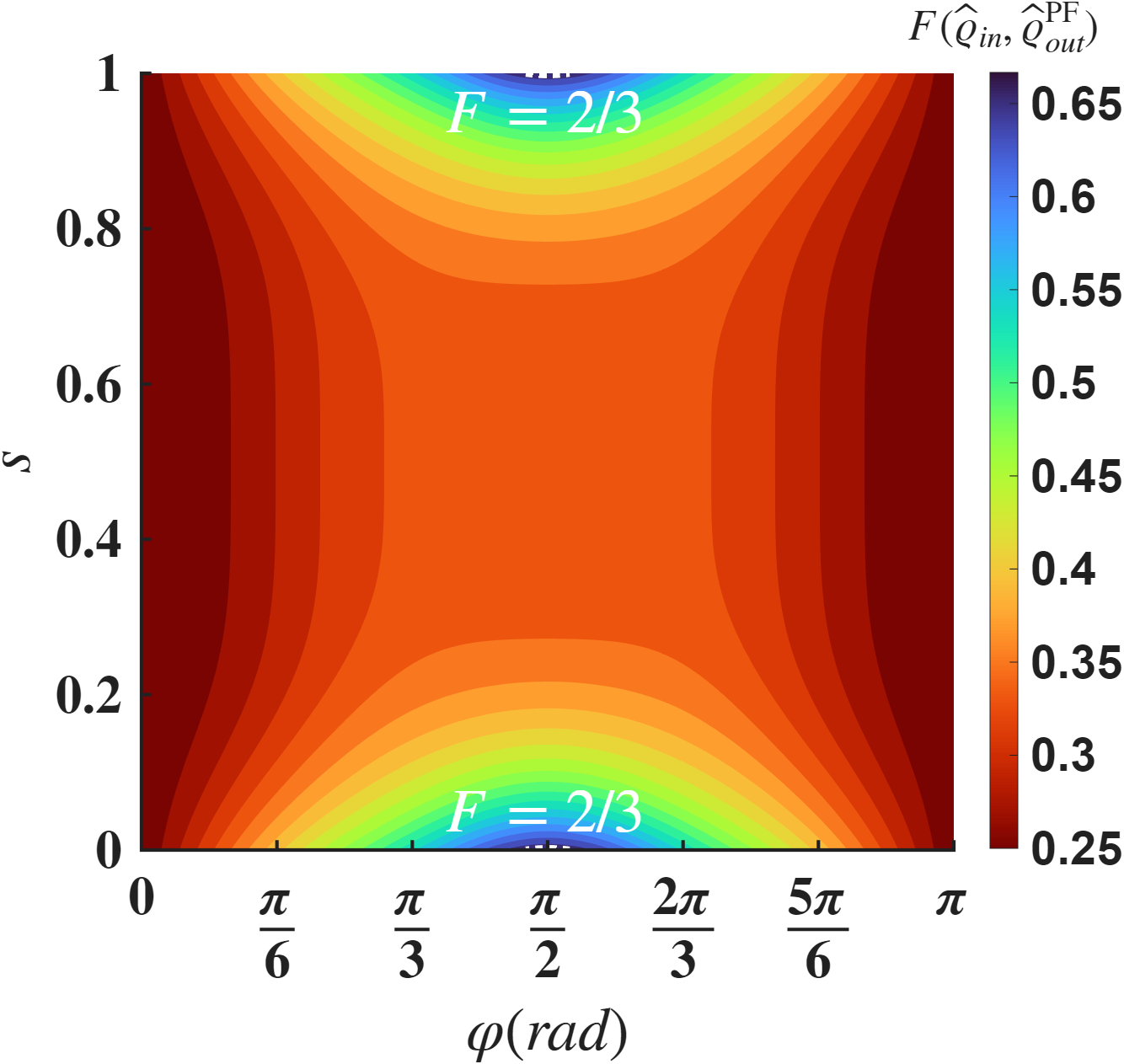}
\includegraphics[scale=0.26]{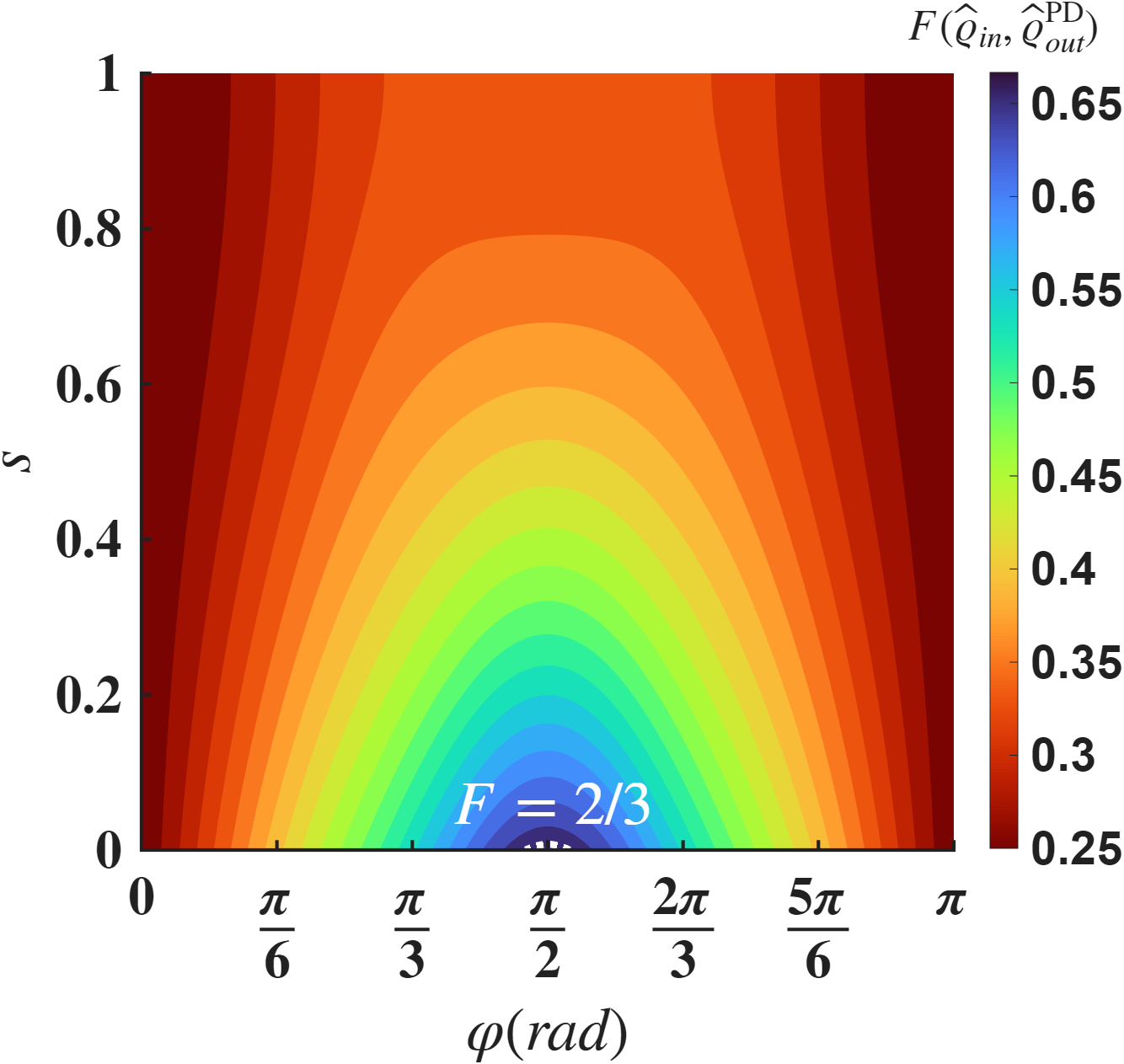}
\put(-485,150){{\bf(a)}}
\put(-315,150){{\bf (b)}}
\put(-145,150){{\bf(c)}}
\end{center}
\caption{Plot of the Fidelity $F$ as a function of the amplitude angle $\theta$ and the the decoherence parameter $s$ in $e^{+}e^{-} \to \Xi^{-}\bar{\Xi}^{+}$ processes, with $\phi=0$ and $\varphi=\pi/2$, for various noisy channels: (a) AD channel, (b) PF channel, (c) PD channel. The white dashed line at $F=2/3$ represents the critical value beyond which the quantum regime begins.. Experimental parameters are taken from Table \ref{t1}.}
\label{fig:FPD}
\end{figure}

In Fig.~\ref{fig:FAD}(a-c), we present the evolution of the fidelity as a function of the phase $\phi$ and the decoherence parameter $s$ for hyperon--antihyperon pairs $\Xi^{-}\bar{\Xi}^{+}$. This figure shows that the fidelity has a sinusoidal dependence on $\phi$, emphasizing the influence of this phase on the entanglement and superposition properties of the $\Xi^{-}\bar{\Xi}^{+}$ pair. We remark that the fidelity attains its maximum values at $\phi = k\pi$ with $k = 0,1,2$, which correspond to optimal configurations of the $\Xi^{-}\bar{\Xi}^{+}$ state. This can be clearly deduced from Eqs.~(\ref{eq:F}), (\ref{eq:FAD}), (\ref{eq:FPF}), and (\ref{eq:FPD}). At these optimal angles, the quantum regime is fully achieved, as the fidelity of the teleported state can be greater than the classical limit of $2/3$. This is observed under both the AD and PD channels when the decoherence $s$ is around 0, as shown in Fig.~\ref{fig:FAD}(a) and Fig.~\ref{fig:FAD}(c). However, under the phase PF channel, the fidelity exceed the classical limit of $2/3$ as the decoherence strength $s$ approaches both 0 and 1, as illustrated in Fig.~\ref{fig:FAD}(b).

The evolution of the fidelity for the three types of noisy channels, as a function of the scattering angle $\varphi$ and the decoherence parameter $s$, is shown in Fig.~\ref{fig:FPF}(a-c). We observe that the fidelity reaches its maximum at $\varphi = \pi/2$ for all three channels, AD, PF, and PD. For this optimal angle, under the effect of the AD channel, the fidelity initially decreases, then increases to reach its maximum value at $s=1$, indicating greater robustness at high values of $s$ for hyperon--antihyperon $\Xi^{-}\bar{\Xi}^{+}$. It is also evident that, in the case of the phase flip (PF) channel, the fidelity initially drops sharply before gradually increasing, returning to its initial values as $s \to 1$, which induces a symmetric fidelity profile around $s = 0.5$. This shows that the presence of noise in this channel can induce a revival of the fidelity. Clearly, in both the AD and PF channels, the teleported state retains a relatively high fidelity even under maximal noise. In contrast, for the PD channel, the fidelity decreases linearly until it reaches its minimum values as $s \to 1$. In the AD and PF channels, a revival of fidelity is observed, where it increases again at high noise levels. Similarly, under the PF channel, the fidelity recovers to its maximum value, $F \geq 2/3$, as $s \to 1$, which corresponds to the quantum regime, as depicted in \ref{fig:FPF}(b).

We explore in Fig.~\ref{fig:FPD}(a--c) the fidelity as a function of the amplitude angle $\theta$ and the decoherence parameter $s$, with $\varphi = \pi/2$. We remark that the fidelity is symmetric around $\theta = \pi/2$ and reaches its maximum value at $\theta = \pi/6$ across the AD, and PD channels. Moreover, the fidelity remains $\geq 2/3$ (in the quantum regime) across the AD and PD channels for $s \leq 0.2$, as depicted in Fig.~\ref{fig:FPD}(a) and (c). Furthermore, for the PF channel, the quantum regime is found in two separate regions of decoherence: approximately for $s \leq 0.2$ and for $s \geq 0.95$, as illustrated in Fig.~\ref{fig:FAD}(b).

\subsection{Quantum correlations : $L_{N}(\hat{\varrho}^{\text{X}}_{out})$, $LQU(\hat{\varrho}^{\text{X}}_{out})$, and $LQFI(\hat{\varrho}^{\text{X}}_{out})$}

In this section, we present a detailed investigation into the behavior of quantum correlations, specifically $L_{N}(\hat{\varrho}^{\text{X}}_{out})$, $LQU(\hat{\varrho}^{\text{X}}_{out})$, and $LQFI(\hat{\varrho}^{\text{X}}_{out})$, within the teleported $\text{Y}\bar{\text{Y}}$ state across the amplitude damping (AD), phase damping (PD), and phase flip (PF) channels, as described in equations (\ref{eq:OFAD}), (\ref{eq:OFPD}), and (\ref{eq:OFPF}). Our analysis explores the influence of the decoherence parameter, system parameters, and input state parameters on the $\text{Y}\bar{\text{Y}}$ state, shedding light on their impact on quantum correlation dynamics. Furthermore, we evaluate the logarithmic negativity of the teleported states for the ideal channel ($\hat{\varrho}^{\text{X}}_{out}$) (without dephasing effect, i.e s=0) as well as the amplitude damping ($\hat{\varrho}^{\text{AD}}_{out}$), phase damping ($\hat{\varrho}^{\text{PD}}_{out}$), and phase flip ($\hat{\varrho}^{\text{PF}}_{out}$) channels. To achieve this, the density matrix $\rho^{\text{X}}_{\text{Y}\bar{\text{Y}}}$ in Eq.~(\ref{eq:N3}) is substituted with the respective matrices $\hat{\varrho}^{\text{X}}$, $\hat{\varrho}^{\text{AD}}$, $\hat{\varrho}^{\text{PD}}$, and $\hat{\varrho}^{\text{PF}}$, as specified in equations (\ref{eq:OF}), (\ref{eq:OFAD}), (\ref{eq:OFPD}), and (\ref{eq:OFPF}). The logarithmic negativity (LN) for the teleported state is defined as
\begin{equation} 
L_{N}(\hat{\varrho}^{\epsilon}_{out}) = \max\left\lbrace 0, -2\hat{\mu}_{\min}\right\rbrace, 
\label{eq:Nout} 
\end{equation}
where $\hat{\mu}_{\min}$ is the smallest eigenvalue, given by
\begin{equation*} 
\hat{\mu}_{\min} = \min\left\lbrace \hat{e}_{1}^{\epsilon}, \hat{e}_{2}^{\epsilon}, \hat{e}_{3}^{\epsilon}, \hat{e}_{4}^{\epsilon}\right\rbrace, \end{equation*}
and the corresponding eigenvalues for each channel are expressed as
\begin{equation} 
\begin{aligned} 
\hat{e}_{1,2}^{\epsilon} &= \frac{\big(\hat{\varrho}^{\epsilon}_{1,1} + \hat{\varrho}^{\epsilon}_{4,4}\big) \pm \sqrt{\big(\hat{\varrho}^{\epsilon}_{1,1} - \hat{\varrho}^{\epsilon}_{4,4}\big)^{2} + 4|\hat{\varrho}^{\epsilon}_{2,3}|^{2}}}{2},\\ 
\hat{e}_{3,4}^{\epsilon} &= \frac{\big(\hat{\varrho}^{\epsilon}_{2,2} + \hat{\varrho}^{\epsilon}_{3,3}\big) \pm \sqrt{\big(\hat{\varrho}^{\epsilon}_{2,2} - \hat{\varrho}^{\epsilon}_{3,3}\big)^{2} + 4|\hat{\varrho}^{\epsilon}_{1,4}|^{2}}}{2}, 
\end{aligned} 
\label{eq:NoutXi} 
\end{equation}
where $\epsilon \in \big({\text{X}, \text{AD}, \text{PF}, \text{PD}}\big)$. Similarly, the LQU and LQFI of the teleported state for the ideal ($\varrho_{\text{Y}\bar{\text{Y}}}^{\text{X}}$) (without dephasing effect, i.e s=0), amplitude damping (AD), phase damping (PD), and phase flip (PF) channels are computed by substituting the density matrix $\rho^{\text{X}}_{\text{Y}\bar{\text{Y}}}$ in Eqs. (\ref{eq:WLQU}) and (\ref{eq:F}) with the matrices $\hat{\varrho}^{\text{X}}_{out}$, $\hat{\varrho}^{\text{AD}}_{out}$, $\hat{\varrho}^{\text{PD}}_{out}$, and $\hat{\varrho}^{\text{PF}}_{out}$, as specified in Eqs. (\ref{eq:OF}), (\ref{eq:OFAD}), (\ref{eq:OFPD}), and (\ref{eq:OFPF}). The expression for the LQU is given by
\begin{equation}
LQU(\hat{\varrho}^{\epsilon}_{out}) = 1 - \max\left\lbrace\mathcal{W}^{\epsilon}_{1,1}, \mathcal{W}_{3,3}^{\epsilon}\right\rbrace, 
\label{eq:WOUT}
\end{equation}
with
\begin{equation*} 
\begin{aligned} 
\mathcal{W}^{\epsilon}_{1,1} &= \left(\sqrt{\lambda_{1}}+\sqrt{\lambda_{2}}\right)\left(\sqrt{\lambda_{3}}+\sqrt{\lambda_{4}}\right) + \frac{\big(\hat{\varrho}^{\epsilon}_{2,2}-\hat{\varrho}^{\epsilon}_{3,3}\big)\big(\hat{\varrho}^{\epsilon}_{4,4}-\hat{\varrho}^{\epsilon}_{1,1}\big)+4|\hat{\varrho}^{\epsilon}_{1,4}\hat{\varrho}^{\epsilon}_{2,3}|}{\left(\sqrt{\lambda_{1}}+\sqrt{\lambda_{2}}\right)\left(\sqrt{\lambda_{3}}+\sqrt{\lambda_{4}}\right)}, \\[6pt] 
\mathcal{W}^{\epsilon}_{3,3} &= \frac{\left(\sqrt{\lambda_{1}}+\sqrt{\lambda_{2}}\right)^{2} + \left(\sqrt{\lambda_{3}}+\sqrt{\lambda_{4}}\right)^{2}}{2} + \frac{\big(\hat{\varrho}^{\epsilon}_{4,4}-\hat{\varrho}^{\epsilon}_{1,1}\big)^{2}-4|\hat{\varrho}^{\epsilon}_{1,4}|^{2}}{2\left(\sqrt{\lambda_{1}}+\sqrt{\lambda_{2}}\right)^{2}} + \frac{\big(\hat{\varrho}^{\epsilon}_{2,2}-\hat{\varrho}^{\epsilon}_{3,3}\big)^{2}-4|\hat{\varrho}^{\epsilon}_{2,3}|^{2}}{2\left(\sqrt{\lambda_{3}}+\sqrt{\lambda_{4}}\right)^{2}}.
\end{aligned} 
\end{equation*}
where the $\lambda_{i}$ $(i=1,2,3,4)$ are the eigenvalues of the matrix $\hat{\varrho}^{\epsilon}_{out}$, given by
\begin{equation*}
\begin{aligned}
\lambda_{1,2}&= \frac{\big(\hat{\varrho}^{\epsilon}_{1,1} + \hat{\varrho}^{\epsilon}_{4,4}\big) \pm \sqrt{\big(\hat{\varrho}^{\epsilon}_{1,1} - \hat{\varrho}^{\epsilon}_{4,4}\big)^{2} + 4|\hat{\varrho}^{\epsilon}_{1,4}|^{2}}}{2},\\
\lambda_{3,4}&= \frac{\hat{\varrho}^{\epsilon}_{2,2} + \hat{\varrho}^{\epsilon}_{3,3}) \pm \sqrt{(\hat{\varrho}^{\epsilon}_{2,2} - \hat{\varrho}^{\epsilon}_{3,3})^2 + 4|\hat{\varrho}^{\epsilon}_{2,3}|^2}}{2},
\end{aligned}
\label{eq:VPout}
\end{equation*}
The expression for the LQFI is given by
\begin{equation}
LQFI(\hat{\varrho}^{\epsilon}_{out}) = 1-\max\left\lbrace \mathcal{M}^{\epsilon}_{x,x}, \mathcal{M}^{\epsilon}_{z,z}\right\rbrace, 
\label{eq:Fout} 
\end{equation}
with
\begin{equation*} 
\begin{aligned} 
\mathcal{M}^{\epsilon}_{x,x} &= \frac{64\mathcal{M}^{\epsilon}_{1}\mathcal{M}^{\epsilon}_{2}}{\mathcal{M}^{\epsilon}_{3}}, \qquad \mathcal{M}^{\epsilon}_{z,z} = 1 - \left(\frac{|\hat{\varrho}^{\epsilon}_{1,4}|}{\hat{\varrho}^{\epsilon}_{1,1}+\hat{\varrho}^{\epsilon}_{4,4}} + \frac{|\hat{\varrho}^{\epsilon}_{2,3}|}{\hat{\varrho}^{\epsilon}_{2,2}+\hat{\varrho}^{\epsilon}_{3,3}}\right), 
\end{aligned} 
\end{equation*}
and
\begin{equation*}
\begin{aligned}
\mathcal{M}^{\epsilon}_{1} &= \hat{\varrho}^{\epsilon}_{1,1}\hat{\varrho}^{\epsilon}_{3,3} + \hat{\varrho}^{\epsilon}_{2,2}\hat{\varrho}^{\epsilon}_{4,4} + \lambda_{1}\lambda_{2} + \lambda_{3}\lambda_{4} + 2|\hat{\varrho}^{\epsilon}_{1,4}\hat{\varrho}^{\epsilon}_{2,3}|,\\
\mathcal{M}^{\epsilon}_{2} &= (\hat{\varrho}^{\epsilon}_{1,1} + \hat{\varrho}^{\epsilon}_{4,4}) \lambda_{3}\lambda_{4} + (\hat{\varrho}^{\epsilon}_{2,2} + \hat{\varrho}^{\epsilon}_{3,3}) \lambda_{1}\lambda_{2},\\
\mathcal{M}^{\epsilon}_{3} &= \Big[1-(\lambda_{1}-\lambda_{2})^{2}-(\lambda_{3}-\lambda_{4})^{2}\Big]^{2} - 4(\lambda_{1}-\lambda_{2})^{2} (\lambda_{3}-\lambda_{4})^{2}.
\end{aligned}
\end{equation*}

Note that the LN, LQU, and LQFI of the non-teleported state $\rho_{\text{Y}\bar{\text{Y}}}^{\text{X}}$ were analyzed in Section~\ref{sec:3} and are not revisited in this section.
\begin{figure}[!h]
\begin{center}
\includegraphics[scale=0.32]{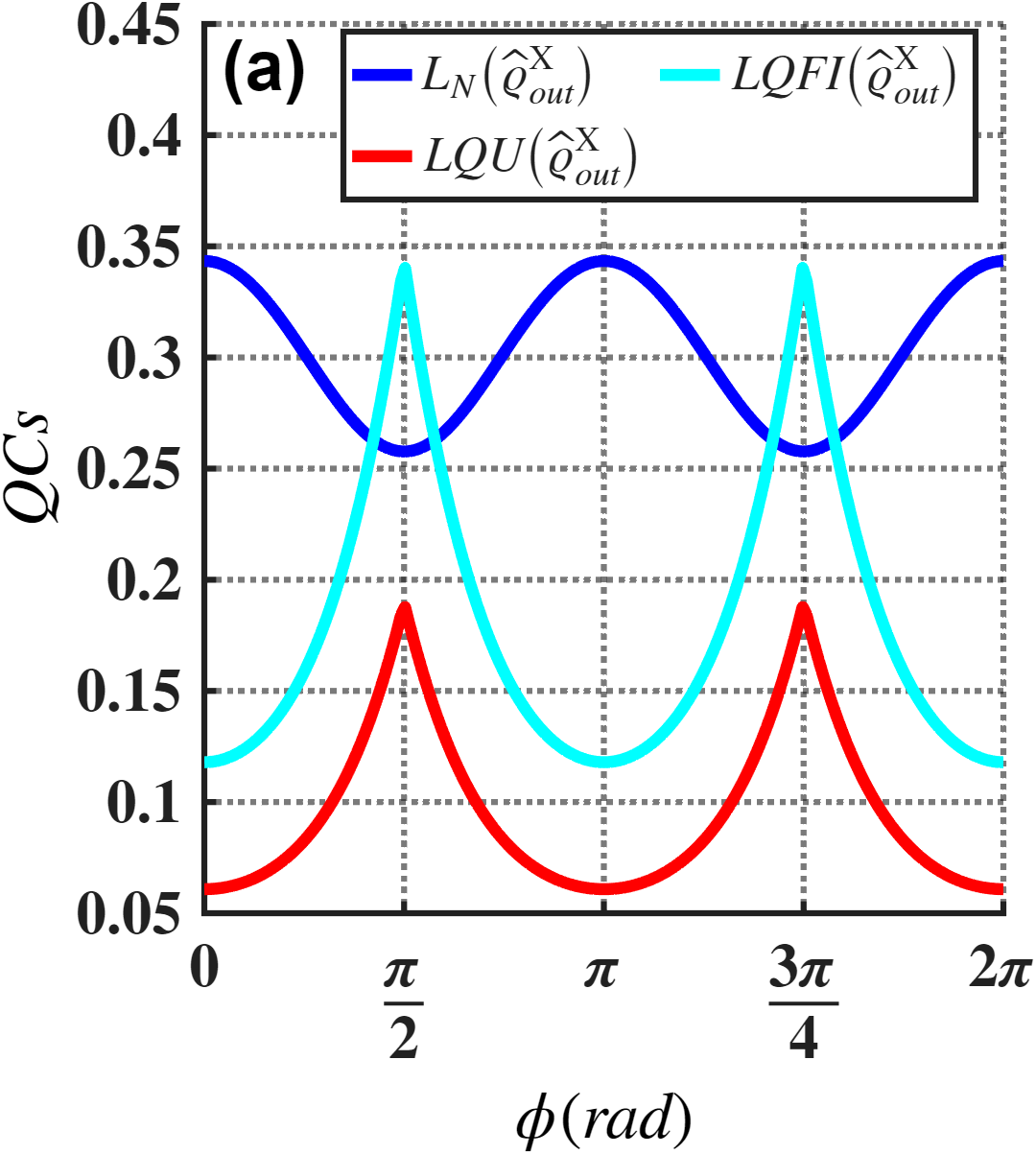}
\includegraphics[scale=0.32]{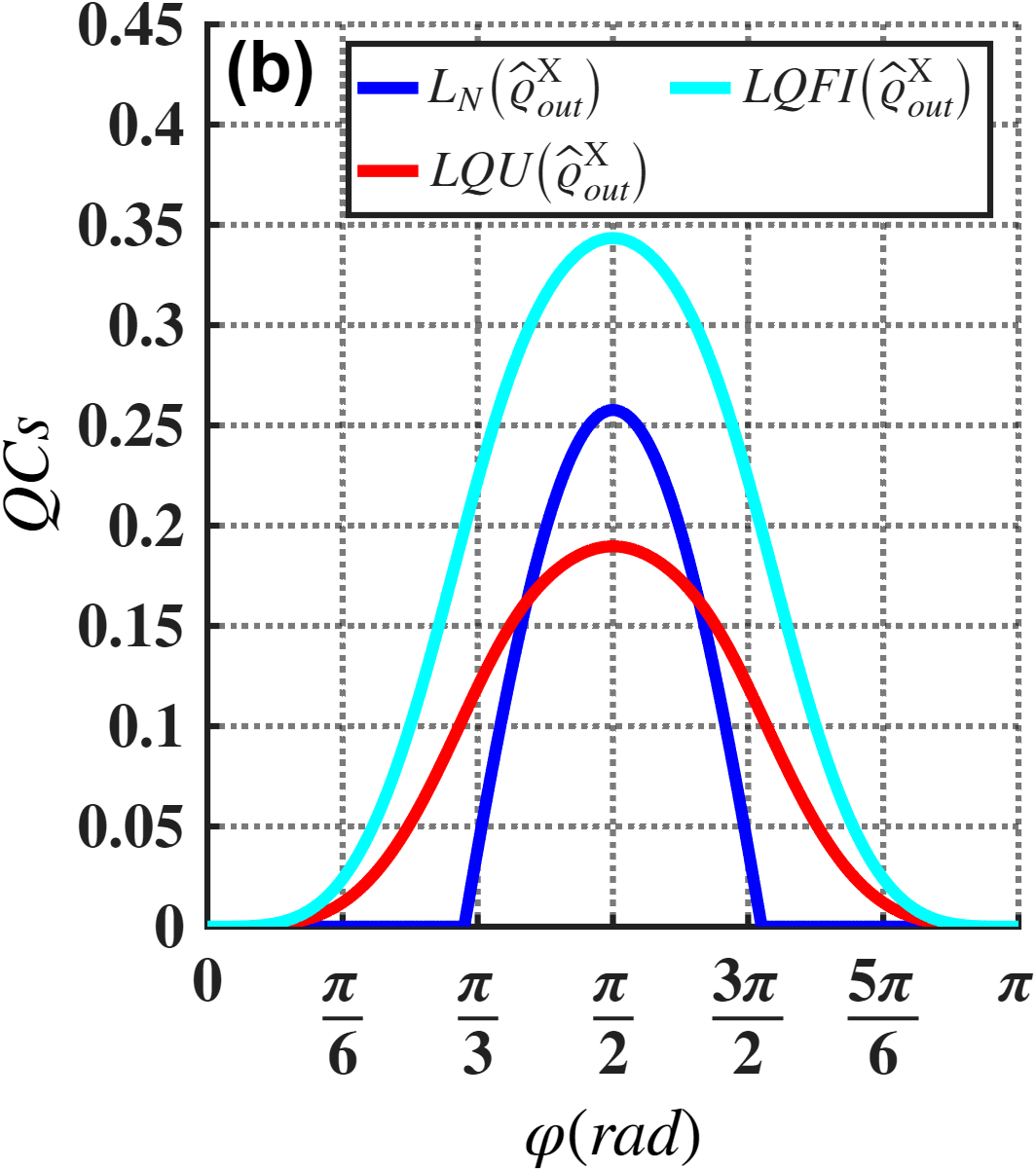}
\includegraphics[scale=0.32]{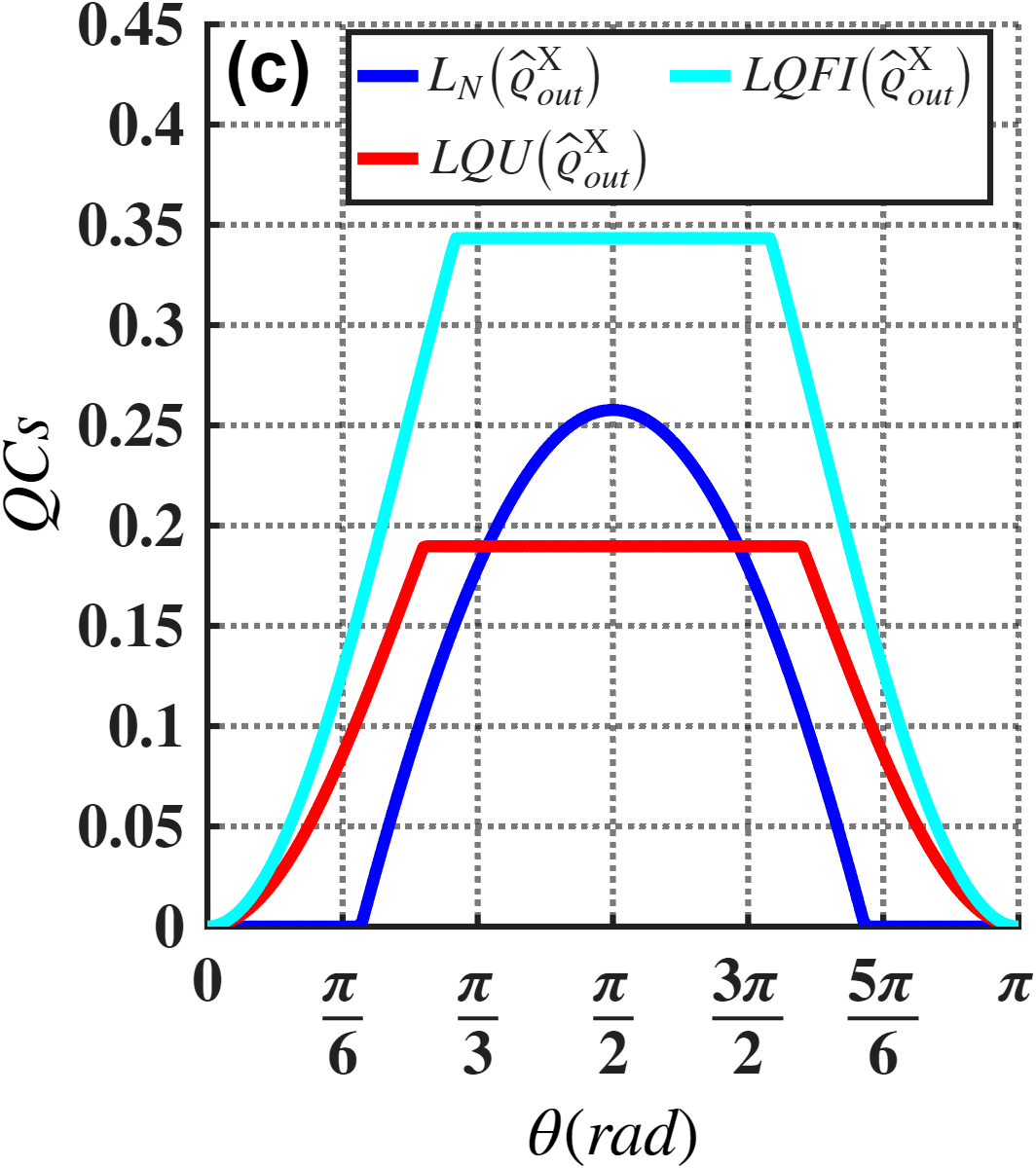}
\end{center}
\caption{Plot of the LN $L_{N}(\hat{\varrho}^{\text{X}}_{out})$, LQU $LQU(\hat{\varrho}^{\text{X}}_{out})$, and LQFI $LQFI(\hat{\varrho}^{\text{X}}_{out})$ as a function of: (a) the phase $\phi$, (b) the scattering angle $\varphi$, and (c) the amplitude angle $\theta$, with $\varphi=\theta=\pi/2$, in $\text{e}^+\text{e}^- \to \Xi^{-} \bar{\Xi}^{+}$ processes, utilizing the experimental parameters listed in Table \ref{tab1}.}
\label{fig:PTout}
\end{figure}

\begin{figure}[!h]
\includegraphics[scale=0.32]{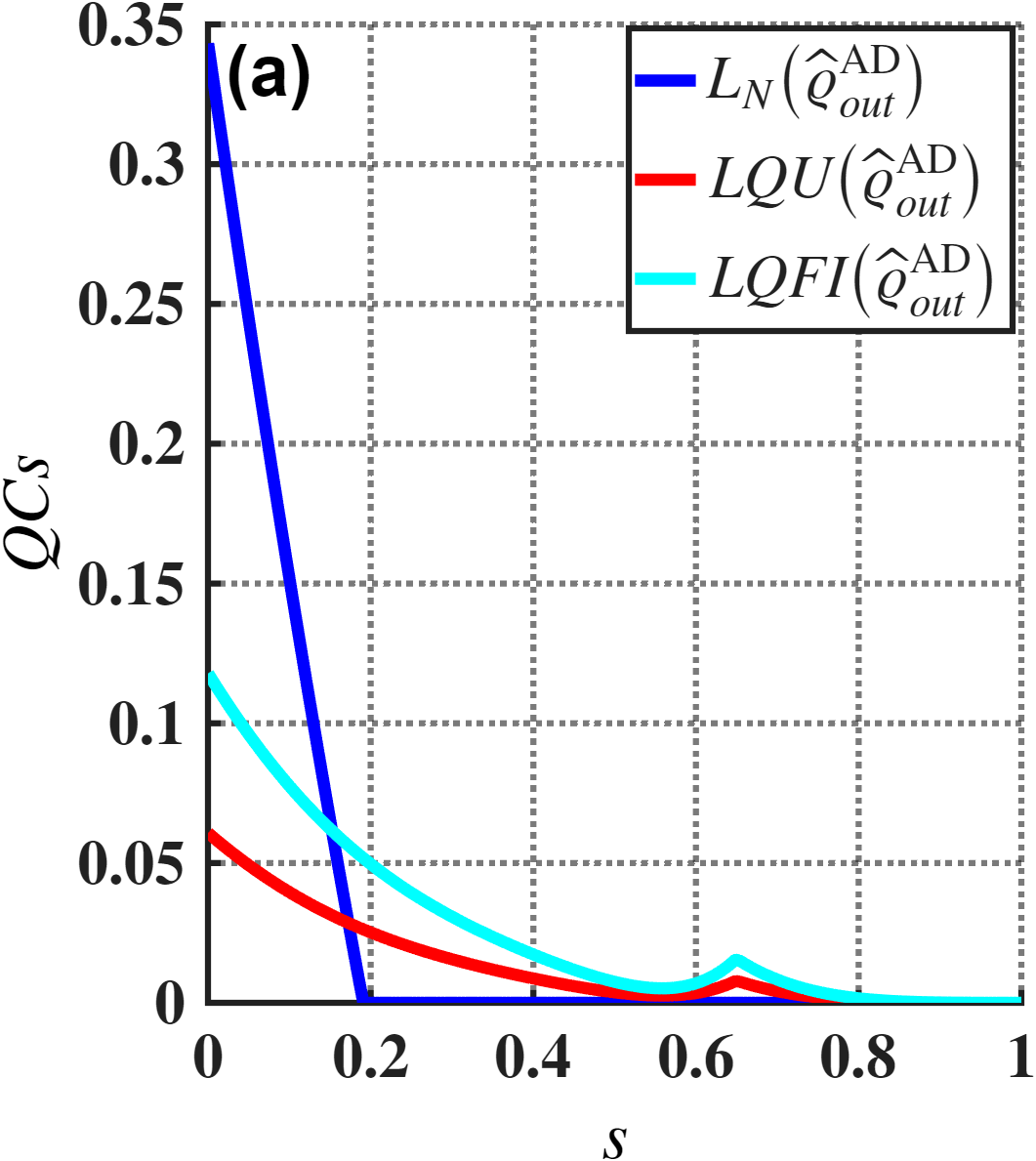}
\includegraphics[scale=0.32]{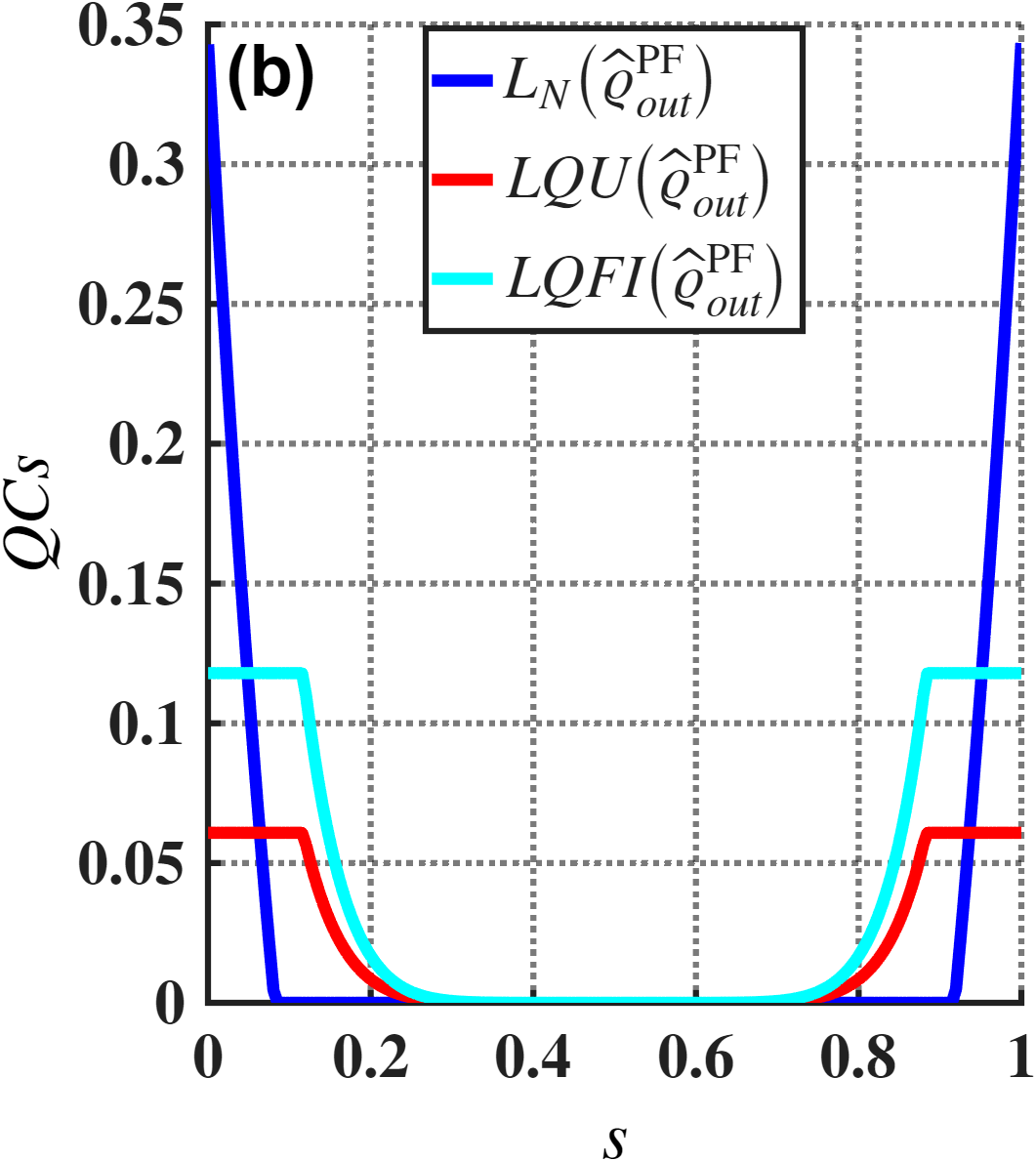}
\includegraphics[scale=0.32]{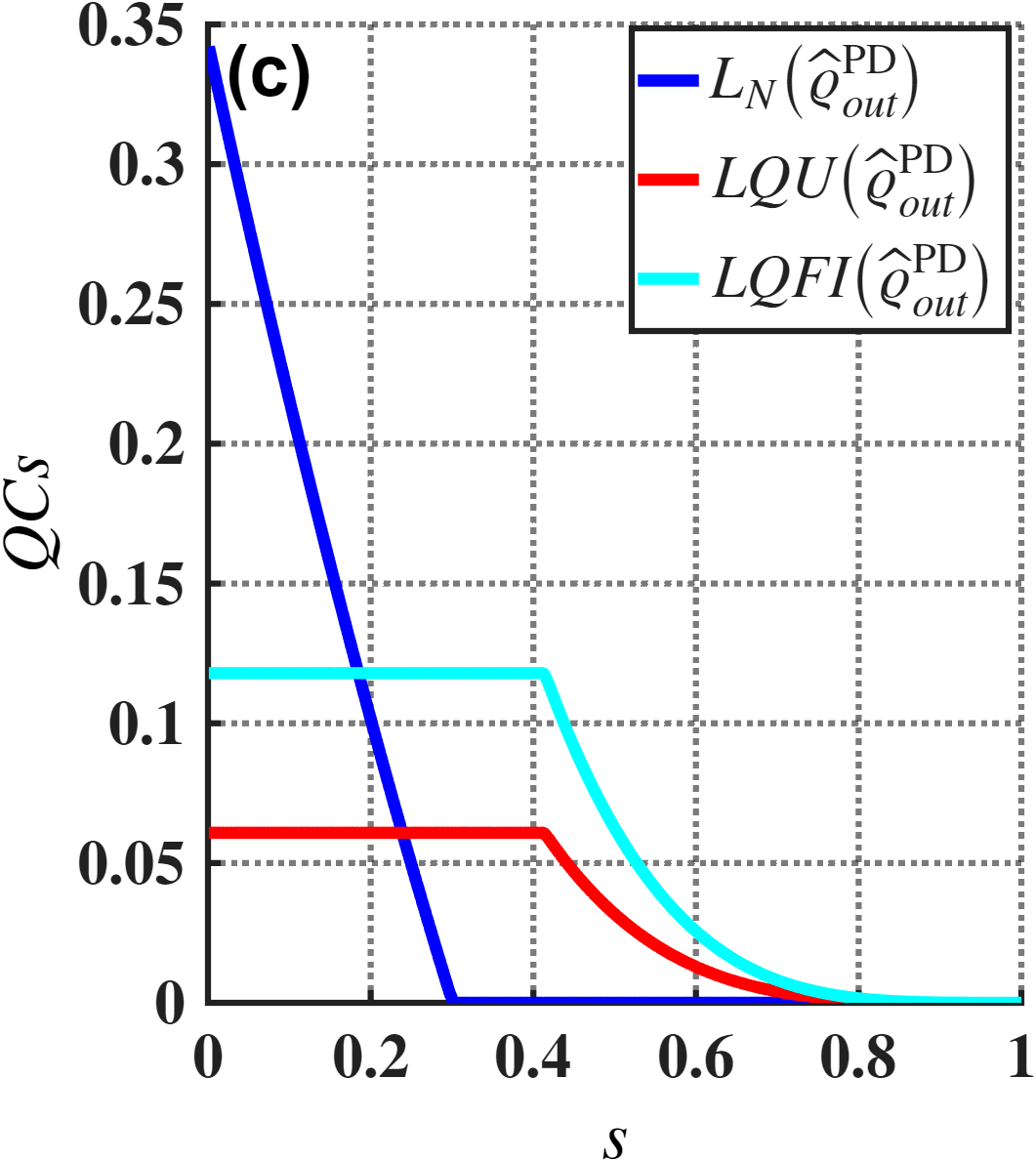}
\caption{Plot of the LN $L_{N}(\hat{\varrho}^{\epsilon}_{out})$, LQU $LQU(\hat{\varrho}^{\epsilon}_{out})$, and LQFI $LQFI(\hat{\varrho}^{\epsilon}_{out})$ versus decoherence parametre $s$, with $\varphi=\theta=\pi/2$ and $\phi=0$, in $\text{e}^+\text{e}^- \to \Xi^{-} \bar{\Xi}^{+}$ processes, under (a) Amplitude Damping (AD), (b) Phase Damping (PD), and (c) Phase Flip (PF) channels, utilizing the experimental parameters listed in Table \ref{tab1}.}
\label{fig:cm}
\end{figure}

In Fig.~\ref{fig:PTout}(a), we explore the hierarchy of quantum correlations between the LN, LQU and LQFI as a function of the phase $\phi$ for hyperon--antihyperon pairs $\Xi^{-}\bar{\Xi^{+}}$. The LN exhibits a sinusoidal and periodic dependence on $\phi$, while the LQU and LQFI also display periodic behavior. An essential feature lies in the phase opposition between the LN and the LQU/LQFI. More specifically, the LN reaches its maximum at $\phi = k\pi$ ($k = 0, 1, 2$), whereas the LQU and LQFI reach their maximum at $\phi = (k + \frac{1}{2})\pi$. This phase opposition, characterized by a shift of $\pi/2$, highlights a complementary relationship between LN, LQU, and LQFI, where the maximum of the $L_N$ coincides with the minimum of the LQU and LQFI, and vice versa.

In Fig.~\ref{fig:PTout}(b), we show the evolution of the LN, LQU and LQFI utilizing the teleported state $\hat{\varrho}_{out}^{\text{X}}$. These three measures peak at $\varphi = \pi/2$, indicating an optimal configuration due to the restructuring of quantum correlations by the teleportation protocol. In contrast, for the non-teleported state, LQU and LQFI reach their maximum at different $\varphi$ values for the $\rho^{\text{X}}_{\text{Y}\bar{\text{Y}}}$ state, as shown in Fig. \ref{fig:L} and Tables \ref{tab:t3} and \ref{tab:t4}. When LN vanishes, signaling separability, however LQU and LQFI remain non-zero, demonstrating their sensitivity to quantum correlations even beyond entanglement. We notice that LQU and LQFI evolve similarly, with their maximum at $\varphi = \pi/2$, and LQU always bounded by LQFI, reflecting their hierarchical relationship.

Fig.~\ref{fig:PTout}(c) illustrates the evolution of the LN, the LQU, and the LQFI as functions of the amplitude angle $\theta$ of the teleported quantum state, with the scattering angle fixed at $\varphi = \pi/2$. A symmetry of the LN is observed around $\theta = \pi/2$, where it reaches its maximum value, indicating maximal quantum entanglement at this state. Moreover, LQU and LQFI show similar behavior with remarkable stability in certain intervals of the amplitude angle $\theta$, as shown in Fig.~\ref{fig:PTout}(c). This stability reflects robustness against variations of the input state, which is a key feature for quantum protocols. Finally, it is worth noting that the LQU and the LQFI remain significant even when the LN vanishes, highlighting their ability to capture quantum resources beyond entanglement. Furthermore, the LQU is always bounded by the LQFI.

In Fig.~\ref{fig:cm}, we analyze the effect of the decoherence parameter \( s \) on the dynamics of quantum correlations (QCs) in the teleported state \( \Xi^{-}\bar{\Xi^{+}} \), across the AD, PF, and PD decoherence channels. It is observed that logarithmic negativity (LN) is more robust under the phase flip (PF) channel than under the amplitude damping (AD) and phase damping (PD) channels when $s \to 1$, whereas LN becomes zero under the AD and PD channels in this regime. Additionally, LN exhibits symmetric behavior under PF channels. Furthermore, when LN is zero, we find that both LQU and LQFI remain $>0$. The influence of the decoherence parameter reveals a hierarchy of quantum correlations, with $N_L \subseteq \text{LQU} \subseteq \text{LQFI}$, highlighting their differential sensitivity. These results indicate that teleported QCs are particularly sensitive to decoherence effects when the amplitude angle $\theta$ and phase $\phi$ are fixed, with maximum correlation observed at $\varphi = \pi/2$.

\section{Conclusion}\label{sec:7}

In summary, we have investigated the quantum correlations in the processes $e^{+}e^{-} \rightarrow \text{Y}\overline{\text{Y}}$ at the BESIII experiment, where $\text{Y}$ and $\overline{\text{Y}}$ represent a spin-1/2 hyperon and its antihyperon ($\Lambda\bar{\Lambda}$, $\Xi^{0}\bar{\Xi}^{0}$, $\Xi^{-}\bar{\Xi}^{+}$, $\Sigma^{+}\bar{\Sigma}^{-}$). We have quantified these correlations using logarithmic negativity (LN), local quantum uncertainty (LQU), and local quantum Fisher information (LQFI). Our analysis also established a hierarchy among these measures and explored their dependence on different parameters. Furthermore, the effects of quantum noise on non-teleported states, specifically LN, LQU, and LQFI, across the Amplitude Damping (AD), Phase Damping (PD), and Phase Flip (PF) channels is analyzed. The AD and PD channels caused a linear decline in LN as the decoherence strength $s$ increased, while the PF channel displayed a unique symmetric behavior around $s=0.5$. For teleported states, we found that LN, LQU, LQFI, and teleportation fidelity all peaked at $\theta=\pi/2$. While the overall performance of these measures was optimal at this angle, their resilience to noise varied. This suggests that the choice of decoherence channel significantly impacts the dynamics of quantum correlations in a teleportation protocol. The PD channel significantly degrades fidelity as $s \to 1$, while the AD channel has a more moderate effect. Remarkably, the PF channel can enhance fidelity beyond the classical limit of $2/3$ as $s$ approaches both 0 and 1, demonstrating that quantum noise, far from being universally detrimental, can in some cases mitigate decoherence and optimize quantum state transfer. These findings highlight the complexity and richness of interactions between quantum noise and quantum correlations. They open promising avenues for manipulating quantum states in noisy environments and underscore the potential of these discoveries for advancing quantum technologies and particle physics.

%\newpage
%----------------------------------------------------------------------------------------
%	REFERENCE LIST
%----------------------------------------------------------------------------------------

%---------Quantum teleportation fidelity---------------------

%\bibliography{references}

\end{document}